%% file: main.tex
 \documentclass[aps,pra,superscriptaddress,notitlepage,twocolumn,showpacs]{revtex4-1}
\usepackage[utf8]{inputenc}
\usepackage[T1]{fontenc}
\usepackage[usenames,dvipsnames]{xcolor}
\usepackage{amsmath}
\usepackage{amssymb}
\usepackage{mathtools}
\usepackage{esint}
\usepackage{dcolumn}
\usepackage{bm}
\usepackage{bbm}
\usepackage{blindtext}
\usepackage{braket}
\usepackage{natbib}
\usepackage{graphicx}

\usepackage[normalem]{ulem}
\usepackage{enumerate}
\usepackage{xpatch} 

\usepackage{multirow}
\usepackage[hidelinks=true,pdfborder={0 0 0},colorlinks=true,linkcolor=blue,urlcolor=blue,citecolor=blue]{hyperref}

\newcommand{\be}{\begin{eqnarray}}
\newcommand{\ee}{\end{eqnarray}}

\newcommand{\btext}[1]{#1}
\newcommand{\otext}[1]{\textcolor{black}{{#1}}}

\definecolor{dgreen}{rgb}{0.0, 0.5, 0.0}

\renewcommand{\vec}[1]{\mathbf {#1}}

\begin{document}

%\title{The energy landscape of Trilobite Trimers}
%\title{The building principle of polyatomic trilobite molecules: from trimers to clusters}
%\title{The building principle of triatomic trilobite molecules}
\title{Ultralong-range Rydberg molecules}

\author{Christian Fey}
\affiliation{Zentrum f\"ur Optische Quantentechnologien, Universit\"at Hamburg, Fachbereich Physik,  22761 Hamburg, Germany}
\affiliation{Max-Planck-Institute of Quantum Optics, 85748 Garching, Germany}
\author{Frederic Hummel}
\affiliation{Zentrum f\"ur Optische Quantentechnologien, Universit\"at Hamburg, Fachbereich Physik,  22761 Hamburg, Germany}
\author{Peter Schmelcher}
\affiliation{Zentrum f\"ur Optische Quantentechnologien, Universit\"at Hamburg, Fachbereich Physik,  22761 Hamburg, Germany}
\affiliation{The Hamburg Centre for Ultrafast Imaging, Luruper Chaussee 149, 22761 Hamburg, Universit\"at Hamburg, Germany}

%Diatomic trilobite molecules are bound states between a ground state atom and a Rydberg atom in an electronic high angular momentum state whose probability density resembles the shape of a trilobite fossil. They posses huge bond lengths on the order of 1000 \AA ngstroms and are extremely sensitive to electric fields. We demonstrate that triatomic trilobite molecules (one Rydberg atom and two ground state atoms)

\date{\today}
\begin{abstract}
We review ultralong-range Rydberg molecules (ULRM), which are bound states between a Rydberg atom and one or more ground-state atoms with bond lengths on the order of thousands of Bohr radii. The binding originates from multiple electron-atom scattering and leads to exotic oscillatory potential energy surfaces that reflect the probability density of the Rydberg electron. This unconventional binding mechanism opens fascinating possibilities to tune molecular properties via weak external fields, to study spin-resolved low-energy electron-atom scattering as well as to control and to probe interatomic forces in few- and many-body systems. Here, we provide an overview on recent theoretical and experimental progress in the field with an emphasis on polyatomic ULRMs, field control and spin interactions.  %Progress of the field is presented in the context of recent theoretical and experimental achievement.   
\end{abstract}
\maketitle
\section{Introduction}
Ultralong-range Rydberg molecules (ULRM) represent an exotic molecular species formed by a Rydberg atom and one or more ground-state atoms. The exaggerated properties of the Rydberg atom equip ULRMs with huge bond lengths on the order of thousands of Angstroms and exceptionally large dipole moments up to the kiloDebye regime, while their binding energies are relatively weak and on the order of hundreds of neV. The origin of their bond lies in low-energy scattering between the Rydberg electron and the ground-state atom, which is substantially different from covalent or ionic bonding mechanisms. As a consequence, ULRMs exhibit oscillatory Born-Oppenheimer potential energy curves (PECs) that reflect the density of the Rydberg wave function.

Since their theoretical prediction in 2000 by C. H Greene, A. S. Dickinson and H. R. Sadeghpour \cite{greene_creation_2000} and their experimental discovery in 2009 \cite{bendkowsky_observation_2009}, many fascinating properties and applications have been revealed.
The main research directions include:
(1) control of molecular properties such as dipole moments or alignment via weak  external electric and magnetic fields \cite{lesanovsky_ultra-long-range_2006,kurz_electrically_2013,krupp_alignment_2014,gaj_hybridization_2015,niederprum_observation_2016,
Hummel_Fey_Schmelcher_spin_dstate_2018, Hummel_Fey_Schmelcher_2019_s_state_alignment}, (2) study of polyatomic ULRMs, where a Rydberg atom binds not only one but several ground-state atoms \cite{liu_polyatomic_2006, liu_ultra-long-range_2009,bendkowsky_rydberg_2010,gaj_molecular_2014,eiles_ultracold_2016,fey_stretching_2016, Luukko_Rost_2017,Fey_2019_effective_three_body, Fey_Hummel_Schmelcher_2019_trilobite} and forms polaron states in the many-body limit  \cite{schmidt_mesoscopic_2016,schlagmuller_probing_2016,camargo_creation_2018, Kleinbach_2018}, 
%(3) realization and investigation of polarons consisting of single Rydberg atoms or a single ion impurities immersed in a sea of ground-state atoms \cite{schmidt_mesoscopic_2016,schlagmuller_probing_2016,camargo_creation_2018, Kleinbach_2018},
(3) probing of spatial correlations in ultracold atomic gases \cite{manthey_dynamically_2015,Whalen_2019, Whalen_Schmidt_Wagner_2019}, and (4) characterization of low-energy electron-atom collisions \cite{anderson_photoassociation_2014,sasmannshausen_experimental_2015,bottcher_observation_2016,MacLennan_Chen_Raithel_2018,
Engel_Fey_Meinert_2019}. In this review we provide a brief introduction of the molecular binding mechanism and present key findings of the different research directions in the context of experimental observations and theoretical models with a focus on molecular aspects. For alternative reviews on ULRMs, see \cite{Shaffer_2018,marcassa_interactions_2014}, the tutorial \cite{Eiles_2019_exotic_specimen} as well as reviews adressing particular subtopics of ULRMs  
\cite{Gaj_Review_2016, Liebisch_2016,Sassmannshausen_Review_2016, Lippe_Eichert_Review_2019}.
% provided in , in the earlier review \cite{}, as well as in the tutorial \cite{} that has a pronounced focus on the theoretical description. Other reviews discuss particular aspects such a field control \cite{Gaj_Review_2016}, ULRM in dense gases \cite{Liebisch_2016} or summarize works performed in particular groups \cite{Sassmannshausen_Review_2016, Lippe_Eichert_Review_2019}.
After a short discussion of the historical precursors of ULRMs in the early 20th century in Sec. I, we present basic molecular properties of ULRMs in Sec. II. Thereafter, we focus on polyatomic ULRMs in Sec. III, field control in Sec. IV and spin-interactions in 
Sec. V. We end the review with our conclusions in Sec. VI that contain open questions as well as future perspectives.

\section{Historical context}
 
In the 1930s E. Amaldi and E. Segr\'e investigated spectra of highly excited alkali atoms in the presence of perturbing noble gas atoms \cite{Amaldi_1934, Amaldi_1934_nature} in Rome. Similar experiments were conducted simultaneously by C. F\"uchtbauer and coworkers in Rostock \cite{Fuechtbauer_1934,Fuechtbauer_1934_2}.  
While it was expected that the polarizable medium would cause a red shift of the atomic resonances, the experiments revealed that both red and blue shifts are possible depending on the atomic buffer gas species and its density.
E. Fermi successfully explained this effect by describing the interactions between the Rydberg electron and the buffer gas atom, as a low-energy scattering process \cite{fermi_sopra_1934,Takeo_1957}.  According to this model the frequency shift is proportional to the buffer gas concentration and the zero-energy $s$-wave scattering length of the electron-atom system. This scattering length can be negative or positive and explains, hence, the occurence of blue shifts. The theory of Fermi was not only in very good agreement with the observations but also allowed experimentalists to extract scattering lengths from measured line shifts \cite{Fuechtbauer_1934_2,Takeo_1957}. 
Since then the Fermi pseudopotential has been an important ingeredient for studies of collisional processes of Rydberg atoms in gaseous environments \cite{Takeo_1957,Allard_1982,Beigman_Lebedev_1995} with applications such as the reconstruction of interatomic potentials or the analysis of the composition of interstellar atmospheres \cite{Allard_1982,Stebbings_Dunning_2011}.    

%These first studies of interactions between Rydberg atoms and ground-state atoms laid the foundation for intense investigations of collisional processes of Rydberg atoms in gaseous environments carried out in the second half of the 20th century\cite{Takeo_1957,Allard_1982,Beigman_Lebedev_1995}. 
% Fostered by the development of tunable laser systems, this research focused on collisional broadening and the variation of line profiles as well as collisional transfer of Rydberg states, e.g.  $(n,l)\rightarrow (n',l')$.
%Applications of these findings cover the experimental reconstruction of interatomic potentials, the development of wavelength standards as well as the analysis of the composition of interstellar atmospheres \cite{Allard_1982,Stebbings_Dunning_2011}. The Fermi pseudopotential was in this context broadly applied and further generalized by A. Omont in 1977  \cite{omont_theory_1977} in order to capture also higher-partial wave scattering, e.g., $p$-wave interaction.

\begin{figure*}
\includegraphics[width= 0.8 \linewidth]{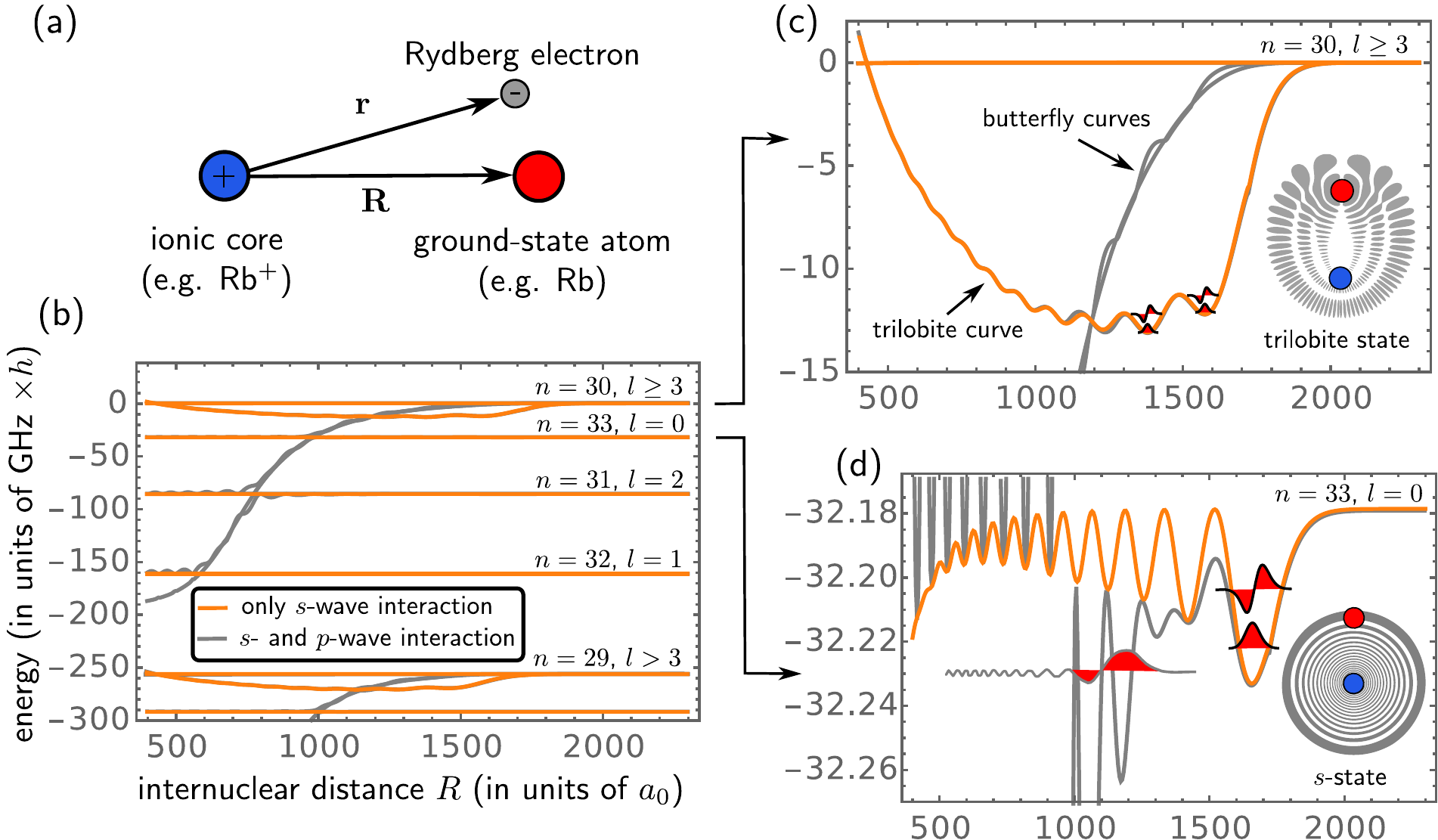}
\caption{(a) Schematic illustration of an ultralong-range Rydberg molecule that consists of a Rydberg atom (ionic core and valence electron) and a ground state atom. (b) The Born-Oppenheimer PEC for two $^{87}$Rb atoms illustrate the impact of the ground state atom on the electronic Rydberg levels with quantum numbers $n$ and $l$ as a function of the internuclear distance $R$. The underlying interaction between the electron and the ground-state atom is modeled via $s$-wave interactions (orange curves) or $s$- and $p$-wave interactions (gray curves). The zero energy is set to the energy of the hydrogen-like $n=30$ manifold. (c) and (d) are magnifications of the regions close to the atomic asymptote of $n=30$, $l>3$ and  $n=33$, $l=1$, respectively. Minima of the PEC support several bound vibrational states (indicated by black curves with red filling) as well as resonances bound by quantum reflection (gray curve with red filling). Gray contour plots illustrate the typical electronic density of the molecular states.}
\label{fig:dimer_basics}
\end{figure*} 
 
With the advent of novel techniques to cool and to trap atoms \cite{Phillips_1998} at the end of the 20th century it became possible to reach ultracold temperatures below 0.5 $\mu$K for ensembles of thousands,  if not millions, of atoms and to realize Bose-Einstein condensates  \cite{Ketterle_2002}. In the context of these developments C.~H. Greene, A.~S. Dickinson and  H.~R. Sadeghpour revealed an unexpected consequence of the Fermi pseudopotential theory when applied to ultracold atomic gases. 
They realized that the oscillatory probability density of the Rydberg electron can act effectively as a trap on surrounding ground-state atoms and binds atoms to the density maxima. 
%The binding is, however, very weak and works only at temperatures below 1 mK when atomic motion becomes sufficiently slow.
Due to the size of the Rydberg orbit these molecules possess huge bond length, on the order of thousands of Bohr radii $a_0$ and are for this reason called ultralong-range Rydberg molecules. Importantly, ULRMs need to be distinguished
%The binding mechanism is very different from conventional covalent or ionic binding and leads to exotic properties, such as an osillatory shape of the Born Oppenheimer potential energy surfaces (PES) and dipole moments on the order of kilodebye, which impressively surpass dipole moments of ground-state molecules that typically range from 0 to 10 debye.  
from traditional Rydberg molecules \cite{Stebbings_Dunning_2011} that possess highly excited electronic states but only small bond lengths.

\section{Basic molecular properties}
\label{subsec:dimers}

\subsection{Overview}
The basic properties of diatomic ULRMs are illustrated in Fig.~\ref{fig:dimer_basics}(a)-(d). As shown in Fig. \ref{fig:dimer_basics}(a), ULRMs consist of a ground-state atom and a Rydberg atom. The latter possesses an ionic core and an excited valence electron that interacts with the ground-state atom, which is sometimes also called 'perturber'. Exemplary Born-Oppenheimer PECs for the case of a $^{87}$Rb ULRM \btext{in the vicinity of the $^{87}$Rb($33s$) level} are presented in Fig.~\ref{fig:dimer_basics}(b).
%The basic properties of ULRM are illustrated in Fig.~\ref{fig:dimer_basics}, which includes a pictorial representation of the system (a) as well as the typical Born-Oppenheimer potential energy curves (PEC) for an example of a $^{87}$Rb ULRM close to the $n=30$ Rydberg state (b)-(d).
%%%%%%%%%%%%%%%%%%%%%%%%%%%%%%%%%%%%%%%%%%%%%%%%%%%%
At large internuclear separations, where interactions between the atoms are negligible, here for $R>1900$ Bohr radii $a_0$, the \btext{PECs} are nearly flat and correspond to energy levels of the isolated Rydberg atom with principal quantum number $n$ and angular momentum $l$. These energies can be expressed by the Rydberg formula $E_{nl}=-1/(2(n-\delta_l)^2)$ with an $l$-dependent quantum defect $\delta_l$ that describes deviations from the hydrogen spectrum due to core penetration. In this context one distinguishes low-$l$ states with substantial quantum defects from high-$l$ hydrogen-like states, which are shielded from the ionic core by a sufficiently large centrifugal barrier and have almost vanishing quantum defects. For instance the Rydberg spectrum of $^{87}$Rb is well approximated\cite{Gallagher_quantum_defects_2003} by $\delta_0=3.13$, $\delta_1=2.65$, $\delta_2=1.35$, and $\delta_{l\geq3}=0$, which leads to the ordering of energy levels depicted in Fig. \ref{fig:dimer_basics}.~(b).

When the ground state atom overlaps with the electron cloud of the Rydberg atom, the energies of the Rydberg electron become perturbed.
This is visible in the PECs in Fig.~\ref{fig:dimer_basics}(b) for internuclear separations $R<1900$ $a_0$, where electron-atom interaction is modeled via $s$-wave scattering (orange) or via $s$- and $p$-wave scattering (gray).  \btext{In this notation $s$- and $p$-wave refer to the angular momentum of the Rydberg electron with respect to the ground-state atom at position $\vec{R}$ which should not be confused with the angular momentum $l$ with respect to the ionic core, see Sec. \ref{sec:spin effects} and Fig. \ref{fig:spin_sketch} for more details.}  
For pure $s$-wave interactions one distinguishes two types of URLM originally predicted in \cite{greene_creation_2000}: trilobite molecules and low-$l$ ULRMs.
The former are correlated to the hydrogenic manifold and possess typically several GHz$\times h$ deep PEC, see Fig.~\ref{fig:dimer_basics}(c). Their name refers to the shape of their electronic density (gray contour plot) which resembles a trilobite fossil. 
Illustratively speaking, the trilobite function is the superposition of all hydrogenic states which maximizes its density on the ground state atom.
Due to this property, trilobite ULRMs are polar molecules that possess huge permanent dipole moments on the order of kiloDebye. 
The second type, low-$l$ ULRMs, are associated to the corresponding low-$l$ Rydberg levels. They possess approximately thousand times shallower PEC with  vibrational energy spacing on the order of 10 MHz$\times h$. As an example the PEC of $s$-state ULRMs are presented in Fig.\ref{fig:dimer_basics}(d). The corresponding electronic state is to a good approximation isotropic (gray contour plot) but can possess small \btext{permanent} dipole moments on the order of one Debye upon admixture of trilobite character \cite{li_homonuclear_2011}. Accordingly, the denotation $l$-state ULRM does not imply that the Rydberg electron is in a pure $l$ state but only that the molecule correlates asymptotically to the atomic $l$-state Rydberg line. The PEC of both types of ULRMs support several quantized vibrational states (indicated by red filled wave functions). In spite of the comparably slow electronic motion of the highly excited Rydberg electron, the Born-Oppenheimer approximation remains typically valid, since the energy spacing of adjacent Rydberg levels is usually orders of magnitude larger than the vibrational energy spacing of the nuclei.

The inclusion of \btext{additional} $p$-wave interactions has two main impacts. Firstly, it gives rise to yet another type of polar ULRM, which are called butterfly molecules \cite{hamilton_shape-resonance-induced_2002,chibisov_2002_pwave}. Again, the name refers to the shape of the electron density. Remarkably, their dipole moments are slightly larger than those of trilobite molecules, while their bond lengths are much shorter, e.g. $R<500$ $a_0$ for $n=30$ \cite{hamilton_shape-resonance-induced_2002}.
The second consequence of $p$-wave interactions are modifications of the PECs of trilobite molecules and $s$-state ULRMs.
Due to an underlying $p$-wave shape resonance, which is a general feature of electron interaction with alkaline atoms \cite{Thumm_Norcross_1991_evidence_for, bahrim_low-lying_2000,Eiles_2018_two_component}, the associated butterfly PECs cross all other Rydberg levels, see Fig.~\ref{fig:dimer_basics}(b)-(c).
For the trilobite molecule this crossing is rather sharp and the shape of the trilobite PEC remains almost preserved, whereas for $s$-state ULRMs there are substantial modifications at inner radii. In particular the shape resonance induces a steep potential drop, visible at $R\approx1000$ $a_0$ in Fig.~\ref{fig:dimer_basics}(d) that acts on vibrational states effectively as a boundary and leads to the formation of vibrational resonances, see exemplary gray wave function in Fig.~\ref{fig:dimer_basics}(d).  Illustratively speaking, these resonances are bound by internal quantum reflection at the potential drop \cite{bendkowsky_rydberg_2010}.

\subsection{Experimental observations}
All of the three types of ULRMs have been realized experimentally
\cite{bendkowsky_observation_2009,booth_production_2015,niederprum_observation_2016}.  
The first to be observed were $s$-state ULRMs with principal quantum numbers ranging from $n=34$ to $n=40$. \cite{bendkowsky_observation_2009}. Starting from an ultracold sample of trapped $^{87}$Rb ground state atoms, the molecules were created using a two-photon photoassociation scheme and, subsequently, detected via field ionization. 

Measured molecular lifetimes were on the order of tens of $\mu$s and significantly lower than the corresponding atomic lifetimes \cite{bendkowsky_observation_2009,butscher_atommolecule_2010}.
Subsequent studies in samples with variable atomic densities revealed that this reduction 
can be attributed to two decay mechanisms: firstly, molecule-atom collisions, with a collision rate that is proportional to the atomic density and, secondly, leakage of quantum states toward smaller internuclear separations caused by the crossing between the $s$-state PEC and the butterfly PEC \cite{Butscher_2011_lifetimes}.

Dipole moments have been quantified by perfoming Stark spectroscopy in weak electric fields. The observed linear Stark shifts demonstrate that $s$-state ULRMs can possess permanent dipole moments on the order of 1 Debye \cite{li_homonuclear_2011}. The polarization of the electronic wave function is caused by a small admixture of trilobite character to the otherwise isotropic $s$-state. Arguments based on exchange and parity symmetry forbid the existence of such permanent dipole moments for homonuclear molecules and imply a quadratic Stark shift. 
However, for the case of ULRMs the energy splitting of the electronic $^3\Sigma_g$ and $^3\Sigma_u$ orbitals is negligible and the rotational splitting is very small, typically on the order of 10 KHz $\times$ $h$. Compared to the experimentally relevant field strengths on the order of a few V/cm, molecular states of different parity are, therefore, energetically degenerate and exhibit linear Stark shifts \cite{li_homonuclear_2011, sadeghpour_how_2013}.

Following these first experiments with \btext{$^{87}$Rb(n$s$) ULRMs, further ULRMs with different atomic species and isotopes ($^{133}$Cs, $^{84}$Sr, $^{87}$Sr, $^{85}$Rb) as well as different electronic states ($p$ and $d$ states) have been realized \cite{tallant_observation_2012, desalvo_ultra-long-range_2015,camargo_lifetimes_2016,anderson_photoassociation_2014,
krupp_alignment_2014,manthey_dynamically_2015,MacLennan_Chen_Raithel_2018,
Whalen_2019,Whalen_Schmidt_Wagner_2019}. These modifications offer some degree of control of the molecular properties.} For instance there is no $p$-wave shape resonance in Sr ULRMs and the molecular lifetimes are not limited by the above mentioned leakage mechanism and can become comparable to those of their parent Rydberg atoms. Another example is the particularity of Cs atoms that the Rydberg $s$-state lies energetically very close to an adjacent hydrogenic manifold, due to an almost integer quantum defect of $\delta_0=4.05$. As a consequence, there is a crossing between the PECs of the trilobite and the $s$ state in Cs ULRMs. 
Although trilobite states are high angular momentum states and conventionally not accessible by one- or two-photon transitions due to angular momentum conservation, they possess in Cs an increased admixture of $s$-state character due to this crossing. This insight led to the excitation of trilobite ULRMs with huge dipole moments on the order of 2 kiloDebye \cite{booth_production_2015}.
In a similar spirit butterfly molecules in Rb with dipole moments on the order of 500 Debye have been excited close to a crossing with the $25p$-state PEC \cite{niederprum_observation_2016}. Furthermore electric fields were used to align the butterfly molecules and deduce their bond lengths from the observed rovibrational splittings.
\btext{Finally, intriguing isotopic modifications include deformations and splittings of the PES due to different hyperfine structures \cite{MacLennan_Chen_Raithel_2018}, see Sec. \ref{sec:spin effects}, as well as alternating excitation strengths reflecting the fermionic or bosonic correlations of the intial atomic cloud \cite{Whalen_2019, Whalen_Schmidt_Wagner_2019}.}

\subsection{Theoretical description}
\label{sec: dimers_theoretical_description}
A simple yet very accurate effective Hamiltonian for the Rydberg electron is given by
\begin{equation}
H=H_0(r) + V_{ca}(R) + V_{ea}(|\vec{r}-\vec{R}|) ,
\label{eqn:Hamiltonian_dimer}
\end{equation}
 with the internuclear axis $\vec{R}$ and the electron position $\vec{r}$, see Fig. \ref{fig:dimer_basics} (a). $H_0$ describes the isolated Rydberg atom, i.e. the kinetic and potential energy of the Rydberg electron in the field of the ionic core. Working in atomic units (a.u.), its eigenstates have energies given by the Rydberg formula $E_{nl}=-1/(2(n-\delta_l)^2)$ and wave functions $\varphi_{nlm}(\vec{r})$, where $n$, $l$, and $m$ are the principal quantum number, the angular momentum, and the magnetic quantum number, respectively. Using quantum defect theory the wave functions can be expressed in terms of appropriate superpositions of regular and irregular Coulomb wave functions that are related to Whittaker functions \cite{Eiles_2019_exotic_specimen}. Alternatively, one obtains $\varphi_{nlm}(\vec{r})$ by employing suitable model potentials for the electron-ion interaction and by solving the resulting radial Schrödinger equation numerically \cite{Marinescu_Sadeghpour_Dalgarno_1994}.
 
  The long-range interaction between the ionic core and the polarizable ground-state atom is given by $V_{ca}(R)=-\frac{\alpha}{2R^4}$, with the polarizability $\alpha$. For instance the polarizability of the $^{87}$Rb(5$s$) ground state is $\alpha=319$ a.u. \cite{Holmgren_2010}.
Finally, $V_{ea}(|\vec{r}-\vec{R}|)$ is the interaction between the Rydberg electron and the ground-state atom. 
Following Fermi's idea this interaction can be expressed as a zero-range pseudopotential \cite{fermi_sopra_1934,omont_theory_1977,greene_creation_2000,hamilton_shape-resonance-induced_2002} 
\begin{align}
V_{ea}(|\vec{r}-\vec{R}|) =  &2 \pi a_s[k(R)] \delta(\vec{R}-\vec{r}) \nonumber \\ &+ 6 \pi a_p[k(R)] \overleftarrow{\nabla}_{\vec{r}} \cdot \delta(\vec{R}-\vec{r})  \overrightarrow{\nabla}_{\vec{r}},
\label{eqn:fermi_pseudopotential}
\end{align} 
where the arrows indicate that the gradient operators $\overleftarrow{\nabla}$ and $\overrightarrow{\nabla}$ act only on wave functions on the left and right hand side, respectively, see. (\ref{eqn:33s_dimer_energy}).  
The first term describes $s$-wave interactions while the second term includes additional $p$-wave interactions. The strength of these interactions depends on the low-energy scattering length and volume $a_s[k]$ and $a_p[k]$, which are linked to the phase shifts $\delta_s(k)$ and $\delta_p(k)$ of a free electron with wave number $k$ that scatters off the ground state atom via $a_s[k]=-\tan(\delta_s)/k$ and $a_p[k]=-\tan(\delta_p)/k^3$. In a semiclassical approximation the wave number $k$ at the position of the atom is determined by $k^2/2-1/R=E_{nl}$, where $E_{nl}$ is the energy of the Rydberg level close to the molecular PEC.
The phase shifts $\delta_s(k)$ and $\delta_p(k)$ are required as an input for the interaction potential and are typically extracted from computational solutions of two-active electron models \cite{bahrim_low-lying_2000, bahrim_3se_2001, khuskivadze_adiabatic_2002, Eiles_2018_two_component} in combination with analytical low-energy expansions based on modified effective range theory \cite{spruch_modification_1960, omalley_modification_1961, Fabrikant_1986}. The pseudopotential formalism is quite general and applies also to ULRMs consisting of complex multichannel Rydberg atoms such as Ca or Si \cite{eiles_ultracold_2015}.   
%$V_{ea}$ is a good approximation as long as the electronic de Broglie wave length $\lambda=2\pi/k$ and its variations are sufficiently small over the the effective range of the electron-atom interaction. This is typically the case, when the ground state atoms lies close to the outer turning points of the electron in the Coulomb potential.

Born-Oppenheimer PEC as depicted in Fig. \ref{fig:dimer_basics} can be obtained by diagonalizing the Hamiltonian $H_0$ in a finite set of basis states $\varphi_{nlm}(\vec{r})$ that cover a certain spectral window of interest or by using alternative Green's function methods \cite{khuskivadze_adiabatic_2002,hamilton_shape-resonance-induced_2002, bendkowsky_rydberg_2010,fey_comparative_2015,Engel_Fey_Meinert_2019} which can, additionally, be combined with R-matrix methods to take into account the finite range for the electron-atom interaction \cite{tarana_adiabatic_2016}.
An advantage of Green's function methods is their numerical reliability, since they are not subject to convergence issues which do currently limit the accuracy of the diagonalization approach \cite{fey_comparative_2015, eiles_hamiltonian_2017}. However, state-of-the-art Green's function approaches \cite{khuskivadze_adiabatic_2002, bendkowsky_rydberg_2010,tarana_adiabatic_2016, Engel_Fey_Meinert_2019} neglect essential interactions such as the fine structure of the Rydberg atom, the hyperfine structure of the ground-state atom as well as couplings to external fields, which are indispensable for a correct interpretation of high resolution spectra of ULRMs. So far only diagonalization approaches are able to capture all those effects \cite{eiles_hamiltonian_2017}.

To obtain basic molecular features it is sufficient to perform first order (degenerate) perturbation theory.
For instance the PEC close to the Rb($ns$) state is well captured by  
\begin{align}
E(R)= &E_{ns} -\frac{\alpha}{2R^4}+2 \pi a_s[k(R)] |\varphi_{ns0}(R)|^2 \nonumber \\ &+ 6 \pi a_p[k(R)] \vec{\nabla} \varphi_{ns0}(R) \cdot  \vec{\nabla} \varphi_{ns0}(R) . 
\label{eqn:33s_dimer_energy}
\end{align}
At sufficiently large distances $R$, the kinetic energy of the electron is small and $p$-wave interactions are of minor importance, see Fig. \ref{fig:dimer_basics}(d) for $R>1600 \, a_0$. In this regime the PECs are proportional to the electronic density of the Rydberg electron. It has been proposed that this peculiarity could be exploited to imprint and to measure images of electronic orbitals onto a BEC \cite{wang_rydberg_2015,karpiuk_imaging_2015}. 
Based on the properties of the Rydberg wave functions the depth of the $s$-wave dominated outer well scales as $n^{-6}$, which is in good agreement with experiments \cite{gaj_molecular_2014}.
For smaller separations, $p$-wave interaction gains in significance as the kinetic energy of the electron approaches the energy of a Rb$^-$ shape resonance at approximately 26 meV \cite{Thumm_Norcross_1991_evidence_for,Thumm_Norcross_1992, bahrim_low-lying_2000,khuskivadze_adiabatic_2002, Engel_Fey_Meinert_2019}.  Due to the resonance the $p$-wave scattering volume diverges and the first-order approximation (\ref{eqn:33s_dimer_energy}) clearly breaks down. However, when including couplings to adjacent Rydberg states, strong level repulsion stabilizes the PECs, such that the individual adiabatic PECs behave regular, i.e. they do not diverge, see Fig.~\ref{fig:dimer_basics}(b).  
The polarization potential $-\alpha/(2R^4)$ contributes to the PECs only at relatively small internuclear distances. Experimentally, its impact has been verified by exciting a single Rydberg atom in a Bose-Einstein condensate to high $n=190$, where the otherwise dominant electron-atom interaction is suppressed since the Rydberg orbit exceeds the size of the typical internuclear separations \cite{Kleinbach_2018}.

An approximate expression for the trilobite PEC can be obtained within (degenerate) first order perturbation theory \cite{greene_creation_2000}
\begin{equation}
E(R)= -\frac{1}{2n^2}-\frac{\alpha}{2R^4} + 2 \pi a_s[k(R)] \sum_{l>3} \frac{2l+1}{4\pi} R^2_{nl}(R) ,
\label{eqn:trilobite_dimer_energy}
\end{equation}
where $R_{nl}(r)$ is the radial wave function associated to $\varphi_{nlm}(\vec{r})$. Contributions of $p$-wave interactions are neglected in (\ref{eqn:trilobite_dimer_energy}). In addition to the energy curves also the trilobite wave function $\psi(\vec{r};\vec{R})$ can be expressed analytically within first order degenerate perturbation theory. The characteristic nodal structure of these states visible in Fig. \ref{fig:dimer_basics} can be understood semiclassically.
In a path integral picture $\psi(\vec{r};\vec{R})$ results from interference of electron states associated to elliptical Keppler trajectories that pass simultaneously through the points $\vec{r}$ and $\vec{R}$. 
Minima in the PECs arise when these tractories satisfy semiclassical Einstein-Brillouin-Keller quantization conditions which imply that the wave function has a certain integer number of nodes $n_\xi$ and $n_\eta$ along two different elliptical directions     \cite{Granger_2001,kanellopoulos_use_2009}. Interestingly, trilobite wave functions can also be realized by applying a sequence of electric and magnetic field pulses even when the ground-state atom is not present \cite{Eiles_ghost_trilobite_2018}.

Once the PECs $E(R)$ are determined, stationary vibrational states $\chi_\nu(R)$ with energy index $\nu$ can be deduced numerically by means of standard shooting methods or finite difference schemes.
Resonances that are bound by quantum reflection, see gray curve with red filling in Fig. \ref{fig:dimer_basics}(d), are in this context treated similiarly to shape-resonances. However, in the case of ULRMs these resonances result from inward scattering as opposed to conventional outward scattering \cite{bendkowsky_rydberg_2010,anderson_angular-momentum_2014}. 
The molecular line strengths of these states, i.e. the probability to excite $\chi_\nu(R)$ out of an ultracold atomic ensemble is proportional to the vibrational part of the Franck-Condon factor $\Gamma_\nu$. The states $\chi_\nu(R)$ are typically rather localized around the bond length $R_\nu$, in which case $\Gamma_\nu \approx \left|\int dR R^2 \chi(\vec{R})\right|^2 g^{(2)}(R_\nu)$, where $g^{(2)}(R_\nu)$ is the pair correlation function of the atomic gas \cite{desalvo_ultra-long-range_2015, Whalen_2019, Whalen_Schmidt_Wagner_2019}. This simple form of $\Gamma_\nu$ explains two essential properties of ULRMs. First, vibrational states $\chi_\nu(R)$ with an approximate gerade (ungerade) symmetry have in experiments strong (weak) signal \cite{desalvo_ultra-long-range_2015, Deiss_Fey_Hummel_Schmelcher_Denschlag_2019}. Second, excitation spectra of ULRMs provide access to pair correlation functions of the gas \cite{Whalen_2019, Whalen_Schmidt_Wagner_2019} which has for instance been used to dynamically monitor the \btext{Mott}-insulator to superfluid phase transition in an optical lattice \cite{manthey_dynamically_2015}.   

\section{Polyatomic ULRMs}
\label{sec:poly}

Since ULRMs are experimentally created in ultracold samples of trapped atoms, it is a natural situation, depending on the density of the gas, to have on average more than one ground-state atom within the size of the Rydberg cloud. 
In this case the Rydberg atom can bind several of these ground-state atoms to form polyatomic ULRMs \cite{liu_polyatomic_2006}.
The generalization of the dimer Hamiltonian (\ref{eqn:Hamiltonian_dimer}) that describes this situation is
\begin{equation}
H= H_0(r)+ \sum_j V_{ce}(R_j) + V_{ea}(|\vec{r}-\vec{R}_j|), 
\end{equation}
where $\vec{R}_j$ are the coordinates of the ground-state atoms.
Similar to diatomic ULRMs one distinguishes polyatomic high-$l$ ULRMs with trilobite-like wave functions and polyatomic low-$l$ ULRM.
First theoretical explorations of polyatomic ULRMs focused on high-$l$ ULRMs consisting of two, three, and four ground-state atoms \cite{liu_polyatomic_2006}. An important insight in this context was the insight that triatomic ULRMs can form Borromean states, where the trimer is stable although the corresponding dimers interact repulsively \cite{liu_ultra-long-range_2009}. Since angular selection rules hinder the direct one- or two-photon excitation of high-$l$ Rydberg states, most of the subsequent experimental and theoretical work concentrated, however, on low-$l$ ULRMs.
In addition, there is research on a second type of polyatomic ULRM consisting of a Rydberg atom that binds one or more polar molecules, e.g. KRb \cite{rittenhouse_ultracold_2010,rittenhouse_ultralong-range_2011,Gonzalez-Ferez_Sadeghpour_Schmelcher_2015,Aguilera_Gonzalez-Ferez_2015}, which is, however, not in the focus of this review.

\begin{figure}
\includegraphics[width= 0.27 \textwidth]{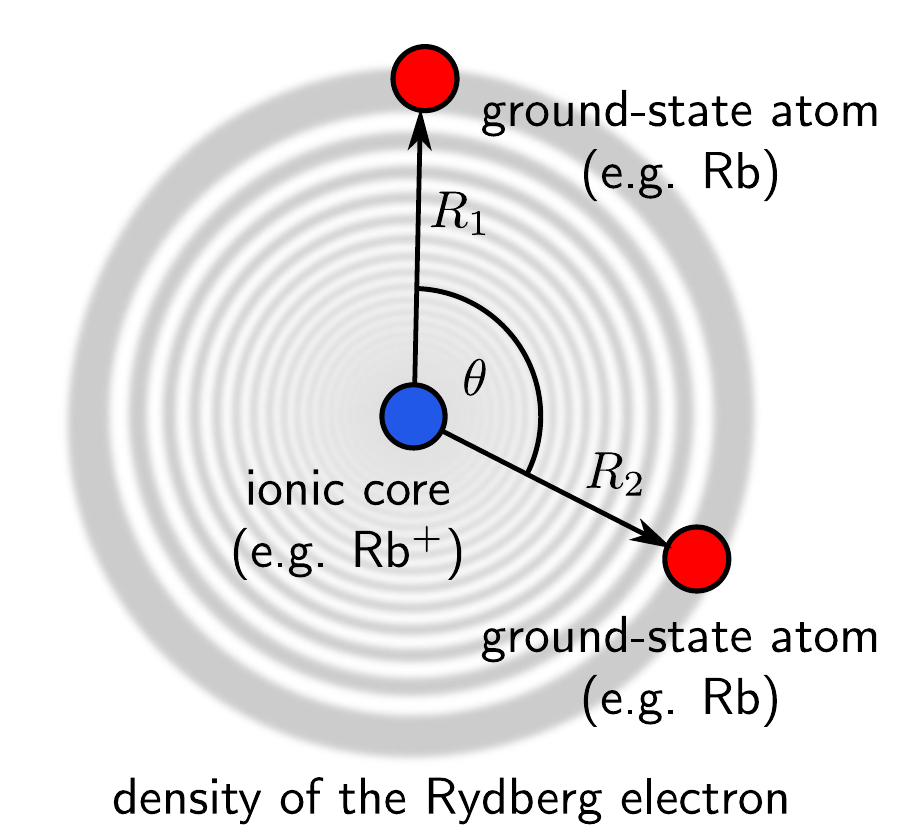}
\caption{Sketch of a triatomic ultralong-range Rydberg molecule in an electronic $s$ state.}
\label{fig:trimer}
\end{figure} 

% ULRM consisting of two, three, and four ground-state atoms were orginally predicted for high-$l$ ULRM \cite{liu_polyatomic_2006}. However, since angular selection rules hinder the direct one- or two-photon excitation of high-$l$ Rydberg states, most of the subsequent experimental and theoretical work focused on low-$l$ ULRM.

\begin{figure}[h!]
\includegraphics[width= 0.47 \textwidth]{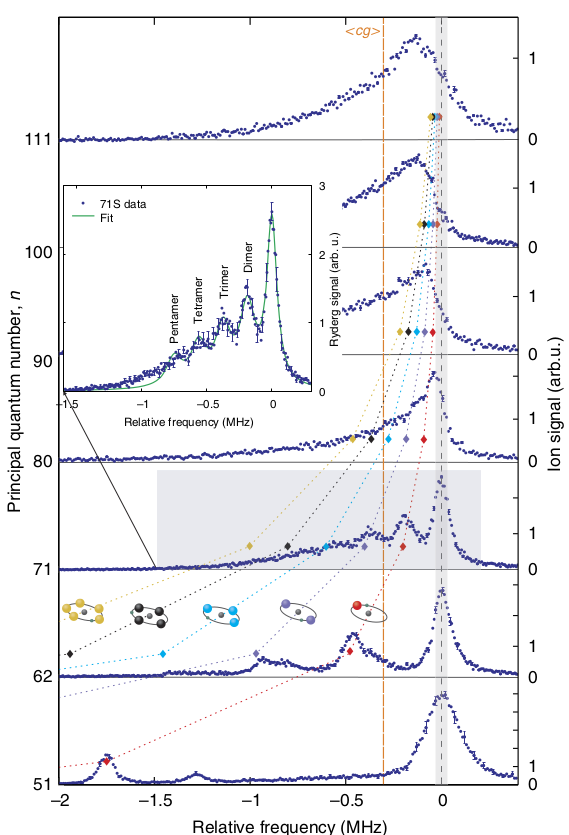}
\caption{Photoassociation spectra of polyatomic \btext{Rb($ns$)} ULRMs for different principal quantum numbers $n$ reprinted from \btext{A. Gaj et al. Nat. Comm. \textbf{5}, 4546 (2014)} \cite{gaj_molecular_2014}, where laser frequencies are expressed relative to the atomic $ns$ line.  Energies of trimers, tetramers, etc. are marked by colored diamonds and appear at integer multiples of dimer energies (red diamonds). While individual polyatomic lines can be resolved at low $n$, see inset for $n=71$, these lines merge for higher $n$ and give rise to a mean shift of the Rydberg line. For even larger $n \gg 111$ the spectrum is expected to become a Gaussian which localizes around the mean center of gravity <cg> that  coincides with the mean shift of the Rydberg line $\Delta E$ in (\ref{eqn: mean_field_shift_fermi}).}
\label{fig:gaj_poly}
\end{figure}

Spectral signatures of \btext{Rb($ns$)} trimers, see Fig. \ref{fig:trimer}, were reported shortly after the experimental discovery of ULRMs \cite{bendkowsky_rydberg_2010}. Ensuing observations confirmed additional tetramers, pentamers and hexamers \cite{gaj_molecular_2014}. Importantly, it was observed that binding energies of these polymers scale linearly with the number of ground-state atoms, i.e. the binding energies are integer multiples of dimer energies. An exemplary photoassociation spectra of polyatomic \btext{Rb($ns$)} ULRMs is presented in Fig. \ref{fig:gaj_poly}, where polyatomic lines are marked by colored diamonds. For instance the trimer (violet) has twice the dimer energy (red) because both ground-state atoms are trapped in the energetically lowest dimer mode.
This building principle has been later generalized to a shell-structure model \cite{schmidt_mesoscopic_2016}, where ground-state atoms are allowed to occupy not only the energetically lowest but also excited vibrational modes, see excited state in Fig. \ref{fig:dimer_basics}(d). Such polymers are not addressed in Fig. \ref{fig:gaj_poly} but have been confirmed experimentally in \cite{camargo_creation_2018}.   

The spectra in Fig. \ref{fig:gaj_poly} were taken at a constant atomic density of $\rho_0=10^{12}$~cm$^{-3}$ but variable principal quantum number $n$. While individual polyatomic few-body states can be well resolved for $n\leq 71$ one approaches for larger $n$ a many-body limit where the average number of ground-state atoms within the Rydberg orbit increases as $n^6$ while binding energies decrease as $n^{-6}$. In this regime the experimental spectra exhibit a single collisionally broadened Rydberg line instead of individual molecular resonances, e.g. for $n=111$ in Fig. \ref{fig:gaj_poly}.  In the mean-field limit, the broadened line will become a Gaussian peak with a certain detuning from the atomic Rydberg state --the so called pressure shift -- that was observed in the pioneering experiments \cite{Amaldi_1934, Amaldi_1934_nature, Fuechtbauer_1934,Fuechtbauer_1934_2} and interpreted by Fermi \cite{fermi_sopra_1934}. This shift is identical to the mean center of gravity <cg> and given by the average interaction energy with all ground-state atoms
\begin{align}
 \Delta E &= \int dR^3 dr^3 V_{ea}(|\vec{r}-\vec{R}|) |\varphi_{ns}(\vec{r})|^2 \rho(\vec{R}) \label{eqn:sampling} \\
 &\approx 2\pi a_s[0] \rho_0 ,
 \label{eqn: mean_field_shift_fermi}
\end{align}
where $\varphi_{ns}(\vec{r})$ is the Rydberg wave function and $\rho(\vec{R})$ the atomic density. Equation (\ref{eqn: mean_field_shift_fermi}) follows from assuming a constant density $\rho(\vec{R})=\rho_0$ and replacing the electron-atom interaction $V_{ea}$ by $s$-wave contact interactions (\ref{eqn:fermi_pseudopotential}) in the zero-energy limit, i.e. constant $s$-wave scattering length $a_s[0]$. Generalized versions of (\ref{eqn: mean_field_shift_fermi}) include energy dependent scattering lengths, $p$-wave contributions as well as inhomogenous densities and have been employed successfully to describe the spectral response of single Rydberg impurities immersed in a dense BEC \cite{balewski_coupling_2013}. An intriguing application of Rydberg impurities in ultracold atomic gases is the production of textbook-like images of atomic orbitals by imprinting the shape of the Rydberg orbital onto the BEC density, which has been proposed theoretically in \cite{wang_rydberg_2015, karpiuk_imaging_2015}.
Another application is the creation of ionic impurities in BECs, which can be excited in the limit of large principal quantum numbers, typically $n \sim 160$, when the Rydberg orbit exceeds the size of the atomic cloud \cite{Kleinbach_2018}.

Spectral lineshapes of Rydberg impurities in atomic gases as shown in Fig. \ref{fig:gaj_poly} can be modelled theoretically using two different approaches. In the first semiclassical approach the spectrum is obtained by sampling the detuning of the Rydberg line  $\Delta E$ in (\ref{eqn:sampling}) over many realizations of $N$ randomly distributed atoms at positions $\vec{R}_i$, such that the atomic density in the sample becomes $\rho(\vec{R})=  \rho_0/N \sum_i \delta(\vec{R}-\vec{R}_i) $ \cite{schlagmuller_probing_2016,perez-rios_mapping_2016}. The resulting histogram of the detuning $\Delta E$ agrees well with the overall lineshape of the pressure shifted Rydberg line in the mean-field limit and can be used to extract information on the $p$-wave scattering from experimental spectra \cite{schlagmuller_probing_2016}. However, the semiclassical approach does not yield molecular resonances which are relevant in the few-body regime.

An alternative fully quantum mechanical method that captures both, few-body and many-body features of polyatomic $s$-state ULRMs, is the functional determinant approach \cite{schmidt_mesoscopic_2016,camargo_creation_2018, schmidt_theory_2018}. The resulting spectra reproduce resonance energies and the lineshape of experimental signals very accurately. Furthermore, they explain beyond mean-field effects related to the collective polaron character of the system, which are not included in the semiclassical sampling approach \cite{schmidt_mesoscopic_2016,camargo_creation_2018, schmidt_theory_2018}.
% These corrections are related to the collective polaron character of the system and have been verified for Rydberg excitations in $^{84}$Sr BEC \cite{camargo_creation_2018}. 
 
%
%Sampling the detuning of the Rydberg line over many realizations of randomly distributed atoms yields not only the mean shift but also a characteristic shape of the Rydberg line, which contains for instance information on the $p$-wave shape resonance \cite{schlagmuller_probing_2016}. 
%%include here wave function imaging proposal...
%Additional higher order corrections that take into account beyond mean-field effects have been in the focus of \cite{schmidt_mesoscopic_2016,camargo_creation_2018, schmidt_theory_2018}. These corrections are related to the collective polaron character of the system and have been verified for Rydberg excitations in $^{84}$Sr BEC \cite{camargo_creation_2018}.
%An intriguing application of polaron systems is the realization of textbook-like images of atomic orbitals by imprinting the shape of the Rydberg orbital onto the BEC density, which has been proposed theoretically in \cite{wang_rydberg_2015, karpiuk_imaging_2015}.
%Furthermore, it has been recently achieved to prepare ionic impurities in the BEC, in the limit when the principal quantum numbers of the Rydberg atom are so high, typicall $n \sim 160$, that the Rydberg orbit exceeds the size of the atomic cloud \cite{Kleinbach_2018}.

\begin{figure}[h]
\includegraphics[width= 0.47 \textwidth]{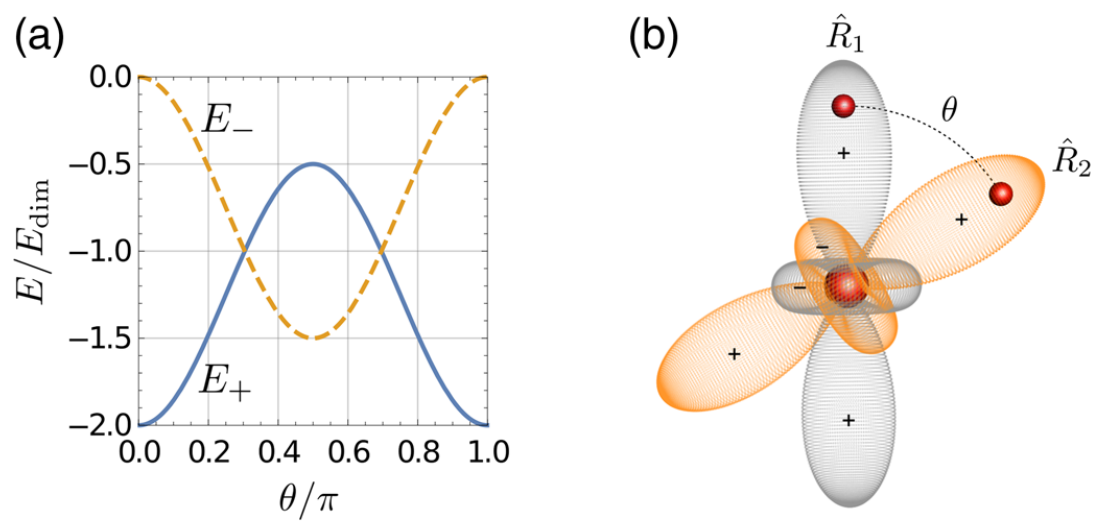}
\caption{Electronic structure of $d$-state trimers for pure $s$-wave interaction reprinted from \btext{C. Fey et al. Phys. Rev. Lett. \textbf{122}, 103001 (2019)} \cite{Fey_2019_effective_three_body}. (a) The trimer possesses two PESs $E_+$ and $E_-$, which are here shown as a function of the trimer angle $\theta$ for fixed radial distances of the ground-state atoms $R_1=R_2=R$. All energies are rescaled by $E_\text{dim}$, which is the energy of the diatomic PEC evaluated at $R$. For instance for the outer equilibrium well of the Cs($34d$) ULRM at $R=1868$ $a_0$ one has $E_\text{dim}=350$ MHz $\times h$. (b) The electronic eigenstates of the trimer are superpositions of primitive diatomic $d$-state orbitals (orange and gray lobes) that localize on the ground-state atoms at position $\vec{R}_1$ and $\vec{R}_2$, respectively.}
\label{fig:d_trimer}
\end{figure}

From a theoretical as well as from an experimental perspective Rydberg systems in $s$-states have certain benefits. The isotropy of the electronic states simplifies theoretical models and gives rise to comparably high experimental excitation efficiencies, due to favorable Franck-Condon factors for one- and two-photon transitions. 
In contrast, $l>0$ states have reduced Franck-Condon factors, are anisotropic and cover due to their degeneracy a larger Hilbert space, which makes ULRMs in those states more difficult to describe but leads also to richer physical properties.  
In particular the Born Oppenheimer potential energy surfaces (PES) of polyatomic ULRMs with $l>0$ become angular dependent, i.e. they depend not only on the radial distances of the ground-state atoms but also on their relative angles \cite{li_homonuclear_2011, eiles_ultracold_2016, fey_stretching_2016,Fey_2019_effective_three_body,Fey_Hummel_Schmelcher_2019_trilobite}. As a consequence of the increased number of nuclear degrees of freedom, many theoretical investigations of the PESs focused either on randomly distributed ground-state atoms \cite{eiles_ultracold_2016,perez-rios_mapping_2016,Luukko_Rost_2017}, or on constrained molecular geometries such as linear, planar-quadratic or cubic configurations \cite{eiles_ultracold_2016, liu_polyatomic_2006, liu_ultra-long-range_2009}.
Most important findings in this context are the prediction of triatomic trilobite molecules akin to borromean states \cite{liu_polyatomic_2006}, the emergence of quantum scars that appear in the form of 'supertrilobite' orbitals when dense and randomly distributed atoms perturb the Rydberg orbital \cite{Luukko_Rost_2017} as well as the insight that computational effort in evaluating the PESs can be reduced significantly by employing basis sets of hybridized dimer orbitals \cite{liu_polyatomic_2006,liu_ultra-long-range_2009,eiles_ultracold_2016}.

The hybridized basis is typically much smaller than the full basis of Rydberg states $\varphi_{nlm}(\vec{r})$ but yields identical PESs.
Additionally, the hybridization of dimer orbitals provides an intuitive understanding of the bond angles $\theta$ of polyatomic URM, which are intrinsically related to the angular momentum $l$ of the Rydberg electron. This has been demonstrated for the case of triatomic ULRMs where the number of degrees of freedom is still small enough to analyze the full electronic and vibrational structure without imposing geometrical constraints \cite{fey_stretching_2016, Fey_2019_effective_three_body, Fey_Hummel_Schmelcher_2019_trilobite}. An exemplary $d$-state trimer is presented in Fig. \ref{fig:d_trimer}. The primitive basis consists of the two diatomic orbitals $\psi(\vec{r};\vec{R}_1)$ and $\psi(\vec{r};\vec{R}_2)$ which are the electronic eigenstates, if there is only a single ground-state atom present at $\vec{R}_1$ or $\vec{R}_2$, respectively. The angular distribution of these two $d$ states is indicated in Fig. \ref{fig:d_trimer} (b) (gray and orange orbitals). When both ground-state atoms are present the electronic eigenstates of the trimer are superpositions of these orbitals and correspond to two PESs $E_+(R_1,R_2,\theta)$ and $E_-(R_1,R_2,\theta)$. Fig. \ref{fig:d_trimer} presents cuts of these PESs for the special situation where $R_1=R_2=R$ is fixed to the radial minimum of the most outer potential well, which is $R=1868$ $a_0$ for a Cs($34d$) trimer. While $E_+$ has minima in the linear ($\theta=\pi$) and antilinear ($\theta=0$) configuration, $E_-$ exhibits a minimum in the perpendicular configuration $\theta=\pi/2$.     
These equilibria exist since the electronic state can maximize its density on both ground-state atoms in these configurations. Due to the angular shape of the primitive $d$ orbitals in Fig. \ref{fig:d_trimer} (b) this is the case for $\theta=0$, $\theta=\pi$ or $\theta=\pi/2$.   
Following the same reasoning triatomic $p$-state ULRMs possess linear ($\theta= \pi$) and anti-linear ($\theta=\pi$) equilibrium geometries \cite{fey_stretching_2016} while an even richer structure of equilibrium angles exist for trilobite trimers where the bond is established by hybridization of two nodal trilobite orbitals \cite{fey_stretching_2016}. The hybridization of orbitals with different angular momenta has two important consequences. First, the binding energies of trimers with $l>0$ are in contrast to $s$-state ULRMs non-additive, since they depend on the bond angle. Second, the shape of the PESs $E_\pm(R_1,R_2,\theta)$ in Fig. \ref{fig:d_trimer} implies that the nuclei are subject to an angular-dependent force which is, hence, an example for a three-body interaction in Rydberg systems. Both effects have been confirmed spectroscopically for $d$-state ULRMs in \cite{Fey_2019_effective_three_body}.  

%Like their diatomic counterparts polyatomic trilobite molecules can possess large dipole moments and are relatively deeply bound. This turns them into an interesting platform for studying field control of molecular properties as well as molecular dynamics.   
%The obvious limitation, that their direct one- or two-photon association is hindered by dipole selection rules, might be overcome in future experiments relying on higher-order photon transitions.

\section{Field control}

ULRMs inherit many of their characteristic features from their parent Rydberg atoms.
This includes their large bond lengths as well as their extreme sensitivity to external fields. Relatively weak electric and magnetic fields on the order of tens of V/cm or tens of Gauss can already have a crucial impact on the electronic structure of ULRMs, such as the deformation of electronic orbitals as well as shifts of their energy levels. As a consequence ULRMs offer unique possibilities to control molecular properties by weak fields which is otherwise impossible for conventional ground state molecules.

\begin{figure}
\includegraphics[width= 0.45 \textwidth]{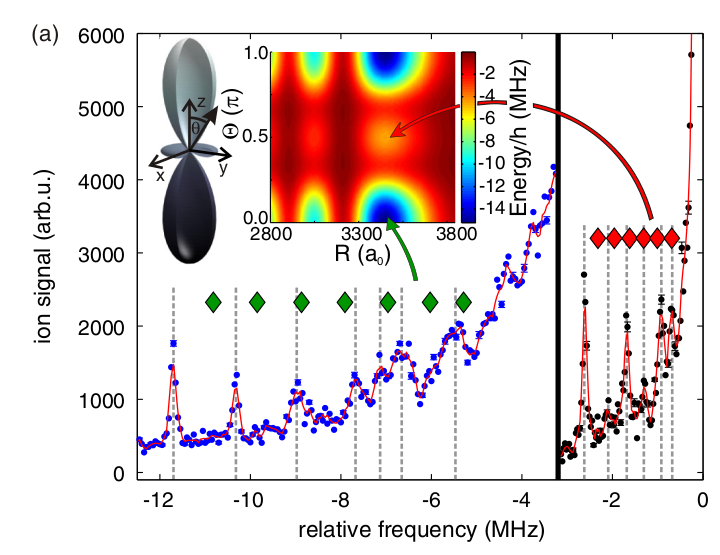}
\caption{Vibrational spectrum of aligned $44d$ ULRMs in a magnetic field of 13.55 Gauss along the $z$ axis reprinted from \btext{A. Krupp et al. Phys. Rev. Lett. \textbf{112}, 143008 (2014)} \cite{krupp_alignment_2014}. The PES $E(R,\theta)$ (inset) depends on the bond lengths $R$ and the orientation of the internuclear axis $\theta$, where the zero energy is set to the atomic 44$d$, $j=5/2$ $m_j=1/2$ level. The angular structure of the PES is dominated by the orbital $|Y_{l=2,m=0}(\theta)|^2$ and supports a series of axial ($\theta=0,\pi$) and toroidal ($\theta=\pi/2$) states. Theoretical energies of the axial (toroidal) states are depicted by green (red) diamonds and compared to the experimental photoassociaton spectrum.}
\label{fig:dimer_b_field}
\end{figure}

A very illustrative example for field control of ULRMs is the alignment of $d$-state ULRMs in magnetic fields observed in \cite{krupp_alignment_2014}.
In the experiment a static magnetic field $\vec{B}=B \vec{e}_z$ with $B=13.55$ Gauss was applied to lift the degeneracy of states with different magnetic quantum numbers $m=-2,...,2$. The associated Zeeman splitting is on the order of 22 MHz$\times h$ and exceeds the binding energy of ULRMs with quantum numbers $n\sim 44$. Consequently, the molecular $s$- and $p$-wave interactions perturb the energetically isolated Zeeman orbitals only weakly and the PES associated to a certain Zeeman level $m$ are for dominant $s$-wave scattering well approximated by $E(\vec{R})= 2\pi a_s[k(R)] |\varphi_{nlm}(\vec{R})|^2$ and proportional to the spherical harmonic $|Y_{lm}(\hat{R})|^2$. For instance the angular distribution of $|Y_{20}(\hat{R})|^2$ is depicted in Fig. \ref{fig:dimer_b_field}. Based on this shape, molecules can be either axially aligned ($\theta=0,\pi$) or perpendicular to the field axis ($\theta=\pi/2$). In the latter case the ground-state atom resides in the toroidal lobe in the $x-y$ plane. The full PES $E(\vec{R})=E(R,\theta)$ is shown in the inset of Fig. \ref{fig:dimer_b_field} and exhibits in addition to the angular structure also oscillations along the $R$ direction which result from the radial probability density of the Rydberg state.        
Theoretical simulations predict that axial and toroidal equilibrium geometries possess both a series of vibrational states with certain theoretical energies (red and green diamonds). As expected from the orbital density for $l=2$ $m=0$, toroidal molecules are bound more weakly than axial molecules and possess smaller energy spacing. Comparison to experimental photoassociation spectra (blue and black dots) confirms the alignment mechanism and allows one to associate the experimental signals to different molecular geometries. 
Quantative corrections due to spin-spin and spin-orbit interactions have been in the focus of \cite{Hummel_Fey_Schmelcher_spin_dstate_2018} and will be discussed in Sec. \ref{sec:spin effects}.  
 
While pure magnetic fields always induce splittings of different $m$ states, 
combining magnetic and electric fields offers the additional freedom to form superpositions of different atomic orbitals, e.g. $m=0$ and $m=2$ states, with mixing ratios that dependent on the field strengths. Since the geometry of ULRMs is closely related to the shape of the Rydberg orbital, this property has been exploited to tune the geometry of ULRMs continuously from axial to toroidal configurations \cite{gaj_hybridization_2015}.

In addition to low-$l$ ULRMs, also field control of trilobite and butterfly ULRMs has been investigated. Early theoretical work focused on the control of trilobite ULRMs via magnetic fields \cite{lesanovsky_ultra-long-range_2006}. 
It has been shown that the topology of the PESs undergoes a qualitative change when the induced Zeeman splitting becomes comparable to the molecular binding energy. Due to the stronger binding of trilobite molecules the required fields are larger than for low-$l$ ULRMs. For instance for an $n=35$ trilobite molecule these are on the order of 500 Gauss.
\btext{Fields below this threshold perturb the trilobite state only weakly and orient the molecular axis perpendicular to the field axis.}
Contrarily, for larger fields, the PESs are characterized by energetically separated $m$ manifolds that originate from the high-$l$ states, each supporting molecular states. Molecules belonging to the energetically lowest PES with $m=-l$ possess a circular electronic wave function and their vibrational states are confined in a 2D harmonic potential perpendicular to the field axis.

The impact of electric fields on trilobite molecules has been studied theoretically in \cite{kurz_electrically_2013}. 
In contrast to magnetic fields, electric fields tend to align the molecules to the field axis. Additionally, the field offers control over the dipole moment as well as the contribution of $p$-wave scattering  and allows one to adjust the relative depth of the potential wells in the trilobite curve.
Furthermore, sufficiently strong fields admix electronic $s$-state character to the polar trilobite wave function and can potentially facilitate two-photon photoassociation schemes for trilobite ULRMs.
%By means of electronic and rovibrational structure calculations it was demonstrated that the electric field are able to align the molecules to the field axis.
In a simple picture, the alignment results from the coupling of the electric field to the permanent dipole moment of the molecule.
%, which converts former rotational degrees of freedom into vibrational ones. 
For butterfly molecules, which behave in this respect similar to trilobite molecules, such an alignment has been realized experimentally \cite{niederprum_observation_2016}.
The measured spectrum can be described very accurately by means of a dipolar rigid rotor model or pendular state model \cite{Rost_1992}. By fitting the model parameters to the experimental spectra it is possible to exctract bond lengths $\sim 250 \, a_0$ and dipole moments $\sim 500 $ Debye. For trilobite molecules or ULRMs with trilobite admixture, bond lengths are larger and the pendular state spectrum is much denser at comparable field strengths. Individual pendular states of trilobite states have, therefore, not been resolved yet. Instead the electric field gives rise to an overall line broadening, which allows one to extract dipole moments \cite{li_homonuclear_2011,booth_production_2015}. Deviations from this line broadening have been reported in \cite{Kleinbach_2017} and are not yet fully understood.
Higher degree of control over trilobite molecules can be achieved by employing parallel electric and magnetic fields. As a function of the field strengths it is possible to tune the angle between the molecular and the fields axis from a perpendicular to a parallel configuration \cite{kurz_ultralong-range_2014}. 

Even richer opportunities of control are expected to exist for polyatomic ULRMs, where fields can tune the internal angular arrangement of the nuclei in addition to the external orientation. However, only few works explored such possibilities so far \cite{Fernandez_2016, Eiles_2019_exotic_specimen}. Apart from the direct impact on the molecular geometry, it is also interesting to consider the impact of fields on the spin degrees of freedom \cite{bottcher_observation_2016, Hummel_Fey_Schmelcher_spin_dstate_2018, Engel_Fey_Meinert_2019}. These effects are important since the relative orientation of spins is crucial for the molecular binding and will be discussed in Sec. \ref{sec:spin effects}.

In addition to trilobite, butterfly and low-$l$ ULRMs, there is yet another species of ULRMs, called ultralong-range giant dipole molecules, that exist exclusively in the presence of crossed electric and magnetic fields \cite{Kurz_Mayle_Schmelcher_2012}.
Sufficiently strong fields, e.g. on the order of 1 T and 50 V/cm, modify the electronic structure of the Rydberg atom and give rise to a decentered harmonic oscillator potential for the Rydberg electron which exists due to the non-separability of the center-of-mass motion and the electronic motion. These states are called giant dipole states since they possess electric dipole moments on the order of many ten thousands of Debye \cite{Burkova_Drukarev_Monozon_1976, Gay_Pendrill_Cagnac_1979, Fauth_Walther_Werner_1987, Baye_Clerbaux_Vincke_1992, Dzyaloshinskii_1992, Raithel_Fauth_Walther_1993, Schmelcher_Cederbaum_1993, Dippel_Schmelcher_Cederbaum_1994,zoellner2005a,zoellner2005b,schmelcher2001,averbukh1999}.
Ultralong-range giant dipole molecules are predicted to form when a ground-state atom interacts with the decentered electronic cloud via the Fermi pseudopotential and becomes thereby bound to the Rydberg atom \cite{Kurz_Mayle_Schmelcher_2012}.
 
\section{Spin structure}
\label{sec:spin effects}
\begin{figure}
\includegraphics[width= 0.3 \textwidth]{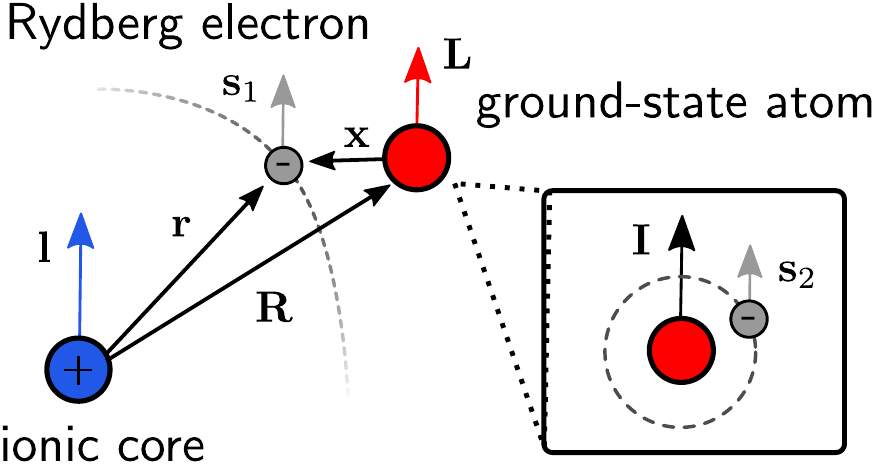}
\caption{Angular momenta in ULRMs. The Rydberg electron carries a spin $\vec{s}_1$ and an angular momentum $\vec{l}$ or $\vec{L}$, depending on the frame of reference which is centered on the ionic core or the ground-state atom, respectively. The insets presents the internal structure of the ground-state atom with nuclear spin $\vec{I}$ and spin $\vec{s}_2$ of its valence electron. }
\label{fig:spin_sketch}
\end{figure}

While the simple Fermi pseudopotential (\ref{eqn:fermi_pseudopotential}) captures many of the essential features of ULRMs, it neglects certain interactions, which are, however, indispensable for an accurate description of experimental high-resolution spectra.
These interactions are related to various spin degrees of freedom present in the system, see Fig. \ref{fig:spin_sketch}.

\begin{figure*}
\includegraphics[width= 0.8 \textwidth]{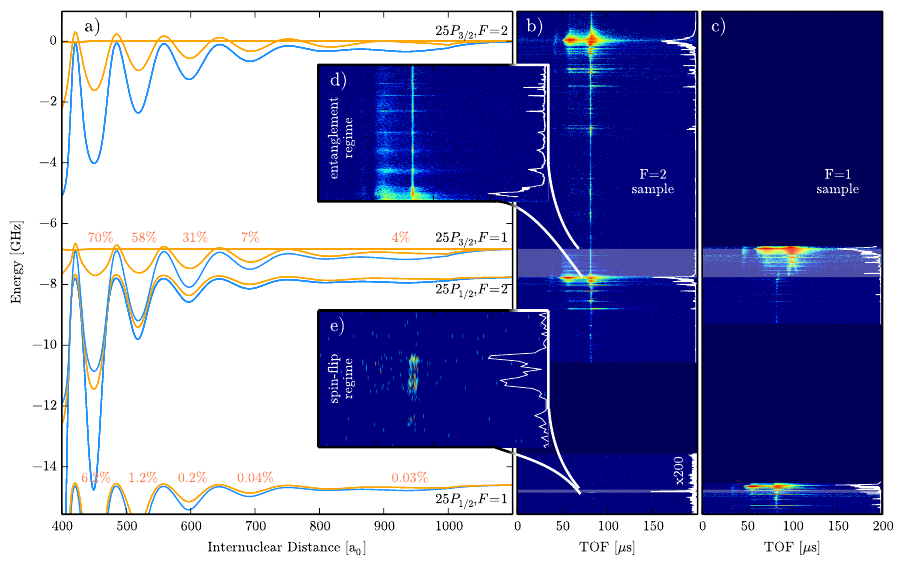}
\caption{Illustration of spin-interaction effects in ULRMs reprinted from \btext{T. Niederprüm et al. Phys. Rev. Lett. \textbf{117}, 123002 (2016)} \cite{Niederpruem_remote_spinflips_2016}. (a) PECs of diatomic $^{87}$Rb that correlate asymptotically to a \btext{$^{87}$Rb($25p_j$) Rydberg atom} and a ground-state atom in hyperfine state $F$. Molecules belonging to deep PECs (blue) are predominately bound by triplet scattering while those associated to shallow PECs (orange) are bound by mixed singlet/triplet scattering. The latter are also of mixed $F$ character with admixtures of opposite spin indicated by the orange numbers. In the experiment, Rydberg states are photoexcited from an atomic BEC and detected via their ionic products Rb$^+$ or Rb$_2^+$ which are produced through field ionization or spontaneous associative ionization. (b) and (c) show the number of detected ions as a function of the laser detuning and the time of flight to the detector (TOF). The hyperfine state of the atoms in the BEC before excitation is either $F=2$ (b) or $F=1$ (c). Magnified regions of these spectra (d) and (e) demonstrate that ULRMs in the shallow potentials can be accessed from both, the $F=1$ and the $F=2$ BEC, which confirms the mixed spin characters of the associated molecules.}
\label{fig:spin_flip}
\end{figure*}

First there is the internal spin structure of the ground state atom as well as of the Rydberg atom.
The ground state atom possesses the nuclear spin $\vec{I}$ and a valence electron with spin $\vec{s}_2$, see inset in Fig. \ref{fig:spin_sketch}, which couples to the hyperfine spin $\vec{F}=\vec{I}+\vec{s}_2$. For instance $^{87}$Rb has two hyperfine levels  $F=1$ and $F=2$ that are separated by 6.8 GHz.
The hyperfine structure of the Rydberg atom is typically negligible since it scales as $n^{-3}$ with the principal quantum number \cite{Arimondo_hyperfine_1977}. However, there is a fine structure splitting of the Rydberg levels described by $\vec {j}=\vec{s}_1+ \vec{l}$ where, $\vec{l}$ and $\vec{s}_1$ are the orbital angular momentum and the spin of the Rydberg electron, respectively.
In the limit of large internuclear separation $R$ the angular momenta $\vec{j}$ and $\vec{F}$ are conserved. Contrarily, at smaller separations the interaction between the electron and the ground-state atom can cause a mixing of $\vec{j}$ and $\vec{F}$.

Early spectral measurements of ULRMs were described succesfully by treating the electron-atom interaction only via a single scattering length $a_s[k]$ for the $s$-wave channel ($L=0$) and a single scattering volume $a_p[k]$ for the $p$-wave channel ($L=1$)  \cite{bendkowsky_observation_2009, bendkowsky_rydberg_2010}, where $\vec{L}$ denotes the orbital angular momentum of the Rydberg electron in the frame centered on the ground-state atom at $\vec{R}$, see Fig. \ref{fig:spin_sketch}.
These scattering lengths and volumes were associated to triplet scattering channels ($S=1$), where $\vec{S}=\vec{s}_1 +\vec{s}_2$ is the total electron spin.
The restriction to $S=1$ is motivated by the fact that interactions in the triplet channels are typically much stronger than in the singlet channels ($S=0$). For instance for Rb the zero-energy scattering length is $\sim -16$ $a_0$ for $S=1$ and only $\sim 0.5$ $a_0$ for $S=0$ \cite{bahrim_3se_2001}. Furthermore, the $p$-wave shape resonance occurs exclusively in the triplet channels \cite{Thumm_Norcross_1991_evidence_for,bahrim_low-lying_2000}.

However, a detailed analysis including singlet and triplet interactions as well as hyperfine and fine structure couplings reveals that singlet scattering can have a crucial impact on the molecular structure \cite{anderson_photoassociation_2014, anderson_angular-momentum_2014}. Due to the interplay of all interactions, the electron spin $S$ is in general not a conserved quantity and one obtains two different classes of ULRMs. The first class is bound by pure triplet scattering and unaffected by singlet interactions, while the second class is bound by mixed singlet and triplet scattering. Typically, this mixing reduces the depth of the corresponding PECs as well as the molecular binding energies. One distinguishes therefore deep PECs belonging to the first class and shallow PECs belonging to the second class. Exemplary PECs are presented in Fig. \ref{fig:spin_flip} close to the \btext{$^{87}$Rb($25p$)} level with deep PECs (blue) and shallow PECs (orange). For experiments with ULRMs this spin mixing has two main consequences.
First, mixed singlet-triplet states contain also contribution from both hyperfine states $F$. Excitation of these molecules can therefore be used to induce remote spin flips of ground-state atoms as well as to entangle the fines tructure state of the Rydberg atom and the hyperfine state of the ground-state atom \cite{Niederpruem_remote_spinflips_2016}. Second, measuring the binding energies of molecules in the deep and shallow PECs provides unique experimental access to low-energy singlet and triplet phase shifts. ULRMs can function therefore as high precision scattering laboratories to prepare and to probe ultra slow electron-atom collision, with electron energies below 10 meV, which is hard to achieve by alternative methods such as electron transmission \cite{Johnston_Burrow_1982}.
Characterization of spin-resolved scattering via ULRMs has been accomplished for $s$-wave phase shifts \cite{bendkowsky_rydberg_2010, sasmannshausen_experimental_2015,bottcher_observation_2016} as well as for $p$-wave phase shifts\cite{bendkowsky_rydberg_2010,schlagmuller_probing_2016, MacLennan_Chen_Raithel_2018, Engel_Fey_Meinert_2019}. 

\btext{An interesting substructure of the Rb($np$) PECs that is not visible  in Fig. \ref{fig:spin_flip} but becomes resovable at lower principal quantum numbers has been reported recently in \cite{Deiss_Fey_Hummel_Schmelcher_Denschlag_2019} when addressing Rb($16p$). It was shown that}
the deep and shallow PECs split into one, two or three subcurves which can be observed in the experiment as singlet, doublets or triplets of molecular resonances. The multiplicity of the splittings can be associated to combinations of quantum numbers $\vec{I}$, $\vec{s}_1$ and $\vec{s}_2$ and allows thereby to correlate experimental spectral signals clearly to certain PECs, even under complicated conditions when several PECs cover the same energy range. A possible explanation for the splitting mechanism is yet another correction of the molecular interaction that arises due to spin-orbit coupling of the electron spin $\vec{S}$ and the orbital angular momentum $\vec{L}$ of the electron with respect to the ground-state atom, see Fig. \ref{fig:spin_sketch}. As a consequence of this $LS$ coupling the triplet phase shifts in the $p$-wave channel split into three subchannels $J=0,1,2$, where $\vec{J}=\vec{L}+\vec{S}$. For each $J$ channel, the $p$-wave shape resonance occurs at a different energy, in accordance with the fine structure splitting of the corresponding negative ion states. It is predicted that this splitting is larger for Cs$^-$ than for Rb$^-$ and in both cases on the order of a few meV \cite{Thumm_Norcross_1991_evidence_for, bahrim_low-lying_2000}. While the $J=1$ shape resonance in Cs$^-$ has been confirmed using photo-detachment spectroscopy \cite{Scheer_1998_cs_shape_resonance}, a measurement of the spin-orbit split shape resonances in Rb has been achieved only recently via spectroscopy of ULRMs \cite{Engel_Fey_Meinert_2019}.

\begin{figure}
\includegraphics[width= 0.99 \linewidth]{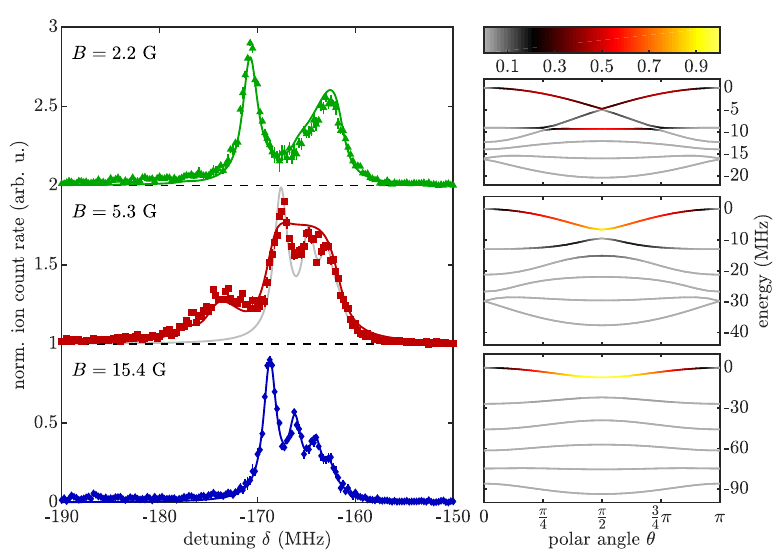}
\caption{Orientation of an $31s$ ULRM with a bond length of ($R\approx 890$ $a_0$) in a magnetic field reprinted from \btext{F. Engel et al. Phys. Rev. Lett \textbf{123}, 073003 (2019)} \cite{Engel_Fey_Meinert_2019}. Left: Molecular spectrum as a function of the laser detuning $\delta$ for different magnetic field strengths. The experimental signal (colored symbols) is compared to theoretical results (solid lines). Right: Molecular PESs as a function of the angle between the field and the molecular axis for fixed bond length $R=890$ $a_0$. Each PES corresponds to a different superposition of spin states $\ket{m_J,F,m_F}$. The color code is the projection onto the state with $m_j=1/2$ and $m_F=2$ which is proportional to the strength of the laser coupling to the different PESs. The shape of the PESs together with the coupling strength determines the lineshape and the spacing of the theoretically predicted spectra (left).}
\label{fig:s_alignment}
\end{figure}

Extraction of these resonances as well as other scattering properties requires accurate theoretical models linking the observable molecular binding energies to the underlying scattering physics. A theoretical pseudo potential Hamiltonian which is well suited for such a purpose has been developed in \cite{eiles_hamiltonian_2017}. It includes all the spin couplings presented in Fig. \ref{fig:spin_sketch} and generalizes previous models \cite{markson_theory_2016,anderson_angular-momentum_2014,khuskivadze_adiabatic_2002,  greene_creation_2000, hamilton_shape-resonance-induced_2002}. Combining this Hamiltonian with additional magnetic fields has been accomplished in \cite{Hummel_Fey_Schmelcher_spin_dstate_2018, Hummel_Fey_Schmelcher_2019_s_state_alignment}. It was shown that couplings between the field and the spins lead to quantitative corrections of binding energies and orientation angles but also to novel opportunities for field control of ULRMs. For instance
addressing molecules with different quantum numbers $m_j$ and $m_F$ allows one to tune the admixture
of singlet and triplet scattering and thereby the depths of the PESs as well as the resulting
molecular alignment \cite{Hummel_Fey_Schmelcher_spin_dstate_2018}. Furthermore $LS$ coupling gives rise to a novel alignment mechanism of ULRMs, which is fundamentally different from the alignment mechanism of $d$-state ULRMs described in Fig. \ref{fig:dimer_b_field}, since it does not rely on the orientation of angular orbitals $Y_{lm}(\theta,\phi)$. Instead, it bases on the orientation of electronic spins in the $B$ field, which in turn couple to spatial degrees of freedom of the molecule via the $LS$-interaction and do thereby orient the molecule. Importantly, this alignment mechanism applies also to ULRMs in $s$ states.
This is illustrated in Fig. \ref{fig:s_alignment} from \cite{Engel_Fey_Meinert_2019}, which compares experimental spectra of \btext{Rb($31s$)} ULRMs in magnetic fields to theoretical simulations.
The signals (left) can be associated to molecules with bond length of $R\approx 890$ $a_0$ that are radially localized in inner potential wells which are strongly affected by $p$-wave interactions, c.f. Fig. \ref{fig:dimer_basics} (d). The simulated signals are determined based on the angular cuts of the PESs for fixed bond length $R = 890$ $a_0$ (Fig. \ref{fig:dimer_b_field} right). These curves exhibit an angular dependence that can be attributed exclusively to $LS$ coupling, i.e. without $LS$ coupling the curves would be flat \cite{Hummel_Fey_Schmelcher_2019_s_state_alignment}. Electronic selection rules enter the simulations by taking into account that the excitation laser couples only to the electronic component with $m_j=1/2$ and $m_F=2$ which is indicated by the color code. Consequently, for low fields (2.2 G) the laser couples to several of the energetically highest PESs, whereas for high fields (15.4 G) only the upmost PES is addressable. In the latter case the peaks in the spectrum can be clearly identified with a series of vibrational states with an anharmonic vibrational spacing on the order of 2-3 MHz that confirms the angular confinement of the molecular axis. Crucially, comparison between the experimental and theoretical data yields information on fundamental scattering properties such as negative-ion resonances. \otext{This relation turns ULRM into a unique laboratory to realize electron-atom collisions with well-defined electronic states of ultra-low collisional energies.}

\section{Conclusions and outlook}
We reviewed basic properties and recent progress in the field of ultralong-range Rydberg molecules. 
Ten years after the first observation of ULRMs and almost twenty years after their prediction,
intense theoretical and experimental research elucidated not only the complex physical structure of the underlying interactions but revealed also many exotic and counterintuitive molecular properties as well as fascinating applications. 
Outstanding applications of ULRMs are field control of molecular geometry and electronic structure \cite{kurz_ultralong-range_2014, gaj_hybridization_2015, niederprum_observation_2016}, experimental characterization of spin-resolved slow electron-atom collisions \cite{anderson_photoassociation_2014, sasmannshausen_experimental_2015, bottcher_observation_2016, Engel_Fey_Meinert_2019} as well as the probe of interatomic correlations in ultracold atomic gases \cite{manthey_dynamically_2015, Whalen_2019, Whalen_Schmidt_Wagner_2019}.
Furthermore, ULRMs possess an extraordinary scalability which allows one to excite not only dimers but also polymers \cite{liu_polyatomic_2006, bendkowsky_rydberg_2010, gaj_molecular_2014, Fey_2019_effective_three_body} and, in the many-body limit, polarons \cite{balewski_coupling_2013, schmidt_mesoscopic_2016, Camargo_2018}.

In this review we focused on the three subjects, polyatomic ULRMs, field control and spin-interactions by providing key experimental results and the corresponding theoretical framework. 
A central finding from research on polyatomic ULRMs is that binding energies of $s$-state ULRMs are to a good approximation additive \cite{gaj_molecular_2014}, i.e. integer multiples of the number of ground-state atoms, whereas binding energies of polyatomic ULRMs with angular momenta $l>0$ are non-additive and involve three- and higher body forces \cite{li_homonuclear_2011, fey_stretching_2016, Fey_2019_effective_three_body, Fey_Hummel_Schmelcher_2019_trilobite}. Since the only experimentally confirmed polyatomic ULRMs with $l>0$ are $d$-state trimers, it is an interesting challenge for future experiments to detect novel polyatomics with more atoms, e.g. $d$-state tetramers, or with larger angular momenta, e.g. trilobite trimers \cite{Fey_Hummel_Schmelcher_2019_trilobite}.

In the context of field control, we presented possibilities to tune bond lengths, orientation as well as the shape of electronic orbitals using relatively weak electric or magnetic fields, which is impossible for conventional molecules close to their ground-state \cite{Koch_Lemeshko_Sugny_2018, Stapelfeldt_Seideman_2003}. A so far unexplored future research direction will be to extent these opportunities to polyatomic ULRMs. In contrast to field control of dimers it is expected that field-impact on polyatomics will manipulate not only the absolute molecular orientation but also the internal configuration of the atoms. For instance, permanent dipole moments in trimers should allow one to tune the system from linear to bent geometries. First steps in this direction have been undertaken in \cite{Fernandez_2016} via exact diagonalization and in \cite{Eiles_2019_exotic_specimen} employing hybridized dimer orbitals which are a useful approximation to reduce computational effort. 

Lastly, we detailed how various spin-degrees of freedom of the Rydberg electron and the ground-state atom impact molecular properties such as binding energies or the orientation in external fields. As a consequence, ULRMs provide a unique experimental platform to probe spin-resolved low-energy electron atom collision. Although it is remarkable, how well diagonalization methods in combination with the Fermi pseudopotential describe such experiments, there is a strong need for a higher level theory in order too improve the accuracy of obtained observables. This is necessary since truncated diagonalization suffers from convergence issues \cite{fey_comparative_2015,eiles_hamiltonian_2017,Engel_2018} with respect to the size of the basis set, while current Green's function methods are intrinsically convergent but neglect the hyperfine structure of the ground state atom. The most promising candiate for an improved theory would be a generalization of the Green's function method presented in \cite{khuskivadze_adiabatic_2002} that includes already spin-orbit-coupled $p$-wave interactions. Alternatively, it might be possible to develop appropriate regularization schemes for the delta potential or to pursue a hybrid approach that deduces PESs based on existing Green's function methods while including additional spin interactions perturbatively.   

Finally, it is noteworthy to mention that research on ULRMs focused so far almost exclusively on stationary properties such as binding energies, bond lengths or dipole moments. 
Hence, a completely novel line of research with rich opportunities is the investigation of dynamical processes in ULRMs, i.e. wave packets propagation in the oscillatory PESs. 
%Such dynamical studies are expected to reveal interesting insights to predissociation and collisional decay processes \cite{schlagmuller_ultracold_2016}.
 Lifetimes of ULRMs $\sim 30$ $\mu s$ are large enough to observe such dynamics. A first interesting application in this direction would be the study of coherent states that perform large amplitude oscillations in the trilobite PECs, e.g. ranging from 500 to 1800 $a_0$ in Fig. \ref{fig:dimer_basics}. Damping or loss rate of wave packet during the oscillations could yield information on decay processes that result from the crossing with the butterfly curves.
\otext{In general this opens the doorway to study dynamics of various chemical reaction processes using ULRM, which is by no means limited to alkali atoms, but applies, due to the general binding mechanism underlying ULRM, to all kinds of single and multichannel Rydberg systems.}

\subsection*{Acknowledgments}
P.S. and F.H. acknowledge support from the German
Research Foundation within the priority program "Giant Interactions in Rydberg Systems" (DFG SPP 1929
GiRyd). The authors thank M. Eiles for fruitful discussions.

%\bibliography{ulrm}
\input{main.bbl}

\onecolumngrid
\newpage
\appendix

\end{document}

%% file: main.bbl

%% file: main.bbl
\begin{thebibliography}{115}%
\makeatletter
\providecommand \@ifxundefined [1]{%
 \@ifx{#1\undefined}
}%
\providecommand \@ifnum [1]{%
 \ifnum #1\expandafter \@firstoftwo
 \else \expandafter \@secondoftwo
 \fi
}%
\providecommand \@ifx [1]{%
 \ifx #1\expandafter \@firstoftwo
 \else \expandafter \@secondoftwo
 \fi
}%
\providecommand \natexlab [1]{#1}%
\providecommand \enquote  [1]{``#1''}%
\providecommand \bibnamefont  [1]{#1}%
\providecommand \bibfnamefont [1]{#1}%
\providecommand \citenamefont [1]{#1}%
\providecommand \href@noop [0]{\@secondoftwo}%
\providecommand \href [0]{\begingroup \@sanitize@url \@href}%
\providecommand \@href[1]{\@@startlink{#1}\@@href}%
\providecommand \@@href[1]{\endgroup#1\@@endlink}%
\providecommand \@sanitize@url [0]{\catcode `\\12\catcode `\$12\catcode
  `\&12\catcode `\#12\catcode `\^12\catcode `\_12\catcode `\%12\relax}%
\providecommand \@@startlink[1]{}%
\providecommand \@@endlink[0]{}%
\providecommand \url  [0]{\begingroup\@sanitize@url \@url }%
\providecommand \@url [1]{\endgroup\@href {#1}{\urlprefix }}%
\providecommand \urlprefix  [0]{URL }%
\providecommand \Eprint [0]{\href }%
\providecommand \doibase [0]{http://dx.doi.org/}%
\providecommand \selectlanguage [0]{\@gobble}%
\providecommand \bibinfo  [0]{\@secondoftwo}%
\providecommand \bibfield  [0]{\@secondoftwo}%
\providecommand \translation [1]{[#1]}%
\providecommand \BibitemOpen [0]{}%
\providecommand \bibitemStop [0]{}%
\providecommand \bibitemNoStop [0]{.\EOS\space}%
\providecommand \EOS [0]{\spacefactor3000\relax}%
\providecommand \BibitemShut  [1]{\csname bibitem#1\endcsname}%
\let\auto@bib@innerbib\@empty
%</preamble>
\bibitem [{\citenamefont {Greene}\ \emph {et~al.}(2000)\citenamefont {Greene},
  \citenamefont {Dickinson},\ and\ \citenamefont
  {Sadeghpour}}]{greene_creation_2000}%
  \BibitemOpen
  \bibfield  {author} {\bibinfo {author} {\bibfnamefont {C.~H.}\ \bibnamefont
  {Greene}}, \bibinfo {author} {\bibfnamefont {A.~S.}\ \bibnamefont
  {Dickinson}}, \ and\ \bibinfo {author} {\bibfnamefont {H.~R.}\ \bibnamefont
  {Sadeghpour}},\ }\href
  {http://journals.aps.org/prl/abstract/10.1103/PhysRevLett.85.2458} {\bibfield
   {journal} {\bibinfo  {journal} {Phys. Rev. Lett.}\ }\textbf {\bibinfo
  {volume} {85}},\ \bibinfo {pages} {2458} (\bibinfo {year}
  {2000})}\BibitemShut {NoStop}%
\bibitem [{\citenamefont {Bendkowsky}\ \emph {et~al.}(2009)\citenamefont
  {Bendkowsky}, \citenamefont {Butscher}, \citenamefont {Nipper}, \citenamefont
  {Shaffer}, \citenamefont {Löw},\ and\ \citenamefont
  {Pfau}}]{bendkowsky_observation_2009}%
  \BibitemOpen
  \bibfield  {author} {\bibinfo {author} {\bibfnamefont {V.}~\bibnamefont
  {Bendkowsky}}, \bibinfo {author} {\bibfnamefont {B.}~\bibnamefont
  {Butscher}}, \bibinfo {author} {\bibfnamefont {J.}~\bibnamefont {Nipper}},
  \bibinfo {author} {\bibfnamefont {J.~P.}\ \bibnamefont {Shaffer}}, \bibinfo
  {author} {\bibfnamefont {R.}~\bibnamefont {Löw}}, \ and\ \bibinfo {author}
  {\bibfnamefont {T.}~\bibnamefont {Pfau}},\ }\href {\doibase
  10.1038/nature07945} {\bibfield  {journal} {\bibinfo  {journal} {Nature}\
  }\textbf {\bibinfo {volume} {458}},\ \bibinfo {pages} {1005} (\bibinfo {year}
  {2009})}\BibitemShut {NoStop}%
\bibitem [{\citenamefont {Lesanovsky}\ \emph {et~al.}(2006)\citenamefont
  {Lesanovsky}, \citenamefont {Schmelcher},\ and\ \citenamefont
  {Sadeghpour}}]{lesanovsky_ultra-long-range_2006}%
  \BibitemOpen
  \bibfield  {author} {\bibinfo {author} {\bibfnamefont {I.}~\bibnamefont
  {Lesanovsky}}, \bibinfo {author} {\bibfnamefont {P.}~\bibnamefont
  {Schmelcher}}, \ and\ \bibinfo {author} {\bibfnamefont {H.~R.}\ \bibnamefont
  {Sadeghpour}},\ }\href {\doibase 10.1088/0953-4075/39/4/L03} {\bibfield
  {journal} {\bibinfo  {journal} {J. Phys. B: At. Mol. Opt. Phys.}\ }\textbf
  {\bibinfo {volume} {39}},\ \bibinfo {pages} {L69} (\bibinfo {year}
  {2006})}\BibitemShut {NoStop}%
\bibitem [{\citenamefont {Kurz}\ and\ \citenamefont
  {Schmelcher}(2013)}]{kurz_electrically_2013}%
  \BibitemOpen
  \bibfield  {author} {\bibinfo {author} {\bibfnamefont {M.}~\bibnamefont
  {Kurz}}\ and\ \bibinfo {author} {\bibfnamefont {P.}~\bibnamefont
  {Schmelcher}},\ }\href {\doibase 10.1103/PhysRevA.88.022501} {\bibfield
  {journal} {\bibinfo  {journal} {Phys. Rev. A}\ }\textbf {\bibinfo {volume}
  {88}},\ \bibinfo {pages} {022501} (\bibinfo {year} {2013})}\BibitemShut
  {NoStop}%
\bibitem [{\citenamefont {Krupp}\ \emph {et~al.}(2014)\citenamefont {Krupp},
  \citenamefont {Gaj}, \citenamefont {Balewski}, \citenamefont {Ilzhöfer},
  \citenamefont {Hofferberth}, \citenamefont {Löw}, \citenamefont {Pfau},
  \citenamefont {Kurz},\ and\ \citenamefont
  {Schmelcher}}]{krupp_alignment_2014}%
  \BibitemOpen
  \bibfield  {author} {\bibinfo {author} {\bibfnamefont {A.}~\bibnamefont
  {Krupp}}, \bibinfo {author} {\bibfnamefont {A.}~\bibnamefont {Gaj}}, \bibinfo
  {author} {\bibfnamefont {J.}~\bibnamefont {Balewski}}, \bibinfo {author}
  {\bibfnamefont {P.}~\bibnamefont {Ilzhöfer}}, \bibinfo {author}
  {\bibfnamefont {S.}~\bibnamefont {Hofferberth}}, \bibinfo {author}
  {\bibfnamefont {R.}~\bibnamefont {Löw}}, \bibinfo {author} {\bibfnamefont
  {T.}~\bibnamefont {Pfau}}, \bibinfo {author} {\bibfnamefont {M.}~\bibnamefont
  {Kurz}}, \ and\ \bibinfo {author} {\bibfnamefont {P.}~\bibnamefont
  {Schmelcher}},\ }\href {\doibase 10.1103/PhysRevLett.112.143008} {\bibfield
  {journal} {\bibinfo  {journal} {Phys. Rev. Lett.}\ }\textbf {\bibinfo
  {volume} {112}},\ \bibinfo {pages} {143008} (\bibinfo {year}
  {2014})}\BibitemShut {NoStop}%
\bibitem [{\citenamefont {Gaj}\ \emph {et~al.}(2015)\citenamefont {Gaj},
  \citenamefont {Krupp}, \citenamefont {Ilzhöfer}, \citenamefont {Löw},
  \citenamefont {Hofferberth},\ and\ \citenamefont
  {Pfau}}]{gaj_hybridization_2015}%
  \BibitemOpen
  \bibfield  {author} {\bibinfo {author} {\bibfnamefont {A.}~\bibnamefont
  {Gaj}}, \bibinfo {author} {\bibfnamefont {A.}~\bibnamefont {Krupp}}, \bibinfo
  {author} {\bibfnamefont {P.}~\bibnamefont {Ilzhöfer}}, \bibinfo {author}
  {\bibfnamefont {R.}~\bibnamefont {Löw}}, \bibinfo {author} {\bibfnamefont
  {S.}~\bibnamefont {Hofferberth}}, \ and\ \bibinfo {author} {\bibfnamefont
  {T.}~\bibnamefont {Pfau}},\ }\href {\doibase 10.1103/PhysRevLett.115.023001}
  {\bibfield  {journal} {\bibinfo  {journal} {Phys. Rev. Lett.}\ }\textbf
  {\bibinfo {volume} {115}},\ \bibinfo {pages} {023001} (\bibinfo {year}
  {2015})}\BibitemShut {NoStop}%
\bibitem [{\citenamefont {Niederprüm}\ \emph
  {et~al.}(2016{\natexlab{a}})\citenamefont {Niederprüm}, \citenamefont
  {Thomas}, \citenamefont {Eichert}, \citenamefont {Lippe}, \citenamefont
  {Pérez-Ríos}, \citenamefont {Greene},\ and\ \citenamefont
  {Ott}}]{niederprum_observation_2016}%
  \BibitemOpen
  \bibfield  {author} {\bibinfo {author} {\bibfnamefont {T.}~\bibnamefont
  {Niederprüm}}, \bibinfo {author} {\bibfnamefont {O.}~\bibnamefont {Thomas}},
  \bibinfo {author} {\bibfnamefont {T.}~\bibnamefont {Eichert}}, \bibinfo
  {author} {\bibfnamefont {C.}~\bibnamefont {Lippe}}, \bibinfo {author}
  {\bibfnamefont {J.}~\bibnamefont {Pérez-Ríos}}, \bibinfo {author}
  {\bibfnamefont {C.~H.}\ \bibnamefont {Greene}}, \ and\ \bibinfo {author}
  {\bibfnamefont {H.}~\bibnamefont {Ott}},\ }\href {\doibase
  10.1038/ncomms12820} {\bibfield  {journal} {\bibinfo  {journal} {Nature
  Comm.}\ }\textbf {\bibinfo {volume} {7}},\ \bibinfo {pages} {12820} (\bibinfo
  {year} {2016}{\natexlab{a}})}\BibitemShut {NoStop}%
\bibitem [{\citenamefont {Hummel}\ \emph {et~al.}(2018)\citenamefont {Hummel},
  \citenamefont {Fey},\ and\ \citenamefont
  {Schmelcher}}]{Hummel_Fey_Schmelcher_spin_dstate_2018}%
  \BibitemOpen
  \bibfield  {author} {\bibinfo {author} {\bibfnamefont {F.}~\bibnamefont
  {Hummel}}, \bibinfo {author} {\bibfnamefont {C.}~\bibnamefont {Fey}}, \ and\
  \bibinfo {author} {\bibfnamefont {P.}~\bibnamefont {Schmelcher}},\ }\href
  {\doibase 10.1103/PhysRevA.97.043422} {\bibfield  {journal} {\bibinfo
  {journal} {Phys. Rev. A}\ }\textbf {\bibinfo {volume} {97}},\ \bibinfo
  {pages} {043422} (\bibinfo {year} {2018})}\BibitemShut {NoStop}%
\bibitem [{\citenamefont {Hummel}\ \emph {et~al.}(2019)\citenamefont {Hummel},
  \citenamefont {Fey},\ and\ \citenamefont
  {Schmelcher}}]{Hummel_Fey_Schmelcher_2019_s_state_alignment}%
  \BibitemOpen
  \bibfield  {author} {\bibinfo {author} {\bibfnamefont {F.}~\bibnamefont
  {Hummel}}, \bibinfo {author} {\bibfnamefont {C.}~\bibnamefont {Fey}}, \ and\
  \bibinfo {author} {\bibfnamefont {P.}~\bibnamefont {Schmelcher}},\ }\href
  {\doibase 10.1103/PhysRevA.99.023401} {\bibfield  {journal} {\bibinfo
  {journal} {Phys. Rev. A}\ }\textbf {\bibinfo {volume} {99}} (\bibinfo {year}
  {2019}),\ 10.1103/PhysRevA.99.023401}\BibitemShut {NoStop}%
\bibitem [{\citenamefont {Liu}\ and\ \citenamefont
  {Rost}(2006)}]{liu_polyatomic_2006}%
  \BibitemOpen
  \bibfield  {author} {\bibinfo {author} {\bibfnamefont {I.~C.}\ \bibnamefont
  {Liu}}\ and\ \bibinfo {author} {\bibfnamefont {J.~M.}\ \bibnamefont {Rost}},\
  }\href {\doibase 10.1140/epjd/e2006-00098-x} {\bibfield  {journal} {\bibinfo
  {journal} {Eur. Phys. J. D}\ }\textbf {\bibinfo {volume} {40}},\ \bibinfo
  {pages} {65} (\bibinfo {year} {2006})}\BibitemShut {NoStop}%
\bibitem [{\citenamefont {Liu}\ \emph {et~al.}(2009)\citenamefont {Liu},
  \citenamefont {Stanojevic},\ and\ \citenamefont
  {Rost}}]{liu_ultra-long-range_2009}%
  \BibitemOpen
  \bibfield  {author} {\bibinfo {author} {\bibfnamefont {I.}~\bibnamefont
  {Liu}}, \bibinfo {author} {\bibfnamefont {J.}~\bibnamefont {Stanojevic}}, \
  and\ \bibinfo {author} {\bibfnamefont {J.}~\bibnamefont {Rost}},\ }\href
  {\doibase 10.1103/PhysRevLett.102.173001} {\bibfield  {journal} {\bibinfo
  {journal} {Phys. Rev. Lett.}\ }\textbf {\bibinfo {volume} {102}},\ \bibinfo
  {pages} {173001} (\bibinfo {year} {2009})}\BibitemShut {NoStop}%
\bibitem [{\citenamefont {Bendkowsky}\ \emph {et~al.}(2010)\citenamefont
  {Bendkowsky}, \citenamefont {Butscher}, \citenamefont {Nipper}, \citenamefont
  {Balewski}, \citenamefont {Shaffer}, \citenamefont {Löw}, \citenamefont
  {Pfau}, \citenamefont {Li}, \citenamefont {Stanojevic}, \citenamefont
  {Pohl},\ and\ \citenamefont {Rost}}]{bendkowsky_rydberg_2010}%
  \BibitemOpen
  \bibfield  {author} {\bibinfo {author} {\bibfnamefont {V.}~\bibnamefont
  {Bendkowsky}}, \bibinfo {author} {\bibfnamefont {B.}~\bibnamefont
  {Butscher}}, \bibinfo {author} {\bibfnamefont {J.}~\bibnamefont {Nipper}},
  \bibinfo {author} {\bibfnamefont {J.~B.}\ \bibnamefont {Balewski}}, \bibinfo
  {author} {\bibfnamefont {J.~P.}\ \bibnamefont {Shaffer}}, \bibinfo {author}
  {\bibfnamefont {R.}~\bibnamefont {Löw}}, \bibinfo {author} {\bibfnamefont
  {T.}~\bibnamefont {Pfau}}, \bibinfo {author} {\bibfnamefont {W.}~\bibnamefont
  {Li}}, \bibinfo {author} {\bibfnamefont {J.}~\bibnamefont {Stanojevic}},
  \bibinfo {author} {\bibfnamefont {T.}~\bibnamefont {Pohl}}, \ and\ \bibinfo
  {author} {\bibfnamefont {J.~M.}\ \bibnamefont {Rost}},\ }\href {\doibase
  10.1103/PhysRevLett.105.163201} {\bibfield  {journal} {\bibinfo  {journal}
  {Phys. Rev. Lett.}\ }\textbf {\bibinfo {volume} {105}},\ \bibinfo {pages}
  {163201} (\bibinfo {year} {2010})}\BibitemShut {NoStop}%
\bibitem [{\citenamefont {Gaj}\ \emph {et~al.}(2014)\citenamefont {Gaj},
  \citenamefont {Krupp}, \citenamefont {Balewski}, \citenamefont {Löw},
  \citenamefont {Hofferberth},\ and\ \citenamefont
  {Pfau}}]{gaj_molecular_2014}%
  \BibitemOpen
  \bibfield  {author} {\bibinfo {author} {\bibfnamefont {A.}~\bibnamefont
  {Gaj}}, \bibinfo {author} {\bibfnamefont {A.~T.}\ \bibnamefont {Krupp}},
  \bibinfo {author} {\bibfnamefont {J.~B.}\ \bibnamefont {Balewski}}, \bibinfo
  {author} {\bibfnamefont {R.}~\bibnamefont {Löw}}, \bibinfo {author}
  {\bibfnamefont {S.}~\bibnamefont {Hofferberth}}, \ and\ \bibinfo {author}
  {\bibfnamefont {T.}~\bibnamefont {Pfau}},\ }\href {\doibase
  10.1038/ncomms5546} {\bibfield  {journal} {\bibinfo  {journal} {Nature
  Comm.}\ }\textbf {\bibinfo {volume} {5}},\ \bibinfo {pages} {4546} (\bibinfo
  {year} {2014})}\BibitemShut {NoStop}%
\bibitem [{\citenamefont {Eiles}\ \emph {et~al.}(2016)\citenamefont {Eiles},
  \citenamefont {Pérez-Ríos}, \citenamefont {Robicheaux},\ and\ \citenamefont
  {Greene}}]{eiles_ultracold_2016}%
  \BibitemOpen
  \bibfield  {author} {\bibinfo {author} {\bibfnamefont {M.~T.}\ \bibnamefont
  {Eiles}}, \bibinfo {author} {\bibfnamefont {J.}~\bibnamefont {Pérez-Ríos}},
  \bibinfo {author} {\bibfnamefont {F.}~\bibnamefont {Robicheaux}}, \ and\
  \bibinfo {author} {\bibfnamefont {C.~H.}\ \bibnamefont {Greene}},\ }\href
  {\doibase 10.1088/0953-4075/49/11/114005} {\bibfield  {journal} {\bibinfo
  {journal} {J. Phys. B: At. Mol. Opt. Phys.}\ }\textbf {\bibinfo {volume}
  {49}},\ \bibinfo {pages} {114005} (\bibinfo {year} {2016})}\BibitemShut
  {NoStop}%
\bibitem [{\citenamefont {Fey}\ \emph {et~al.}(2016)\citenamefont {Fey},
  \citenamefont {Kurz},\ and\ \citenamefont
  {Schmelcher}}]{fey_stretching_2016}%
  \BibitemOpen
  \bibfield  {author} {\bibinfo {author} {\bibfnamefont {C.}~\bibnamefont
  {Fey}}, \bibinfo {author} {\bibfnamefont {M.}~\bibnamefont {Kurz}}, \ and\
  \bibinfo {author} {\bibfnamefont {P.}~\bibnamefont {Schmelcher}},\ }\href
  {\doibase 10.1103/PhysRevA.94.012516} {\bibfield  {journal} {\bibinfo
  {journal} {Phys. Rev. A}\ }\textbf {\bibinfo {volume} {94}},\ \bibinfo
  {pages} {012516} (\bibinfo {year} {2016})}\BibitemShut {NoStop}%
\bibitem [{\citenamefont {Luukko}\ and\ \citenamefont
  {Rost}(2017)}]{Luukko_Rost_2017}%
  \BibitemOpen
  \bibfield  {author} {\bibinfo {author} {\bibfnamefont {P.~J.}\ \bibnamefont
  {Luukko}}\ and\ \bibinfo {author} {\bibfnamefont {J.-M.}\ \bibnamefont
  {Rost}},\ }\href {\doibase 10.1103/PhysRevLett.119.203001} {\bibfield
  {journal} {\bibinfo  {journal} {Phys. Rev. Lett.}\ }\textbf {\bibinfo
  {volume} {119}},\ \bibinfo {pages} {203001} (\bibinfo {year}
  {2017})}\BibitemShut {NoStop}%
\bibitem [{\citenamefont {Fey}\ \emph {et~al.}(2019{\natexlab{a}})\citenamefont
  {Fey}, \citenamefont {Yang}, \citenamefont {Rittenhouse}, \citenamefont
  {Munkes}, \citenamefont {Baluktsian}, \citenamefont {Schmelcher},
  \citenamefont {Sadeghpour},\ and\ \citenamefont
  {Shaffer}}]{Fey_2019_effective_three_body}%
  \BibitemOpen
  \bibfield  {author} {\bibinfo {author} {\bibfnamefont {C.}~\bibnamefont
  {Fey}}, \bibinfo {author} {\bibfnamefont {J.}~\bibnamefont {Yang}}, \bibinfo
  {author} {\bibfnamefont {S.~T.}\ \bibnamefont {Rittenhouse}}, \bibinfo
  {author} {\bibfnamefont {F.}~\bibnamefont {Munkes}}, \bibinfo {author}
  {\bibfnamefont {M.}~\bibnamefont {Baluktsian}}, \bibinfo {author}
  {\bibfnamefont {P.}~\bibnamefont {Schmelcher}}, \bibinfo {author}
  {\bibfnamefont {H.}~\bibnamefont {Sadeghpour}}, \ and\ \bibinfo {author}
  {\bibfnamefont {J.~P.}\ \bibnamefont {Shaffer}},\ }\href {\doibase
  10.1103/PhysRevLett.122.103001} {\bibfield  {journal} {\bibinfo  {journal}
  {Phys. Rev. Lett.}\ }\textbf {\bibinfo {volume} {122}},\ \bibinfo {pages}
  {103001} (\bibinfo {year} {2019}{\natexlab{a}})}\BibitemShut {NoStop}%
\bibitem [{\citenamefont {Fey}\ \emph {et~al.}(2019{\natexlab{b}})\citenamefont
  {Fey}, \citenamefont {Hummel},\ and\ \citenamefont
  {Schmelcher}}]{Fey_Hummel_Schmelcher_2019_trilobite}%
  \BibitemOpen
  \bibfield  {author} {\bibinfo {author} {\bibfnamefont {C.}~\bibnamefont
  {Fey}}, \bibinfo {author} {\bibfnamefont {F.}~\bibnamefont {Hummel}}, \ and\
  \bibinfo {author} {\bibfnamefont {P.}~\bibnamefont {Schmelcher}},\ }\href
  {\doibase 10.1103/PhysRevA.99.022506} {\bibfield  {journal} {\bibinfo
  {journal} {Phys. Rev. A}\ }\textbf {\bibinfo {volume} {99}} (\bibinfo {year}
  {2019}{\natexlab{b}}),\ 10.1103/PhysRevA.99.022506}\BibitemShut {NoStop}%
\bibitem [{\citenamefont {Schmidt}\ \emph {et~al.}(2016)\citenamefont
  {Schmidt}, \citenamefont {Sadeghpour},\ and\ \citenamefont
  {Demler}}]{schmidt_mesoscopic_2016}%
  \BibitemOpen
  \bibfield  {author} {\bibinfo {author} {\bibfnamefont {R.}~\bibnamefont
  {Schmidt}}, \bibinfo {author} {\bibfnamefont {H.}~\bibnamefont {Sadeghpour}},
  \ and\ \bibinfo {author} {\bibfnamefont {E.}~\bibnamefont {Demler}},\ }\href
  {\doibase 10.1103/PhysRevLett.116.105302} {\bibfield  {journal} {\bibinfo
  {journal} {Phys. Rev. Lett.}\ }\textbf {\bibinfo {volume} {116}},\ \bibinfo
  {pages} {105302} (\bibinfo {year} {2016})}\BibitemShut {NoStop}%
\bibitem [{\citenamefont {Schlagmüller}\ \emph {et~al.}(2016)\citenamefont
  {Schlagmüller}, \citenamefont {Liebisch}, \citenamefont {Nguyen},
  \citenamefont {Lochead}, \citenamefont {Engel}, \citenamefont {Böttcher},
  \citenamefont {Westphal}, \citenamefont {Kleinbach}, \citenamefont {Löw},
  \citenamefont {Hofferberth}, \citenamefont {Pfau}, \citenamefont
  {Pérez-Ríos},\ and\ \citenamefont {Greene}}]{schlagmuller_probing_2016}%
  \BibitemOpen
  \bibfield  {author} {\bibinfo {author} {\bibfnamefont {M.}~\bibnamefont
  {Schlagmüller}}, \bibinfo {author} {\bibfnamefont {T.~C.}\ \bibnamefont
  {Liebisch}}, \bibinfo {author} {\bibfnamefont {H.}~\bibnamefont {Nguyen}},
  \bibinfo {author} {\bibfnamefont {G.}~\bibnamefont {Lochead}}, \bibinfo
  {author} {\bibfnamefont {F.}~\bibnamefont {Engel}}, \bibinfo {author}
  {\bibfnamefont {F.}~\bibnamefont {Böttcher}}, \bibinfo {author}
  {\bibfnamefont {K.~M.}\ \bibnamefont {Westphal}}, \bibinfo {author}
  {\bibfnamefont {K.~S.}\ \bibnamefont {Kleinbach}}, \bibinfo {author}
  {\bibfnamefont {R.}~\bibnamefont {Löw}}, \bibinfo {author} {\bibfnamefont
  {S.}~\bibnamefont {Hofferberth}}, \bibinfo {author} {\bibfnamefont
  {T.}~\bibnamefont {Pfau}}, \bibinfo {author} {\bibfnamefont {J.}~\bibnamefont
  {Pérez-Ríos}}, \ and\ \bibinfo {author} {\bibfnamefont {C.~H.}\
  \bibnamefont {Greene}},\ }\href
  {http://link.aps.org/doi/10.1103/PhysRevLett.116.053001} {\bibfield
  {journal} {\bibinfo  {journal} {Phys. Rev. Lett.}\ }\textbf {\bibinfo
  {volume} {116}},\ \bibinfo {pages} {053001} (\bibinfo {year}
  {2016})}\BibitemShut {NoStop}%
\bibitem [{\citenamefont {Camargo}\ \emph
  {et~al.}(2018{\natexlab{a}})\citenamefont {Camargo}, \citenamefont {Schmidt},
  \citenamefont {Whalen}, \citenamefont {Ding}, \citenamefont {Woehl},
  \citenamefont {Yoshida}, \citenamefont {Burgdörfer}, \citenamefont
  {Dunning}, \citenamefont {Sadeghpour}, \citenamefont {Demler},\ and\
  \citenamefont {Killian}}]{camargo_creation_2018}%
  \BibitemOpen
  \bibfield  {author} {\bibinfo {author} {\bibfnamefont {F.}~\bibnamefont
  {Camargo}}, \bibinfo {author} {\bibfnamefont {R.}~\bibnamefont {Schmidt}},
  \bibinfo {author} {\bibfnamefont {J.}~\bibnamefont {Whalen}}, \bibinfo
  {author} {\bibfnamefont {R.}~\bibnamefont {Ding}}, \bibinfo {author}
  {\bibfnamefont {G.}~\bibnamefont {Woehl}}, \bibinfo {author} {\bibfnamefont
  {S.}~\bibnamefont {Yoshida}}, \bibinfo {author} {\bibfnamefont
  {J.}~\bibnamefont {Burgdörfer}}, \bibinfo {author} {\bibfnamefont
  {F.}~\bibnamefont {Dunning}}, \bibinfo {author} {\bibfnamefont
  {H.}~\bibnamefont {Sadeghpour}}, \bibinfo {author} {\bibfnamefont
  {E.}~\bibnamefont {Demler}}, \ and\ \bibinfo {author} {\bibfnamefont
  {T.}~\bibnamefont {Killian}},\ }\href
  {https://link.aps.org/doi/10.1103/PhysRevLett.120.083401} {\bibfield
  {journal} {\bibinfo  {journal} {Phys. Rev. Lett.}\ }\textbf {\bibinfo
  {volume} {120}},\ \bibinfo {pages} {083401} (\bibinfo {year}
  {2018}{\natexlab{a}})}\BibitemShut {NoStop}%
\bibitem [{\citenamefont {Kleinbach}\ \emph {et~al.}(2018)\citenamefont
  {Kleinbach}, \citenamefont {Engel}, \citenamefont {Dieterle}, \citenamefont
  {Löw}, \citenamefont {Pfau},\ and\ \citenamefont
  {Meinert}}]{Kleinbach_2018}%
  \BibitemOpen
  \bibfield  {author} {\bibinfo {author} {\bibfnamefont {K.}~\bibnamefont
  {Kleinbach}}, \bibinfo {author} {\bibfnamefont {F.}~\bibnamefont {Engel}},
  \bibinfo {author} {\bibfnamefont {T.}~\bibnamefont {Dieterle}}, \bibinfo
  {author} {\bibfnamefont {R.}~\bibnamefont {Löw}}, \bibinfo {author}
  {\bibfnamefont {T.}~\bibnamefont {Pfau}}, \ and\ \bibinfo {author}
  {\bibfnamefont {F.}~\bibnamefont {Meinert}},\ }\href {\doibase
  10.1103/PhysRevLett.120.193401} {\bibfield  {journal} {\bibinfo  {journal}
  {Phys. Rev. Lett.}\ }\textbf {\bibinfo {volume} {120}},\ \bibinfo {pages}
  {193401} (\bibinfo {year} {2018})}\BibitemShut {NoStop}%
\bibitem [{\citenamefont {Manthey}\ \emph {et~al.}(2015)\citenamefont
  {Manthey}, \citenamefont {Niederprüm}, \citenamefont {Thomas},\ and\
  \citenamefont {Ott}}]{manthey_dynamically_2015}%
  \BibitemOpen
  \bibfield  {author} {\bibinfo {author} {\bibfnamefont {T.}~\bibnamefont
  {Manthey}}, \bibinfo {author} {\bibfnamefont {T.}~\bibnamefont
  {Niederprüm}}, \bibinfo {author} {\bibfnamefont {O.}~\bibnamefont {Thomas}},
  \ and\ \bibinfo {author} {\bibfnamefont {H.}~\bibnamefont {Ott}},\ }\href
  {\doibase 10.1088/1367-2630/17/10/103024} {\bibfield  {journal} {\bibinfo
  {journal} {New J. Phys.}\ }\textbf {\bibinfo {volume} {17}},\ \bibinfo
  {pages} {103024} (\bibinfo {year} {2015})}\BibitemShut {NoStop}%
\bibitem [{\citenamefont {Whalen}\ \emph
  {et~al.}(2019{\natexlab{a}})\citenamefont {Whalen}, \citenamefont {Ding},
  \citenamefont {Kanungo}, \citenamefont {Killian}, \citenamefont {Yoshida},
  \citenamefont {Burgdörfer},\ and\ \citenamefont {Dunning}}]{Whalen_2019}%
  \BibitemOpen
  \bibfield  {author} {\bibinfo {author} {\bibfnamefont {J.~D.}\ \bibnamefont
  {Whalen}}, \bibinfo {author} {\bibfnamefont {R.}~\bibnamefont {Ding}},
  \bibinfo {author} {\bibfnamefont {S.~K.}\ \bibnamefont {Kanungo}}, \bibinfo
  {author} {\bibfnamefont {T.~C.}\ \bibnamefont {Killian}}, \bibinfo {author}
  {\bibfnamefont {S.}~\bibnamefont {Yoshida}}, \bibinfo {author} {\bibfnamefont
  {J.}~\bibnamefont {Burgdörfer}}, \ and\ \bibinfo {author} {\bibfnamefont
  {F.~B.}\ \bibnamefont {Dunning}},\ }\href {\doibase
  10.1080/00268976.2019.1575485} {\bibfield  {journal} {\bibinfo  {journal}
  {Mol. Phys.}\ ,\ \bibinfo {pages} {1}} (\bibinfo {year}
  {2019}{\natexlab{a}})}\BibitemShut {NoStop}%
\bibitem [{\citenamefont {Whalen}\ \emph
  {et~al.}(2019{\natexlab{b}})\citenamefont {Whalen}, \citenamefont {Kanungo},
  \citenamefont {Ding}, \citenamefont {Wagner}, \citenamefont {Schmidt},
  \citenamefont {Sadeghpour}, \citenamefont {Yoshida}, \citenamefont
  {Burgd\"orfer}, \citenamefont {Dunning},\ and\ \citenamefont
  {Killian}}]{Whalen_Schmidt_Wagner_2019}%
  \BibitemOpen
  \bibfield  {author} {\bibinfo {author} {\bibfnamefont {J.~D.}\ \bibnamefont
  {Whalen}}, \bibinfo {author} {\bibfnamefont {S.~K.}\ \bibnamefont {Kanungo}},
  \bibinfo {author} {\bibfnamefont {R.}~\bibnamefont {Ding}}, \bibinfo {author}
  {\bibfnamefont {M.}~\bibnamefont {Wagner}}, \bibinfo {author} {\bibfnamefont
  {R.}~\bibnamefont {Schmidt}}, \bibinfo {author} {\bibfnamefont {H.~R.}\
  \bibnamefont {Sadeghpour}}, \bibinfo {author} {\bibfnamefont
  {S.}~\bibnamefont {Yoshida}}, \bibinfo {author} {\bibfnamefont
  {J.}~\bibnamefont {Burgd\"orfer}}, \bibinfo {author} {\bibfnamefont {F.~B.}\
  \bibnamefont {Dunning}}, \ and\ \bibinfo {author} {\bibfnamefont {T.~C.}\
  \bibnamefont {Killian}},\ }\href {\doibase 10.1103/PhysRevA.100.011402}
  {\bibfield  {journal} {\bibinfo  {journal} {Phys. Rev. A}\ }\textbf {\bibinfo
  {volume} {100}},\ \bibinfo {pages} {011402} (\bibinfo {year}
  {2019}{\natexlab{b}})}\BibitemShut {NoStop}%
\bibitem [{\citenamefont {Anderson}\ \emph
  {et~al.}(2014{\natexlab{a}})\citenamefont {Anderson}, \citenamefont
  {Miller},\ and\ \citenamefont {Raithel}}]{anderson_photoassociation_2014}%
  \BibitemOpen
  \bibfield  {author} {\bibinfo {author} {\bibfnamefont {D.}~\bibnamefont
  {Anderson}}, \bibinfo {author} {\bibfnamefont {S.}~\bibnamefont {Miller}}, \
  and\ \bibinfo {author} {\bibfnamefont {G.}~\bibnamefont {Raithel}},\ }\href
  {\doibase 10.1103/PhysRevLett.112.163201} {\bibfield  {journal} {\bibinfo
  {journal} {Phys. Rev. Lett.}\ }\textbf {\bibinfo {volume} {112}},\ \bibinfo
  {pages} {163201} (\bibinfo {year} {2014}{\natexlab{a}})}\BibitemShut
  {NoStop}%
\bibitem [{\citenamefont {Saßmannshausen}\ \emph {et~al.}(2015)\citenamefont
  {Saßmannshausen}, \citenamefont {Merkt},\ and\ \citenamefont
  {Deiglmayr}}]{sasmannshausen_experimental_2015}%
  \BibitemOpen
  \bibfield  {author} {\bibinfo {author} {\bibfnamefont {H.}~\bibnamefont
  {Saßmannshausen}}, \bibinfo {author} {\bibfnamefont {F.}~\bibnamefont
  {Merkt}}, \ and\ \bibinfo {author} {\bibfnamefont {J.}~\bibnamefont
  {Deiglmayr}},\ }\href {\doibase 10.1103/PhysRevLett.114.133201} {\bibfield
  {journal} {\bibinfo  {journal} {Phys. Rev. Lett.}\ }\textbf {\bibinfo
  {volume} {114}},\ \bibinfo {pages} {133201} (\bibinfo {year}
  {2015})}\BibitemShut {NoStop}%
\bibitem [{\citenamefont {Böttcher}\ \emph {et~al.}(2016)\citenamefont
  {Böttcher}, \citenamefont {Gaj}, \citenamefont {Westphal}, \citenamefont
  {Schlagmüller}, \citenamefont {Kleinbach}, \citenamefont {Löw},
  \citenamefont {Liebisch}, \citenamefont {Pfau},\ and\ \citenamefont
  {Hofferberth}}]{bottcher_observation_2016}%
  \BibitemOpen
  \bibfield  {author} {\bibinfo {author} {\bibfnamefont {F.}~\bibnamefont
  {Böttcher}}, \bibinfo {author} {\bibfnamefont {A.}~\bibnamefont {Gaj}},
  \bibinfo {author} {\bibfnamefont {K.~M.}\ \bibnamefont {Westphal}}, \bibinfo
  {author} {\bibfnamefont {M.}~\bibnamefont {Schlagmüller}}, \bibinfo {author}
  {\bibfnamefont {K.~S.}\ \bibnamefont {Kleinbach}}, \bibinfo {author}
  {\bibfnamefont {R.}~\bibnamefont {Löw}}, \bibinfo {author} {\bibfnamefont
  {T.~C.}\ \bibnamefont {Liebisch}}, \bibinfo {author} {\bibfnamefont
  {T.}~\bibnamefont {Pfau}}, \ and\ \bibinfo {author} {\bibfnamefont
  {S.}~\bibnamefont {Hofferberth}},\ }\href {\doibase
  10.1103/PhysRevA.93.032512} {\bibfield  {journal} {\bibinfo  {journal} {Phys.
  Rev. A}\ }\textbf {\bibinfo {volume} {93}},\ \bibinfo {pages} {032512}
  (\bibinfo {year} {2016})}\BibitemShut {NoStop}%
\bibitem [{\citenamefont {MacLennan}\ \emph {et~al.}(2019)\citenamefont
  {MacLennan}, \citenamefont {Chen},\ and\ \citenamefont
  {Raithel}}]{MacLennan_Chen_Raithel_2018}%
  \BibitemOpen
  \bibfield  {author} {\bibinfo {author} {\bibfnamefont {J.~L.}\ \bibnamefont
  {MacLennan}}, \bibinfo {author} {\bibfnamefont {Y.-J.}\ \bibnamefont {Chen}},
  \ and\ \bibinfo {author} {\bibfnamefont {G.}~\bibnamefont {Raithel}},\ }\href
  {\doibase 10.1103/PhysRevA.99.033407} {\bibfield  {journal} {\bibinfo
  {journal} {Phys. Rev. A}\ }\textbf {\bibinfo {volume} {99}},\ \bibinfo
  {pages} {033407} (\bibinfo {year} {2019})}\BibitemShut {NoStop}%
\bibitem [{\citenamefont {Engel}\ \emph {et~al.}(2019)\citenamefont {Engel},
  \citenamefont {Dieterle}, \citenamefont {Hummel}, \citenamefont {Fey},
  \citenamefont {Schmelcher}, \citenamefont {Löw}, \citenamefont {Pfau},\ and\
  \citenamefont {Meinert}}]{Engel_Fey_Meinert_2019}%
  \BibitemOpen
  \bibfield  {author} {\bibinfo {author} {\bibfnamefont {F.}~\bibnamefont
  {Engel}}, \bibinfo {author} {\bibfnamefont {T.}~\bibnamefont {Dieterle}},
  \bibinfo {author} {\bibfnamefont {F.}~\bibnamefont {Hummel}}, \bibinfo
  {author} {\bibfnamefont {C.}~\bibnamefont {Fey}}, \bibinfo {author}
  {\bibfnamefont {P.}~\bibnamefont {Schmelcher}}, \bibinfo {author}
  {\bibfnamefont {R.}~\bibnamefont {Löw}}, \bibinfo {author} {\bibfnamefont
  {T.}~\bibnamefont {Pfau}}, \ and\ \bibinfo {author} {\bibfnamefont
  {F.}~\bibnamefont {Meinert}},\ }\href
  {https://journals.aps.org/prl/abstract/10.1103/PhysRevLett.123.073003}
  {\bibfield  {journal} {\bibinfo  {journal} {Phys. Rev. Lett}\ }\textbf
  {\bibinfo {volume} {123}},\ \bibinfo {pages} {073003} (\bibinfo {year}
  {2019})}\BibitemShut {NoStop}%
\bibitem [{\citenamefont {Shaffer}\ \emph {et~al.}(2018)\citenamefont
  {Shaffer}, \citenamefont {Rittenhouse},\ and\ \citenamefont
  {Sadeghpour}}]{Shaffer_2018}%
  \BibitemOpen
  \bibfield  {author} {\bibinfo {author} {\bibfnamefont {J.~P.}\ \bibnamefont
  {Shaffer}}, \bibinfo {author} {\bibfnamefont {S.~T.}\ \bibnamefont
  {Rittenhouse}}, \ and\ \bibinfo {author} {\bibfnamefont {H.~R.}\ \bibnamefont
  {Sadeghpour}},\ }\href {\doibase 10.1038/s41467-018-04135-6} {\bibfield
  {journal} {\bibinfo  {journal} {Nature Comm.}\ }\textbf {\bibinfo {volume}
  {9}},\ \bibinfo {pages} {1965} (\bibinfo {year} {2018})}\BibitemShut
  {NoStop}%
\bibitem [{\citenamefont {Marcassa}\ and\ \citenamefont
  {Shaffer}(2014)}]{marcassa_interactions_2014}%
  \BibitemOpen
  \bibfield  {author} {\bibinfo {author} {\bibfnamefont {L.~G.}\ \bibnamefont
  {Marcassa}}\ and\ \bibinfo {author} {\bibfnamefont {J.~P.}\ \bibnamefont
  {Shaffer}},\ }\href
  {http://linkinghub.elsevier.com/retrieve/pii/B978012800129500002X} {\bibfield
   {journal} {\bibinfo  {journal} {Adv. At. Mol. Opt. Phys.}\ }\textbf
  {\bibinfo {volume} {63}},\ \bibinfo {pages} {47} (\bibinfo {year}
  {2014})}\BibitemShut {NoStop}%
\bibitem [{\citenamefont {Eiles}(2019)}]{Eiles_2019_exotic_specimen}%
  \BibitemOpen
  \bibfield  {author} {\bibinfo {author} {\bibfnamefont {M.~T.}\ \bibnamefont
  {Eiles}},\ }\href {http://arxiv.org/abs/1902.10803} {\bibfield  {journal}
  {\bibinfo  {journal} {arxiv preprint: 1902.10803}\ } (\bibinfo {year}
  {2019})}\BibitemShut {NoStop}%
\bibitem [{\citenamefont {Gaj}(2016)}]{Gaj_Review_2016}%
  \BibitemOpen
  \bibfield  {author} {\bibinfo {author} {\bibfnamefont {A.}~\bibnamefont
  {Gaj}},\ }\href {\doibase 10.1140/epjst/e2016-60008-6} {\bibfield  {journal}
  {\bibinfo  {journal} {Eur. Phys. J. Special Topics}\ }\textbf {\bibinfo
  {volume} {225}},\ \bibinfo {pages} {2919} (\bibinfo {year}
  {2016})}\BibitemShut {NoStop}%
\bibitem [{\citenamefont {Liebisch}\ \emph {et~al.}(2016)\citenamefont
  {Liebisch}, \citenamefont {Schlagmüller}, \citenamefont {Engel},
  \citenamefont {Nguyen}, \citenamefont {Balewski}, \citenamefont {Lochead},
  \citenamefont {Böttcher}, \citenamefont {Westphal}, \citenamefont
  {Kleinbach}, \citenamefont {Schmid}, \citenamefont {Gaj}, \citenamefont
  {Löw}, \citenamefont {Hofferberth}, \citenamefont {Pérez-Ríos},\ and\
  \citenamefont {Greene}}]{Liebisch_2016}%
  \BibitemOpen
  \bibfield  {author} {\bibinfo {author} {\bibfnamefont {T.~C.}\ \bibnamefont
  {Liebisch}}, \bibinfo {author} {\bibfnamefont {M.}~\bibnamefont
  {Schlagmüller}}, \bibinfo {author} {\bibfnamefont {F.}~\bibnamefont
  {Engel}}, \bibinfo {author} {\bibfnamefont {H.}~\bibnamefont {Nguyen}},
  \bibinfo {author} {\bibfnamefont {J.}~\bibnamefont {Balewski}}, \bibinfo
  {author} {\bibfnamefont {G.}~\bibnamefont {Lochead}}, \bibinfo {author}
  {\bibfnamefont {F.}~\bibnamefont {Böttcher}}, \bibinfo {author}
  {\bibfnamefont {K.~M.}\ \bibnamefont {Westphal}}, \bibinfo {author}
  {\bibfnamefont {K.~S.}\ \bibnamefont {Kleinbach}}, \bibinfo {author}
  {\bibfnamefont {T.}~\bibnamefont {Schmid}}, \bibinfo {author} {\bibfnamefont
  {A.}~\bibnamefont {Gaj}}, \bibinfo {author} {\bibfnamefont {R.}~\bibnamefont
  {Löw}}, \bibinfo {author} {\bibfnamefont {T.}~\bibnamefont {Hofferberth},
  \bibfnamefont {Sebastian anf~Pfau}}, \bibinfo {author} {\bibfnamefont
  {J.}~\bibnamefont {Pérez-Ríos}}, \ and\ \bibinfo {author} {\bibfnamefont
  {C.~H.}\ \bibnamefont {Greene}},\ }\href {\doibase
  10.1088/0953-4075/49/18/182001} {\bibfield  {journal} {\bibinfo  {journal}
  {J. Phys. B: At. Mol. Opt. Phys.}\ }\textbf {\bibinfo {volume} {49}},\
  \bibinfo {pages} {182001} (\bibinfo {year} {2016})}\BibitemShut {NoStop}%
\bibitem [{\citenamefont {Saßmannshausen}\ \emph {et~al.}(2016)\citenamefont
  {Saßmannshausen}, \citenamefont {Deiglmayr},\ and\ \citenamefont
  {Merkt}}]{Sassmannshausen_Review_2016}%
  \BibitemOpen
  \bibfield  {author} {\bibinfo {author} {\bibfnamefont {H.}~\bibnamefont
  {Saßmannshausen}}, \bibinfo {author} {\bibfnamefont {J.}~\bibnamefont
  {Deiglmayr}}, \ and\ \bibinfo {author} {\bibfnamefont {F.}~\bibnamefont
  {Merkt}},\ }\href {\doibase 10.1140/epjst/e2016-60124-9} {\bibfield
  {journal} {\bibinfo  {journal} {Eur. Phys. J. Special Topics}\ }\textbf
  {\bibinfo {volume} {225}},\ \bibinfo {pages} {2891} (\bibinfo {year}
  {2016})}\BibitemShut {NoStop}%
\bibitem [{\citenamefont {Lippe}\ \emph {et~al.}(2019)\citenamefont {Lippe},
  \citenamefont {Eichert}, \citenamefont {Thomas}, \citenamefont
  {Niederprüm},\ and\ \citenamefont {Ott}}]{Lippe_Eichert_Review_2019}%
  \BibitemOpen
  \bibfield  {author} {\bibinfo {author} {\bibfnamefont {C.}~\bibnamefont
  {Lippe}}, \bibinfo {author} {\bibfnamefont {T.}~\bibnamefont {Eichert}},
  \bibinfo {author} {\bibfnamefont {O.}~\bibnamefont {Thomas}}, \bibinfo
  {author} {\bibfnamefont {T.}~\bibnamefont {Niederprüm}}, \ and\ \bibinfo
  {author} {\bibfnamefont {H.}~\bibnamefont {Ott}},\ }\href {\doibase
  10.1002/pssb.201800654} {\bibfield  {journal} {\bibinfo  {journal} {physica
  status solidi (b)}\ ,\ \bibinfo {pages} {1800654}} (\bibinfo {year}
  {2019})}\BibitemShut {NoStop}%
\bibitem [{\citenamefont {Amaldi}\ and\ \citenamefont
  {Segré}(1934{\natexlab{a}})}]{Amaldi_1934}%
  \BibitemOpen
  \bibfield  {author} {\bibinfo {author} {\bibfnamefont {E.}~\bibnamefont
  {Amaldi}}\ and\ \bibinfo {author} {\bibfnamefont {E.}~\bibnamefont
  {Segré}},\ }\href {\doibase 10.1007/BF02959828} {\bibfield  {journal}
  {\bibinfo  {journal} {Il Nuovo Cimento}\ }\textbf {\bibinfo {volume} {11}},\
  \bibinfo {pages} {145} (\bibinfo {year} {1934}{\natexlab{a}})}\BibitemShut
  {NoStop}%
\bibitem [{\citenamefont {Amaldi}\ and\ \citenamefont
  {Segré}(1934{\natexlab{b}})}]{Amaldi_1934_nature}%
  \BibitemOpen
  \bibfield  {author} {\bibinfo {author} {\bibfnamefont {E.}~\bibnamefont
  {Amaldi}}\ and\ \bibinfo {author} {\bibfnamefont {E.}~\bibnamefont
  {Segré}},\ }\href {\doibase 10.1038/133141a0} {\bibfield  {journal}
  {\bibinfo  {journal} {Nature}\ }\textbf {\bibinfo {volume} {133}},\ \bibinfo
  {pages} {141} (\bibinfo {year} {1934}{\natexlab{b}})}\BibitemShut {NoStop}%
\bibitem [{\citenamefont {Füchtbauer}\ and\ \citenamefont
  {Gössler}(1934)}]{Fuechtbauer_1934}%
  \BibitemOpen
  \bibfield  {author} {\bibinfo {author} {\bibfnamefont {C.}~\bibnamefont
  {Füchtbauer}}\ and\ \bibinfo {author} {\bibfnamefont {F.}~\bibnamefont
  {Gössler}},\ }\href@noop {} {\bibfield  {journal} {\bibinfo  {journal}
  {Zeitschrift für Physik}\ }\textbf {\bibinfo {volume} {87}},\ \bibinfo
  {pages} {89} (\bibinfo {year} {1934})}\BibitemShut {NoStop}%
\bibitem [{\citenamefont {Füchtbauer}\ \emph {et~al.}(1934)\citenamefont
  {Füchtbauer}, \citenamefont {Schulz},\ and\ \citenamefont
  {Brandt}}]{Fuechtbauer_1934_2}%
  \BibitemOpen
  \bibfield  {author} {\bibinfo {author} {\bibfnamefont {C.}~\bibnamefont
  {Füchtbauer}}, \bibinfo {author} {\bibfnamefont {P.}~\bibnamefont {Schulz}},
  \ and\ \bibinfo {author} {\bibfnamefont {A.~F.}\ \bibnamefont {Brandt}},\
  }\href {\doibase 10.1007/BF01334059} {\bibfield  {journal} {\bibinfo
  {journal} {Zeitschrift für Physik}\ }\textbf {\bibinfo {volume} {90}},\
  \bibinfo {pages} {403} (\bibinfo {year} {1934})}\BibitemShut {NoStop}%
\bibitem [{\citenamefont {Fermi}(1934)}]{fermi_sopra_1934}%
  \BibitemOpen
  \bibfield  {author} {\bibinfo {author} {\bibfnamefont {E.}~\bibnamefont
  {Fermi}},\ }\href {http://www.springerlink.com/index/04267161T7608557.pdf}
  {\bibfield  {journal} {\bibinfo  {journal} {Il Nuovo Cimento (1924-1942)}\
  }\textbf {\bibinfo {volume} {11}},\ \bibinfo {pages} {157} (\bibinfo {year}
  {1934})}\BibitemShut {NoStop}%
\bibitem [{\citenamefont {Ch’en}\ and\ \citenamefont
  {Takeo}(1957)}]{Takeo_1957}%
  \BibitemOpen
  \bibfield  {author} {\bibinfo {author} {\bibfnamefont {S.-y.}\ \bibnamefont
  {Ch’en}}\ and\ \bibinfo {author} {\bibfnamefont {M.}~\bibnamefont
  {Takeo}},\ }\href {\doibase 10.1103/RevModPhys.29.20} {\bibfield  {journal}
  {\bibinfo  {journal} {Rev. Mod. Phys.}\ }\textbf {\bibinfo {volume} {29}},\
  \bibinfo {pages} {20} (\bibinfo {year} {1957})}\BibitemShut {NoStop}%
\bibitem [{\citenamefont {Allard}\ and\ \citenamefont
  {Kielkopf}(1982)}]{Allard_1982}%
  \BibitemOpen
  \bibfield  {author} {\bibinfo {author} {\bibfnamefont {N.}~\bibnamefont
  {Allard}}\ and\ \bibinfo {author} {\bibfnamefont {J.}~\bibnamefont
  {Kielkopf}},\ }\href {\doibase 10.1103/RevModPhys.54.1103} {\bibfield
  {journal} {\bibinfo  {journal} {Rev. Mod. Phys.}\ }\textbf {\bibinfo {volume}
  {54}},\ \bibinfo {pages} {1103} (\bibinfo {year} {1982})}\BibitemShut
  {NoStop}%
\bibitem [{\citenamefont {Beigman}\ and\ \citenamefont
  {Lebedev}(1995)}]{Beigman_Lebedev_1995}%
  \BibitemOpen
  \bibfield  {author} {\bibinfo {author} {\bibfnamefont {I.}~\bibnamefont
  {Beigman}}\ and\ \bibinfo {author} {\bibfnamefont {V.}~\bibnamefont
  {Lebedev}},\ }\href {\doibase 10.1016/0370-1573(95)00074-Q} {\bibfield
  {journal} {\bibinfo  {journal} {Phys. Rep.}\ }\textbf {\bibinfo {volume}
  {250}},\ \bibinfo {pages} {95} (\bibinfo {year} {1995})}\BibitemShut
  {NoStop}%
\bibitem [{\citenamefont {Stebbings}\ and\ \citenamefont
  {Dunning}(1983)}]{Stebbings_Dunning_2011}%
  \BibitemOpen
  \bibfield  {author} {\bibinfo {author} {\bibfnamefont {R.~F.}\ \bibnamefont
  {Stebbings}}\ and\ \bibinfo {author} {\bibfnamefont {F.~B.}\ \bibnamefont
  {Dunning}},\ }\href@noop {} {\emph {\bibinfo {title} {Rydberg states of atoms
  and molecules}}}\ (\bibinfo  {publisher} {Cambridge University Press},\
  \bibinfo {year} {1983})\BibitemShut {NoStop}%
\bibitem [{\citenamefont {Phillips}(1998)}]{Phillips_1998}%
  \BibitemOpen
  \bibfield  {author} {\bibinfo {author} {\bibfnamefont {W.~D.}\ \bibnamefont
  {Phillips}},\ }\href {\doibase 10.1103/RevModPhys.70.721} {\bibfield
  {journal} {\bibinfo  {journal} {Rev. Mod. Phys.}\ }\textbf {\bibinfo {volume}
  {70}},\ \bibinfo {pages} {721} (\bibinfo {year} {1998})}\BibitemShut
  {NoStop}%
\bibitem [{\citenamefont {Ketterle}(2002)}]{Ketterle_2002}%
  \BibitemOpen
  \bibfield  {author} {\bibinfo {author} {\bibfnamefont {W.}~\bibnamefont
  {Ketterle}},\ }\href {\doibase 10.1103/RevModPhys.74.1131} {\bibfield
  {journal} {\bibinfo  {journal} {Rev. Mod. Phys.}\ }\textbf {\bibinfo {volume}
  {74}},\ \bibinfo {pages} {1131} (\bibinfo {year} {2002})}\BibitemShut
  {NoStop}%
\bibitem [{\citenamefont {Li}\ \emph {et~al.}(2003)\citenamefont {Li},
  \citenamefont {Mourachko}, \citenamefont {Noel},\ and\ \citenamefont
  {Gallagher}}]{Gallagher_quantum_defects_2003}%
  \BibitemOpen
  \bibfield  {author} {\bibinfo {author} {\bibfnamefont {W.}~\bibnamefont
  {Li}}, \bibinfo {author} {\bibfnamefont {I.}~\bibnamefont {Mourachko}},
  \bibinfo {author} {\bibfnamefont {M.~W.}\ \bibnamefont {Noel}}, \ and\
  \bibinfo {author} {\bibfnamefont {T.~F.}\ \bibnamefont {Gallagher}},\ }\href
  {\doibase 10.1103/PhysRevA.67.052502} {\bibfield  {journal} {\bibinfo
  {journal} {Phys. Rev. A}\ }\textbf {\bibinfo {volume} {67}},\ \bibinfo
  {pages} {052502} (\bibinfo {year} {2003})}\BibitemShut {NoStop}%
\bibitem [{\citenamefont {Li}\ \emph {et~al.}(2011)\citenamefont {Li},
  \citenamefont {Pohl}, \citenamefont {Rost}, \citenamefont {Rittenhouse},
  \citenamefont {Sadeghpour}, \citenamefont {Nipper}, \citenamefont {Butscher},
  \citenamefont {Balewski}, \citenamefont {Bendkowsky}, \citenamefont {Low},\
  and\ \citenamefont {Pfau}}]{li_homonuclear_2011}%
  \BibitemOpen
  \bibfield  {author} {\bibinfo {author} {\bibfnamefont {W.}~\bibnamefont
  {Li}}, \bibinfo {author} {\bibfnamefont {T.}~\bibnamefont {Pohl}}, \bibinfo
  {author} {\bibfnamefont {J.~M.}\ \bibnamefont {Rost}}, \bibinfo {author}
  {\bibfnamefont {S.~T.}\ \bibnamefont {Rittenhouse}}, \bibinfo {author}
  {\bibfnamefont {H.~R.}\ \bibnamefont {Sadeghpour}}, \bibinfo {author}
  {\bibfnamefont {J.}~\bibnamefont {Nipper}}, \bibinfo {author} {\bibfnamefont
  {B.}~\bibnamefont {Butscher}}, \bibinfo {author} {\bibfnamefont {J.~B.}\
  \bibnamefont {Balewski}}, \bibinfo {author} {\bibfnamefont {V.}~\bibnamefont
  {Bendkowsky}}, \bibinfo {author} {\bibfnamefont {R.}~\bibnamefont {Low}}, \
  and\ \bibinfo {author} {\bibfnamefont {T.}~\bibnamefont {Pfau}},\ }\href
  {\doibase 10.1126/science.1211255} {\bibfield  {journal} {\bibinfo  {journal}
  {Science}\ }\textbf {\bibinfo {volume} {334}},\ \bibinfo {pages} {1110}
  (\bibinfo {year} {2011})}\BibitemShut {NoStop}%
\bibitem [{\citenamefont {Hamilton}\ \emph {et~al.}(2002)\citenamefont
  {Hamilton}, \citenamefont {Greene},\ and\ \citenamefont
  {Sadeghpour}}]{hamilton_shape-resonance-induced_2002}%
  \BibitemOpen
  \bibfield  {author} {\bibinfo {author} {\bibfnamefont {E.~L.}\ \bibnamefont
  {Hamilton}}, \bibinfo {author} {\bibfnamefont {C.~H.}\ \bibnamefont
  {Greene}}, \ and\ \bibinfo {author} {\bibfnamefont {H.~R.}\ \bibnamefont
  {Sadeghpour}},\ }\href {http://iopscience.iop.org/0953-4075/35/10/102}
  {\bibfield  {journal} {\bibinfo  {journal} {J. Phys. B: At. Mol. Opt. Phys.}\
  }\textbf {\bibinfo {volume} {35}},\ \bibinfo {pages} {L199} (\bibinfo {year}
  {2002})}\BibitemShut {NoStop}%
\bibitem [{\citenamefont {Chibisov}\ \emph {et~al.}(2002)\citenamefont
  {Chibisov}, \citenamefont {Khuskivadze},\ and\ \citenamefont
  {Fabrikant}}]{chibisov_2002_pwave}%
  \BibitemOpen
  \bibfield  {author} {\bibinfo {author} {\bibfnamefont {M.~I.}\ \bibnamefont
  {Chibisov}}, \bibinfo {author} {\bibfnamefont {A.~A.}\ \bibnamefont
  {Khuskivadze}}, \ and\ \bibinfo {author} {\bibfnamefont {I.~I.}\ \bibnamefont
  {Fabrikant}},\ }\href@noop {} {\bibfield  {journal} {\bibinfo  {journal} {J.
  Phys. B: At. Mol. Opt. Phys.}\ }\textbf {\bibinfo {volume} {35}},\ \bibinfo
  {pages} {L193} (\bibinfo {year} {2002})}\BibitemShut {NoStop}%
\bibitem [{\citenamefont {Thumm}\ and\ \citenamefont
  {Norcross}(1991)}]{Thumm_Norcross_1991_evidence_for}%
  \BibitemOpen
  \bibfield  {author} {\bibinfo {author} {\bibfnamefont {U.}~\bibnamefont
  {Thumm}}\ and\ \bibinfo {author} {\bibfnamefont {D.~W.}\ \bibnamefont
  {Norcross}},\ }\href {\doibase 10.1103/PhysRevLett.67.3495} {\bibfield
  {journal} {\bibinfo  {journal} {Phys. Rev. Lett.}\ }\textbf {\bibinfo
  {volume} {67}},\ \bibinfo {pages} {3495} (\bibinfo {year}
  {1991})}\BibitemShut {NoStop}%
\bibitem [{\citenamefont {Bahrim}\ and\ \citenamefont
  {Thumm}(2000)}]{bahrim_low-lying_2000}%
  \BibitemOpen
  \bibfield  {author} {\bibinfo {author} {\bibfnamefont {C.}~\bibnamefont
  {Bahrim}}\ and\ \bibinfo {author} {\bibfnamefont {U.}~\bibnamefont {Thumm}},\
  }\href {http://journals.aps.org/pra/abstract/10.1103/PhysRevA.61.022722}
  {\bibfield  {journal} {\bibinfo  {journal} {Phys. Rev. A}\ }\textbf {\bibinfo
  {volume} {61}},\ \bibinfo {pages} {022722} (\bibinfo {year}
  {2000})}\BibitemShut {NoStop}%
\bibitem [{\citenamefont {Eiles}(2018)}]{Eiles_2018_two_component}%
  \BibitemOpen
  \bibfield  {author} {\bibinfo {author} {\bibfnamefont {M.~T.}\ \bibnamefont
  {Eiles}},\ }\href {\doibase 10.1103/PhysRevA.98.042706} {\bibfield  {journal}
  {\bibinfo  {journal} {Phys. Rev. A}\ }\textbf {\bibinfo {volume} {98}},\
  \bibinfo {pages} {042706} (\bibinfo {year} {2018})}\BibitemShut {NoStop}%
\bibitem [{\citenamefont {Booth}\ \emph {et~al.}(2015)\citenamefont {Booth},
  \citenamefont {Rittenhouse}, \citenamefont {Yang}, \citenamefont
  {Sadeghpour},\ and\ \citenamefont {Shaffer}}]{booth_production_2015}%
  \BibitemOpen
  \bibfield  {author} {\bibinfo {author} {\bibfnamefont {D.}~\bibnamefont
  {Booth}}, \bibinfo {author} {\bibfnamefont {S.~T.}\ \bibnamefont
  {Rittenhouse}}, \bibinfo {author} {\bibfnamefont {J.}~\bibnamefont {Yang}},
  \bibinfo {author} {\bibfnamefont {H.~R.}\ \bibnamefont {Sadeghpour}}, \ and\
  \bibinfo {author} {\bibfnamefont {J.~P.}\ \bibnamefont {Shaffer}},\ }\href
  {http://www.sciencemag.org/content/348/6230/99.short} {\bibfield  {journal}
  {\bibinfo  {journal} {Science}\ }\textbf {\bibinfo {volume} {348}},\ \bibinfo
  {pages} {99} (\bibinfo {year} {2015})}\BibitemShut {NoStop}%
\bibitem [{\citenamefont {Butscher}\ \emph {et~al.}(2010)\citenamefont
  {Butscher}, \citenamefont {Nipper}, \citenamefont {Balewski}, \citenamefont
  {Kukota}, \citenamefont {Bendkowsky}, \citenamefont {Löw},\ and\
  \citenamefont {Pfau}}]{butscher_atommolecule_2010}%
  \BibitemOpen
  \bibfield  {author} {\bibinfo {author} {\bibfnamefont {B.}~\bibnamefont
  {Butscher}}, \bibinfo {author} {\bibfnamefont {J.}~\bibnamefont {Nipper}},
  \bibinfo {author} {\bibfnamefont {J.~B.}\ \bibnamefont {Balewski}}, \bibinfo
  {author} {\bibfnamefont {L.}~\bibnamefont {Kukota}}, \bibinfo {author}
  {\bibfnamefont {V.}~\bibnamefont {Bendkowsky}}, \bibinfo {author}
  {\bibfnamefont {R.}~\bibnamefont {Löw}}, \ and\ \bibinfo {author}
  {\bibfnamefont {T.}~\bibnamefont {Pfau}},\ }\href {\doibase
  10.1038/nphys1828} {\bibfield  {journal} {\bibinfo  {journal} {Nature
  Physics}\ }\textbf {\bibinfo {volume} {6}},\ \bibinfo {pages} {970} (\bibinfo
  {year} {2010})}\BibitemShut {NoStop}%
\bibitem [{\citenamefont {Butscher}\ \emph {et~al.}(2011)\citenamefont
  {Butscher}, \citenamefont {Bendkowsky}, \citenamefont {Nipper}, \citenamefont
  {Balewski}, \citenamefont {Kukota}, \citenamefont {Löw}, \citenamefont
  {Pfau}, \citenamefont {Li}, \citenamefont {Pohl},\ and\ \citenamefont
  {Rost}}]{Butscher_2011_lifetimes}%
  \BibitemOpen
  \bibfield  {author} {\bibinfo {author} {\bibfnamefont {B.}~\bibnamefont
  {Butscher}}, \bibinfo {author} {\bibfnamefont {V.}~\bibnamefont
  {Bendkowsky}}, \bibinfo {author} {\bibfnamefont {J.}~\bibnamefont {Nipper}},
  \bibinfo {author} {\bibfnamefont {J.~B.}\ \bibnamefont {Balewski}}, \bibinfo
  {author} {\bibfnamefont {L.}~\bibnamefont {Kukota}}, \bibinfo {author}
  {\bibfnamefont {R.}~\bibnamefont {Löw}}, \bibinfo {author} {\bibfnamefont
  {T.}~\bibnamefont {Pfau}}, \bibinfo {author} {\bibfnamefont {W.}~\bibnamefont
  {Li}}, \bibinfo {author} {\bibfnamefont {T.}~\bibnamefont {Pohl}}, \ and\
  \bibinfo {author} {\bibfnamefont {J.~M.}\ \bibnamefont {Rost}},\ }\href
  {\doibase 10.1088/0953-4075/44/18/184004} {\bibfield  {journal} {\bibinfo
  {journal} {J. Phys. B: At. Mol. Opt. Phys.}\ }\textbf {\bibinfo {volume}
  {44}},\ \bibinfo {pages} {184004} (\bibinfo {year} {2011})}\BibitemShut
  {NoStop}%
\bibitem [{\citenamefont {Sadeghpour}\ and\ \citenamefont
  {Rittenhouse}(2013)}]{sadeghpour_how_2013}%
  \BibitemOpen
  \bibfield  {author} {\bibinfo {author} {\bibfnamefont {H.~R.}\ \bibnamefont
  {Sadeghpour}}\ and\ \bibinfo {author} {\bibfnamefont {S.~T.}\ \bibnamefont
  {Rittenhouse}},\ }\href {\doibase 10.1080/00268976.2013.811555} {\bibfield
  {journal} {\bibinfo  {journal} {Mol. Phys.}\ }\textbf {\bibinfo {volume}
  {111}},\ \bibinfo {pages} {1902} (\bibinfo {year} {2013})}\BibitemShut
  {NoStop}%
\bibitem [{\citenamefont {Tallant}\ \emph {et~al.}(2012)\citenamefont
  {Tallant}, \citenamefont {Rittenhouse}, \citenamefont {Booth}, \citenamefont
  {Sadeghpour},\ and\ \citenamefont {Shaffer}}]{tallant_observation_2012}%
  \BibitemOpen
  \bibfield  {author} {\bibinfo {author} {\bibfnamefont {J.}~\bibnamefont
  {Tallant}}, \bibinfo {author} {\bibfnamefont {S.~T.}\ \bibnamefont
  {Rittenhouse}}, \bibinfo {author} {\bibfnamefont {D.}~\bibnamefont {Booth}},
  \bibinfo {author} {\bibfnamefont {H.~R.}\ \bibnamefont {Sadeghpour}}, \ and\
  \bibinfo {author} {\bibfnamefont {J.~P.}\ \bibnamefont {Shaffer}},\ }\href
  {\doibase 10.1103/PhysRevLett.109.173202} {\bibfield  {journal} {\bibinfo
  {journal} {Phys. Rev. Lett.}\ }\textbf {\bibinfo {volume} {109}},\ \bibinfo
  {pages} {173202} (\bibinfo {year} {2012})}\BibitemShut {NoStop}%
\bibitem [{\citenamefont {DeSalvo}\ \emph {et~al.}(2015)\citenamefont
  {DeSalvo}, \citenamefont {Aman}, \citenamefont {Dunning}, \citenamefont
  {Killian}, \citenamefont {Sadeghpour}, \citenamefont {Yoshida},\ and\
  \citenamefont {Burgdörfer}}]{desalvo_ultra-long-range_2015}%
  \BibitemOpen
  \bibfield  {author} {\bibinfo {author} {\bibfnamefont {B.~J.}\ \bibnamefont
  {DeSalvo}}, \bibinfo {author} {\bibfnamefont {J.~A.}\ \bibnamefont {Aman}},
  \bibinfo {author} {\bibfnamefont {F.~B.}\ \bibnamefont {Dunning}}, \bibinfo
  {author} {\bibfnamefont {T.~C.}\ \bibnamefont {Killian}}, \bibinfo {author}
  {\bibfnamefont {H.~R.}\ \bibnamefont {Sadeghpour}}, \bibinfo {author}
  {\bibfnamefont {S.}~\bibnamefont {Yoshida}}, \ and\ \bibinfo {author}
  {\bibfnamefont {J.}~\bibnamefont {Burgdörfer}},\ }\href {\doibase
  10.1103/PhysRevA.92.031403} {\bibfield  {journal} {\bibinfo  {journal} {Phys.
  Rev. A}\ }\textbf {\bibinfo {volume} {92}},\ \bibinfo {pages} {031403}
  (\bibinfo {year} {2015})}\BibitemShut {NoStop}%
\bibitem [{\citenamefont {Camargo}\ \emph {et~al.}(2016)\citenamefont
  {Camargo}, \citenamefont {Whalen}, \citenamefont {Ding}, \citenamefont
  {Sadeghpour}, \citenamefont {Yoshida}, \citenamefont {Burgdörfer},
  \citenamefont {Dunning},\ and\ \citenamefont
  {Killian}}]{camargo_lifetimes_2016}%
  \BibitemOpen
  \bibfield  {author} {\bibinfo {author} {\bibfnamefont {F.}~\bibnamefont
  {Camargo}}, \bibinfo {author} {\bibfnamefont {J.~D.}\ \bibnamefont {Whalen}},
  \bibinfo {author} {\bibfnamefont {R.}~\bibnamefont {Ding}}, \bibinfo {author}
  {\bibfnamefont {H.~R.}\ \bibnamefont {Sadeghpour}}, \bibinfo {author}
  {\bibfnamefont {S.}~\bibnamefont {Yoshida}}, \bibinfo {author} {\bibfnamefont
  {J.}~\bibnamefont {Burgdörfer}}, \bibinfo {author} {\bibfnamefont {F.~B.}\
  \bibnamefont {Dunning}}, \ and\ \bibinfo {author} {\bibfnamefont {T.~C.}\
  \bibnamefont {Killian}},\ }\href
  {http://link.aps.org/doi/10.1103/PhysRevA.93.022702} {\bibfield  {journal}
  {\bibinfo  {journal} {Phys. Rev. A}\ }\textbf {\bibinfo {volume} {93}},\
  \bibinfo {pages} {022702} (\bibinfo {year} {2016})}\BibitemShut {NoStop}%
\bibitem [{\citenamefont {Marinescu}\ \emph {et~al.}(1994)\citenamefont
  {Marinescu}, \citenamefont {Sadeghpour},\ and\ \citenamefont
  {Dalgarno}}]{Marinescu_Sadeghpour_Dalgarno_1994}%
  \BibitemOpen
  \bibfield  {author} {\bibinfo {author} {\bibfnamefont {M.}~\bibnamefont
  {Marinescu}}, \bibinfo {author} {\bibfnamefont {H.~R.}\ \bibnamefont
  {Sadeghpour}}, \ and\ \bibinfo {author} {\bibfnamefont {A.}~\bibnamefont
  {Dalgarno}},\ }\href@noop {} {\bibfield  {journal} {\bibinfo  {journal}
  {Phys. Rev. A}\ }\textbf {\bibinfo {volume} {49}},\ \bibinfo {pages} {982}
  (\bibinfo {year} {1994})}\BibitemShut {NoStop}%
\bibitem [{\citenamefont {Holmgren}\ \emph {et~al.}(2010)\citenamefont
  {Holmgren}, \citenamefont {Revelle}, \citenamefont {Lonij},\ and\
  \citenamefont {Cronin}}]{Holmgren_2010}%
  \BibitemOpen
  \bibfield  {author} {\bibinfo {author} {\bibfnamefont {W.~F.}\ \bibnamefont
  {Holmgren}}, \bibinfo {author} {\bibfnamefont {M.~C.}\ \bibnamefont
  {Revelle}}, \bibinfo {author} {\bibfnamefont {V.~P.~A.}\ \bibnamefont
  {Lonij}}, \ and\ \bibinfo {author} {\bibfnamefont {A.~D.}\ \bibnamefont
  {Cronin}},\ }\href {\doibase 10.1103/PhysRevA.81.053607} {\bibfield
  {journal} {\bibinfo  {journal} {Phys. Rev. A}\ }\textbf {\bibinfo {volume}
  {81}},\ \bibinfo {pages} {053607} (\bibinfo {year} {2010})}\BibitemShut
  {NoStop}%
\bibitem [{\citenamefont {Omont}(1977)}]{omont_theory_1977}%
  \BibitemOpen
  \bibfield  {author} {\bibinfo {author} {\bibfnamefont {A.}~\bibnamefont
  {Omont}},\ }\href {\doibase 10.1051/jphys:0197700380110134300} {\bibfield
  {journal} {\bibinfo  {journal} {J. Phys. (Paris)}\ }\textbf {\bibinfo
  {volume} {38}},\ \bibinfo {pages} {1343} (\bibinfo {year}
  {1977})}\BibitemShut {NoStop}%
\bibitem [{\citenamefont {Bahrim}\ \emph {et~al.}(2001)\citenamefont {Bahrim},
  \citenamefont {Thumm},\ and\ \citenamefont {Fabrikant}}]{bahrim_3se_2001}%
  \BibitemOpen
  \bibfield  {author} {\bibinfo {author} {\bibfnamefont {C.}~\bibnamefont
  {Bahrim}}, \bibinfo {author} {\bibfnamefont {U.}~\bibnamefont {Thumm}}, \
  and\ \bibinfo {author} {\bibfnamefont {I.~I.}\ \bibnamefont {Fabrikant}},\
  }\href {http://iopscience.iop.org/0953-4075/34/6/107} {\bibfield  {journal}
  {\bibinfo  {journal} {J. Phys. B: At. Mol. Opt. Phys.}\ }\textbf {\bibinfo
  {volume} {34}},\ \bibinfo {pages} {L195} (\bibinfo {year}
  {2001})}\BibitemShut {NoStop}%
\bibitem [{\citenamefont {Khuskivadze}\ \emph {et~al.}(2002)\citenamefont
  {Khuskivadze}, \citenamefont {Chibisov},\ and\ \citenamefont
  {Fabrikant}}]{khuskivadze_adiabatic_2002}%
  \BibitemOpen
  \bibfield  {author} {\bibinfo {author} {\bibfnamefont {A.}~\bibnamefont
  {Khuskivadze}}, \bibinfo {author} {\bibfnamefont {M.}~\bibnamefont
  {Chibisov}}, \ and\ \bibinfo {author} {\bibfnamefont {I.}~\bibnamefont
  {Fabrikant}},\ }\href {\doibase 10.1103/PhysRevA.66.042709} {\bibfield
  {journal} {\bibinfo  {journal} {Phys. Rev. A}\ }\textbf {\bibinfo {volume}
  {66}},\ \bibinfo {pages} {042709} (\bibinfo {year} {2002})}\BibitemShut
  {NoStop}%
\bibitem [{\citenamefont {Spruch}\ \emph {et~al.}(1960)\citenamefont {Spruch},
  \citenamefont {O'Malley},\ and\ \citenamefont
  {Rosenberg}}]{spruch_modification_1960}%
  \BibitemOpen
  \bibfield  {author} {\bibinfo {author} {\bibfnamefont {L.}~\bibnamefont
  {Spruch}}, \bibinfo {author} {\bibfnamefont {T.~F.}\ \bibnamefont
  {O'Malley}}, \ and\ \bibinfo {author} {\bibfnamefont {L.}~\bibnamefont
  {Rosenberg}},\ }\href {\doibase 10.1103/PhysRevLett.5.375} {\bibfield
  {journal} {\bibinfo  {journal} {Phys. Rev. Lett.}\ }\textbf {\bibinfo
  {volume} {5}},\ \bibinfo {pages} {375} (\bibinfo {year} {1960})}\BibitemShut
  {NoStop}%
\bibitem [{\citenamefont {O'Malley}\ \emph {et~al.}(1961)\citenamefont
  {O'Malley}, \citenamefont {Spruch},\ and\ \citenamefont
  {Rosenberg}}]{omalley_modification_1961}%
  \BibitemOpen
  \bibfield  {author} {\bibinfo {author} {\bibfnamefont {T.~F.}\ \bibnamefont
  {O'Malley}}, \bibinfo {author} {\bibfnamefont {L.}~\bibnamefont {Spruch}}, \
  and\ \bibinfo {author} {\bibfnamefont {L.}~\bibnamefont {Rosenberg}},\
  }\href@noop {} {\bibfield  {journal} {\bibinfo  {journal} {Journal of
  Mathematical Physics}\ }\textbf {\bibinfo {volume} {2}},\ \bibinfo {pages}
  {491} (\bibinfo {year} {1961})}\BibitemShut {NoStop}%
\bibitem [{\citenamefont {Fabrikant}(1986)}]{Fabrikant_1986}%
  \BibitemOpen
  \bibfield  {author} {\bibinfo {author} {\bibfnamefont {I.~I.}\ \bibnamefont
  {Fabrikant}},\ }\href {\doibase 10.1088/0022-3700/19/10/021} {\bibfield
  {journal} {\bibinfo  {journal} {J. Phys. B: At. Mol. Opt. Phys.}\ }\textbf
  {\bibinfo {volume} {19}},\ \bibinfo {pages} {1527} (\bibinfo {year}
  {1986})}\BibitemShut {NoStop}%
\bibitem [{\citenamefont {Eiles}\ and\ \citenamefont
  {Greene}(2015)}]{eiles_ultracold_2015}%
  \BibitemOpen
  \bibfield  {author} {\bibinfo {author} {\bibfnamefont {M.~T.}\ \bibnamefont
  {Eiles}}\ and\ \bibinfo {author} {\bibfnamefont {C.~H.}\ \bibnamefont
  {Greene}},\ }\href {\doibase 10.1103/PhysRevLett.115.193201} {\bibfield
  {journal} {\bibinfo  {journal} {Phys. Rev. Lett.}\ }\textbf {\bibinfo
  {volume} {115}},\ \bibinfo {pages} {193201} (\bibinfo {year}
  {2015})}\BibitemShut {NoStop}%
\bibitem [{\citenamefont {Fey}\ \emph {et~al.}(2015)\citenamefont {Fey},
  \citenamefont {Kurz}, \citenamefont {Schmelcher}, \citenamefont
  {Rittenhouse},\ and\ \citenamefont {Sadeghpour}}]{fey_comparative_2015}%
  \BibitemOpen
  \bibfield  {author} {\bibinfo {author} {\bibfnamefont {C.}~\bibnamefont
  {Fey}}, \bibinfo {author} {\bibfnamefont {M.}~\bibnamefont {Kurz}}, \bibinfo
  {author} {\bibfnamefont {P.}~\bibnamefont {Schmelcher}}, \bibinfo {author}
  {\bibfnamefont {S.~T.}\ \bibnamefont {Rittenhouse}}, \ and\ \bibinfo {author}
  {\bibfnamefont {H.~R.}\ \bibnamefont {Sadeghpour}},\ }\href {\doibase
  10.1088/1367-2630/17/5/055010} {\bibfield  {journal} {\bibinfo  {journal}
  {New J. Phys.}\ }\textbf {\bibinfo {volume} {17}},\ \bibinfo {pages} {055010}
  (\bibinfo {year} {2015})}\BibitemShut {NoStop}%
\bibitem [{\citenamefont {Tarana}\ and\ \citenamefont
  {Čurík}(2016)}]{tarana_adiabatic_2016}%
  \BibitemOpen
  \bibfield  {author} {\bibinfo {author} {\bibfnamefont {M.}~\bibnamefont
  {Tarana}}\ and\ \bibinfo {author} {\bibfnamefont {R.}~\bibnamefont
  {Čurík}},\ }\href {\doibase 10.1103/PhysRevA.93.012515} {\bibfield
  {journal} {\bibinfo  {journal} {Phy. Rev. A}\ }\textbf {\bibinfo {volume}
  {93}},\ \bibinfo {pages} {012515} (\bibinfo {year} {2016})}\BibitemShut
  {NoStop}%
\bibitem [{\citenamefont {Eiles}\ and\ \citenamefont
  {Greene}(2017)}]{eiles_hamiltonian_2017}%
  \BibitemOpen
  \bibfield  {author} {\bibinfo {author} {\bibfnamefont {M.~T.}\ \bibnamefont
  {Eiles}}\ and\ \bibinfo {author} {\bibfnamefont {C.~H.}\ \bibnamefont
  {Greene}},\ }\href {http://link.aps.org/doi/10.1103/PhysRevA.95.042515}
  {\bibfield  {journal} {\bibinfo  {journal} {Phys. Rev. A}\ }\textbf {\bibinfo
  {volume} {95}},\ \bibinfo {pages} {042515} (\bibinfo {year}
  {2017})}\BibitemShut {NoStop}%
\bibitem [{\citenamefont {Wang}\ \emph {et~al.}(2015)\citenamefont {Wang},
  \citenamefont {Gacesa},\ and\ \citenamefont {Côté}}]{wang_rydberg_2015}%
  \BibitemOpen
  \bibfield  {author} {\bibinfo {author} {\bibfnamefont {J.}~\bibnamefont
  {Wang}}, \bibinfo {author} {\bibfnamefont {M.}~\bibnamefont {Gacesa}}, \ and\
  \bibinfo {author} {\bibfnamefont {R.}~\bibnamefont {Côté}},\ }\href
  {\doibase 10.1103/PhysRevLett.114.243003} {\bibfield  {journal} {\bibinfo
  {journal} {Phys. Rev. Lett.}\ }\textbf {\bibinfo {volume} {114}},\ \bibinfo
  {pages} {243003} (\bibinfo {year} {2015})}\BibitemShut {NoStop}%
\bibitem [{\citenamefont {Karpiuk}\ \emph {et~al.}(2015)\citenamefont
  {Karpiuk}, \citenamefont {Brewczyk}, \citenamefont {Rzążewski},
  \citenamefont {Gaj}, \citenamefont {Balewski}, \citenamefont {Krupp},
  \citenamefont {Schlagmüller}, \citenamefont {Löw}, \citenamefont
  {Hofferberth},\ and\ \citenamefont {Pfau}}]{karpiuk_imaging_2015}%
  \BibitemOpen
  \bibfield  {author} {\bibinfo {author} {\bibfnamefont {T.}~\bibnamefont
  {Karpiuk}}, \bibinfo {author} {\bibfnamefont {M.}~\bibnamefont {Brewczyk}},
  \bibinfo {author} {\bibfnamefont {K.}~\bibnamefont {Rzążewski}}, \bibinfo
  {author} {\bibfnamefont {A.}~\bibnamefont {Gaj}}, \bibinfo {author}
  {\bibfnamefont {J.~B.}\ \bibnamefont {Balewski}}, \bibinfo {author}
  {\bibfnamefont {A.~T.}\ \bibnamefont {Krupp}}, \bibinfo {author}
  {\bibfnamefont {M.}~\bibnamefont {Schlagmüller}}, \bibinfo {author}
  {\bibfnamefont {R.}~\bibnamefont {Löw}}, \bibinfo {author} {\bibfnamefont
  {S.}~\bibnamefont {Hofferberth}}, \ and\ \bibinfo {author} {\bibfnamefont
  {T.}~\bibnamefont {Pfau}},\ }\href {\doibase 10.1088/1367-2630/17/5/053046}
  {\bibfield  {journal} {\bibinfo  {journal} {New J. Phys.}\ }\textbf {\bibinfo
  {volume} {17}},\ \bibinfo {pages} {053046} (\bibinfo {year}
  {2015})}\BibitemShut {NoStop}%
\bibitem [{\citenamefont {Thumm}\ and\ \citenamefont
  {Norcross}(1992)}]{Thumm_Norcross_1992}%
  \BibitemOpen
  \bibfield  {author} {\bibinfo {author} {\bibfnamefont {U.}~\bibnamefont
  {Thumm}}\ and\ \bibinfo {author} {\bibfnamefont {D.~W.}\ \bibnamefont
  {Norcross}},\ }\href {\doibase 10.1103/PhysRevA.45.6349} {\bibfield
  {journal} {\bibinfo  {journal} {Phys. Rev. A}\ }\textbf {\bibinfo {volume}
  {45}},\ \bibinfo {pages} {6349} (\bibinfo {year} {1992})}\BibitemShut
  {NoStop}%
\bibitem [{\citenamefont {Granger}\ \emph {et~al.}(2001)\citenamefont
  {Granger}, \citenamefont {Hamilton},\ and\ \citenamefont
  {Greene}}]{Granger_2001}%
  \BibitemOpen
  \bibfield  {author} {\bibinfo {author} {\bibfnamefont {B.}~\bibnamefont
  {Granger}}, \bibinfo {author} {\bibfnamefont {E.}~\bibnamefont {Hamilton}}, \
  and\ \bibinfo {author} {\bibfnamefont {C.}~\bibnamefont {Greene}},\ }\href
  {\doibase 10.1103/PhysRevA.64.042508} {\bibfield  {journal} {\bibinfo
  {journal} {Phys. Rev. A}\ }\textbf {\bibinfo {volume} {64}},\ \bibinfo
  {pages} {042508} (\bibinfo {year} {2001})}\BibitemShut {NoStop}%
\bibitem [{\citenamefont {Kanellopoulos}\ \emph {et~al.}(2009)\citenamefont
  {Kanellopoulos}, \citenamefont {Kleber},\ and\ \citenamefont
  {Kramer}}]{kanellopoulos_use_2009}%
  \BibitemOpen
  \bibfield  {author} {\bibinfo {author} {\bibfnamefont {V.}~\bibnamefont
  {Kanellopoulos}}, \bibinfo {author} {\bibfnamefont {M.}~\bibnamefont
  {Kleber}}, \ and\ \bibinfo {author} {\bibfnamefont {T.}~\bibnamefont
  {Kramer}},\ }\href {\doibase 10.1103/PhysRevA.80.012101} {\bibfield
  {journal} {\bibinfo  {journal} {Phys. Rev. A}\ }\textbf {\bibinfo {volume}
  {80}},\ \bibinfo {pages} {012101} (\bibinfo {year} {2009})}\BibitemShut
  {NoStop}%
\bibitem [{\citenamefont {Eiles}\ \emph {et~al.}(2018)\citenamefont {Eiles},
  \citenamefont {Tong},\ and\ \citenamefont
  {Greene}}]{Eiles_ghost_trilobite_2018}%
  \BibitemOpen
  \bibfield  {author} {\bibinfo {author} {\bibfnamefont {M.~T.}\ \bibnamefont
  {Eiles}}, \bibinfo {author} {\bibfnamefont {Z.}~\bibnamefont {Tong}}, \ and\
  \bibinfo {author} {\bibfnamefont {C.~H.}\ \bibnamefont {Greene}},\ }\href
  {\doibase 10.1103/PhysRevLett.121.113203} {\bibfield  {journal} {\bibinfo
  {journal} {Phys. Rev. Lett.}\ }\textbf {\bibinfo {volume} {121}},\ \bibinfo
  {pages} {113203} (\bibinfo {year} {2018})}\BibitemShut {NoStop}%
\bibitem [{\citenamefont {Anderson}\ \emph
  {et~al.}(2014{\natexlab{b}})\citenamefont {Anderson}, \citenamefont
  {Miller},\ and\ \citenamefont {Raithel}}]{anderson_angular-momentum_2014}%
  \BibitemOpen
  \bibfield  {author} {\bibinfo {author} {\bibfnamefont {D.~A.}\ \bibnamefont
  {Anderson}}, \bibinfo {author} {\bibfnamefont {S.~A.}\ \bibnamefont
  {Miller}}, \ and\ \bibinfo {author} {\bibfnamefont {G.}~\bibnamefont
  {Raithel}},\ }\href {\doibase 10.1103/PhysRevA.90.062518} {\bibfield
  {journal} {\bibinfo  {journal} {Phys. Rev. A}\ }\textbf {\bibinfo {volume}
  {90}},\ \bibinfo {pages} {062518} (\bibinfo {year}
  {2014}{\natexlab{b}})}\BibitemShut {NoStop}%
\bibitem [{\citenamefont {Deiß}\ \emph {et~al.}(2019)\citenamefont {Deiß},
  \citenamefont {Haze}, \citenamefont {Wolf}, \citenamefont {Wang},
  \citenamefont {Meinert}, \citenamefont {Fey}, \citenamefont {Hummel},
  \citenamefont {Schmelcher},\ and\ \citenamefont
  {Denschlag}}]{Deiss_Fey_Hummel_Schmelcher_Denschlag_2019}%
  \BibitemOpen
  \bibfield  {author} {\bibinfo {author} {\bibfnamefont {M.}~\bibnamefont
  {Deiß}}, \bibinfo {author} {\bibfnamefont {S.}~\bibnamefont {Haze}},
  \bibinfo {author} {\bibfnamefont {J.}~\bibnamefont {Wolf}}, \bibinfo {author}
  {\bibfnamefont {L.}~\bibnamefont {Wang}}, \bibinfo {author} {\bibfnamefont
  {F.}~\bibnamefont {Meinert}}, \bibinfo {author} {\bibfnamefont
  {C.}~\bibnamefont {Fey}}, \bibinfo {author} {\bibfnamefont {F.}~\bibnamefont
  {Hummel}}, \bibinfo {author} {\bibfnamefont {P.}~\bibnamefont {Schmelcher}},
  \ and\ \bibinfo {author} {\bibfnamefont {J.~H.}\ \bibnamefont {Denschlag}},\
  }\href {http://arxiv.org/abs/1901.08792} {\bibfield  {journal} {\bibinfo
  {journal} {arXiv preprint: 1901.08792}\ } (\bibinfo {year}
  {2019})}\BibitemShut {NoStop}%
\bibitem [{\citenamefont {Rittenhouse}\ and\ \citenamefont
  {Sadeghpour}(2010)}]{rittenhouse_ultracold_2010}%
  \BibitemOpen
  \bibfield  {author} {\bibinfo {author} {\bibfnamefont {S.~T.}\ \bibnamefont
  {Rittenhouse}}\ and\ \bibinfo {author} {\bibfnamefont {H.~R.}\ \bibnamefont
  {Sadeghpour}},\ }\href {\doibase 10.1103/PhysRevLett.104.243002} {\bibfield
  {journal} {\bibinfo  {journal} {Phys. Rev. Lett.}\ }\textbf {\bibinfo
  {volume} {104}},\ \bibinfo {pages} {243002} (\bibinfo {year}
  {2010})}\BibitemShut {NoStop}%
\bibitem [{\citenamefont {Rittenhouse}\ \emph {et~al.}(2011)\citenamefont
  {Rittenhouse}, \citenamefont {Mayle}, \citenamefont {Schmelcher},\ and\
  \citenamefont {Sadeghpour}}]{rittenhouse_ultralong-range_2011}%
  \BibitemOpen
  \bibfield  {author} {\bibinfo {author} {\bibfnamefont {S.~T.}\ \bibnamefont
  {Rittenhouse}}, \bibinfo {author} {\bibfnamefont {M.}~\bibnamefont {Mayle}},
  \bibinfo {author} {\bibfnamefont {P.}~\bibnamefont {Schmelcher}}, \ and\
  \bibinfo {author} {\bibfnamefont {H.~R.}\ \bibnamefont {Sadeghpour}},\ }\href
  {\doibase 10.1088/0953-4075/44/18/184005} {\bibfield  {journal} {\bibinfo
  {journal} {J. Phys. B: At. Mol. Opt. Phys.}\ }\textbf {\bibinfo {volume}
  {44}},\ \bibinfo {pages} {184005} (\bibinfo {year} {2011})}\BibitemShut
  {NoStop}%
\bibitem [{\citenamefont {Gonz{\'{a}}lez-F{\'{e}}rez}\ \emph
  {et~al.}(2015)\citenamefont {Gonz{\'{a}}lez-F{\'{e}}rez}, \citenamefont
  {Sadeghpour},\ and\ \citenamefont
  {Schmelcher}}]{Gonzalez-Ferez_Sadeghpour_Schmelcher_2015}%
  \BibitemOpen
  \bibfield  {author} {\bibinfo {author} {\bibfnamefont {R.}~\bibnamefont
  {Gonz{\'{a}}lez-F{\'{e}}rez}}, \bibinfo {author} {\bibfnamefont {H.~R.}\
  \bibnamefont {Sadeghpour}}, \ and\ \bibinfo {author} {\bibfnamefont
  {P.}~\bibnamefont {Schmelcher}},\ }\href {\doibase
  10.1088/1367-2630/17/1/013021} {\bibfield  {journal} {\bibinfo  {journal}
  {New J. Phys.}\ }\textbf {\bibinfo {volume} {17}},\ \bibinfo {pages} {013021}
  (\bibinfo {year} {2015})}\BibitemShut {NoStop}%
\bibitem [{\citenamefont {Aguilera-Fernández}\ \emph
  {et~al.}(2015)\citenamefont {Aguilera-Fernández}, \citenamefont
  {Sadeghpour}, \citenamefont {Schmelcher},\ and\ \citenamefont
  {González-Férez}}]{Aguilera_Gonzalez-Ferez_2015}%
  \BibitemOpen
  \bibfield  {author} {\bibinfo {author} {\bibfnamefont {J.}~\bibnamefont
  {Aguilera-Fernández}}, \bibinfo {author} {\bibfnamefont {H.~R.}\
  \bibnamefont {Sadeghpour}}, \bibinfo {author} {\bibfnamefont
  {P.}~\bibnamefont {Schmelcher}}, \ and\ \bibinfo {author} {\bibfnamefont
  {R.}~\bibnamefont {González-Férez}},\ }\href {\doibase
  10.1088/1742-6596/635/1/012023} {\bibfield  {journal} {\bibinfo  {journal}
  {J. Phys.: Conf. Ser.}\ }\textbf {\bibinfo {volume} {635}},\ \bibinfo {pages}
  {012023} (\bibinfo {year} {2015})}\BibitemShut {NoStop}%
\bibitem [{\citenamefont {Balewski}\ \emph {et~al.}(2013)\citenamefont
  {Balewski}, \citenamefont {Krupp}, \citenamefont {Gaj}, \citenamefont
  {Peter}, \citenamefont {Büchler}, \citenamefont {Löw}, \citenamefont
  {Hofferberth},\ and\ \citenamefont {Pfau}}]{balewski_coupling_2013}%
  \BibitemOpen
  \bibfield  {author} {\bibinfo {author} {\bibfnamefont {J.~B.}\ \bibnamefont
  {Balewski}}, \bibinfo {author} {\bibfnamefont {A.~T.}\ \bibnamefont {Krupp}},
  \bibinfo {author} {\bibfnamefont {A.}~\bibnamefont {Gaj}}, \bibinfo {author}
  {\bibfnamefont {D.}~\bibnamefont {Peter}}, \bibinfo {author} {\bibfnamefont
  {H.~P.}\ \bibnamefont {Büchler}}, \bibinfo {author} {\bibfnamefont
  {R.}~\bibnamefont {Löw}}, \bibinfo {author} {\bibfnamefont {S.}~\bibnamefont
  {Hofferberth}}, \ and\ \bibinfo {author} {\bibfnamefont {T.}~\bibnamefont
  {Pfau}},\ }\href {\doibase 10.1038/nature12592} {\bibfield  {journal}
  {\bibinfo  {journal} {Nature}\ }\textbf {\bibinfo {volume} {502}},\ \bibinfo
  {pages} {664} (\bibinfo {year} {2013})}\BibitemShut {NoStop}%
\bibitem [{\citenamefont {Pérez-Ríos}\ \emph {et~al.}(2016)\citenamefont
  {Pérez-Ríos}, \citenamefont {Eiles},\ and\ \citenamefont
  {Greene}}]{perez-rios_mapping_2016}%
  \BibitemOpen
  \bibfield  {author} {\bibinfo {author} {\bibfnamefont {J.}~\bibnamefont
  {Pérez-Ríos}}, \bibinfo {author} {\bibfnamefont {M.~T.}\ \bibnamefont
  {Eiles}}, \ and\ \bibinfo {author} {\bibfnamefont {C.~H.}\ \bibnamefont
  {Greene}},\ }\href
  {http://stacks.iop.org/0953-4075/49/i=14/a=14LT01?key=crossref.2f6200f4a64cdd03b33bf41e5d9e1589}
  {\bibfield  {journal} {\bibinfo  {journal} {J. Phys. B: At. Mol. Opt. Phys.}\
  }\textbf {\bibinfo {volume} {49}},\ \bibinfo {pages} {14LT01} (\bibinfo
  {year} {2016})}\BibitemShut {NoStop}%
\bibitem [{\citenamefont {Schmidt}\ \emph {et~al.}(2018)\citenamefont
  {Schmidt}, \citenamefont {Whalen}, \citenamefont {Ding}, \citenamefont
  {Camargo}, \citenamefont {Woehl}, \citenamefont {Yoshida}, \citenamefont
  {Burgdörfer}, \citenamefont {Dunning}, \citenamefont {Demler}, \citenamefont
  {Sadeghpour},\ and\ \citenamefont {Killian}}]{schmidt_theory_2018}%
  \BibitemOpen
  \bibfield  {author} {\bibinfo {author} {\bibfnamefont {R.}~\bibnamefont
  {Schmidt}}, \bibinfo {author} {\bibfnamefont {J.~D.}\ \bibnamefont {Whalen}},
  \bibinfo {author} {\bibfnamefont {R.}~\bibnamefont {Ding}}, \bibinfo {author}
  {\bibfnamefont {F.}~\bibnamefont {Camargo}}, \bibinfo {author} {\bibfnamefont
  {G.}~\bibnamefont {Woehl}}, \bibinfo {author} {\bibfnamefont
  {S.}~\bibnamefont {Yoshida}}, \bibinfo {author} {\bibfnamefont
  {J.}~\bibnamefont {Burgdörfer}}, \bibinfo {author} {\bibfnamefont {F.~B.}\
  \bibnamefont {Dunning}}, \bibinfo {author} {\bibfnamefont {E.}~\bibnamefont
  {Demler}}, \bibinfo {author} {\bibfnamefont {H.~R.}\ \bibnamefont
  {Sadeghpour}}, \ and\ \bibinfo {author} {\bibfnamefont {T.~C.}\ \bibnamefont
  {Killian}},\ }\href {https://link.aps.org/doi/10.1103/PhysRevA.97.022707}
  {\bibfield  {journal} {\bibinfo  {journal} {Phys. Rev. A}\ }\textbf {\bibinfo
  {volume} {97}},\ \bibinfo {pages} {022707} (\bibinfo {year}
  {2018})}\BibitemShut {NoStop}%
\bibitem [{\citenamefont {Rost}\ \emph {et~al.}(1992)\citenamefont {Rost},
  \citenamefont {Griffin}, \citenamefont {Friedrich},\ and\ \citenamefont
  {Herschbach}}]{Rost_1992}%
  \BibitemOpen
  \bibfield  {author} {\bibinfo {author} {\bibfnamefont {J.~M.}\ \bibnamefont
  {Rost}}, \bibinfo {author} {\bibfnamefont {J.~C.}\ \bibnamefont {Griffin}},
  \bibinfo {author} {\bibfnamefont {B.}~\bibnamefont {Friedrich}}, \ and\
  \bibinfo {author} {\bibfnamefont {D.~R.}\ \bibnamefont {Herschbach}},\ }\href
  {\doibase 10.1103/PhysRevLett.68.1299} {\bibfield  {journal} {\bibinfo
  {journal} {Phys. Rev. Lett.}\ }\textbf {\bibinfo {volume} {68}},\ \bibinfo
  {pages} {1299} (\bibinfo {year} {1992})}\BibitemShut {NoStop}%
\bibitem [{\citenamefont {Kleinbach}\ \emph {et~al.}(2017)\citenamefont
  {Kleinbach}, \citenamefont {Meinert}, \citenamefont {Engel}, \citenamefont
  {Kwon}, \citenamefont {Löw}, \citenamefont {Pfau},\ and\ \citenamefont
  {Raithel}}]{Kleinbach_2017}%
  \BibitemOpen
  \bibfield  {author} {\bibinfo {author} {\bibfnamefont {K.}~\bibnamefont
  {Kleinbach}}, \bibinfo {author} {\bibfnamefont {F.}~\bibnamefont {Meinert}},
  \bibinfo {author} {\bibfnamefont {F.}~\bibnamefont {Engel}}, \bibinfo
  {author} {\bibfnamefont {W.}~\bibnamefont {Kwon}}, \bibinfo {author}
  {\bibfnamefont {R.}~\bibnamefont {Löw}}, \bibinfo {author} {\bibfnamefont
  {T.}~\bibnamefont {Pfau}}, \ and\ \bibinfo {author} {\bibfnamefont
  {G.}~\bibnamefont {Raithel}},\ }\href {\doibase
  10.1103/PhysRevLett.118.223001} {\bibfield  {journal} {\bibinfo  {journal}
  {Phys. Rev. Lett.}\ }\textbf {\bibinfo {volume} {118}},\ \bibinfo {pages}
  {223001} (\bibinfo {year} {2017})}\BibitemShut {NoStop}%
\bibitem [{\citenamefont {Kurz}\ and\ \citenamefont
  {Schmelcher}(2014)}]{kurz_ultralong-range_2014}%
  \BibitemOpen
  \bibfield  {author} {\bibinfo {author} {\bibfnamefont {M.}~\bibnamefont
  {Kurz}}\ and\ \bibinfo {author} {\bibfnamefont {P.}~\bibnamefont
  {Schmelcher}},\ }\href {\doibase 10.1088/0953-4075/47/16/165101} {\bibfield
  {journal} {\bibinfo  {journal} {J. Phys. B: At. Mol. Opt. Phys.}\ }\textbf
  {\bibinfo {volume} {47}},\ \bibinfo {pages} {165101} (\bibinfo {year}
  {2014})}\BibitemShut {NoStop}%
\bibitem [{\citenamefont {Fernández}\ \emph {et~al.}(2016)\citenamefont
  {Fernández}, \citenamefont {Schmelcher},\ and\ \citenamefont
  {González-Férez}}]{Fernandez_2016}%
  \BibitemOpen
  \bibfield  {author} {\bibinfo {author} {\bibfnamefont {J.~A.}\ \bibnamefont
  {Fernández}}, \bibinfo {author} {\bibfnamefont {P.}~\bibnamefont
  {Schmelcher}}, \ and\ \bibinfo {author} {\bibfnamefont {R.}~\bibnamefont
  {González-Férez}},\ }\href {\doibase 10.1088/0953-4075/49/12/124002}
  {\bibfield  {journal} {\bibinfo  {journal} {J. Phys. B: At. Mol. Opt. Phys.}\
  }\textbf {\bibinfo {volume} {49}},\ \bibinfo {pages} {124002} (\bibinfo
  {year} {2016})}\BibitemShut {NoStop}%
\bibitem [{\citenamefont {Kurz}\ \emph {et~al.}(2012)\citenamefont {Kurz},
  \citenamefont {Mayle},\ and\ \citenamefont
  {Schmelcher}}]{Kurz_Mayle_Schmelcher_2012}%
  \BibitemOpen
  \bibfield  {author} {\bibinfo {author} {\bibfnamefont {M.}~\bibnamefont
  {Kurz}}, \bibinfo {author} {\bibfnamefont {M.}~\bibnamefont {Mayle}}, \ and\
  \bibinfo {author} {\bibfnamefont {P.}~\bibnamefont {Schmelcher}},\ }\href
  {\doibase 10.1209/0295-5075/97/43001} {\bibfield  {journal} {\bibinfo
  {journal} {Europhys. Lett.}\ }\textbf {\bibinfo {volume} {97}},\ \bibinfo
  {pages} {43001} (\bibinfo {year} {2012})}\BibitemShut {NoStop}%
\bibitem [{\citenamefont {Burkova}\ \emph {et~al.}(1976)\citenamefont
  {Burkova}, \citenamefont {Drukarev},\ and\ \citenamefont
  {Monozon}}]{Burkova_Drukarev_Monozon_1976}%
  \BibitemOpen
  \bibfield  {author} {\bibinfo {author} {\bibfnamefont {A.}~\bibnamefont
  {Burkova}}, \bibinfo {author} {\bibfnamefont {D.~G.~F.}\ \bibnamefont
  {Drukarev}}, \ and\ \bibinfo {author} {\bibfnamefont {B.~S.}\ \bibnamefont
  {Monozon}},\ }\href@noop {} {\bibfield  {journal} {\bibinfo  {journal}
  {JETP}\ }\textbf {\bibinfo {volume} {44}},\ \bibinfo {pages} {276} (\bibinfo
  {year} {1976})}\BibitemShut {NoStop}%
\bibitem [{\citenamefont {Gay}\ \emph {et~al.}(1979)\citenamefont {Gay},
  \citenamefont {Pendrill},\ and\ \citenamefont
  {Cagnac}}]{Gay_Pendrill_Cagnac_1979}%
  \BibitemOpen
  \bibfield  {author} {\bibinfo {author} {\bibfnamefont {J.}~\bibnamefont
  {Gay}}, \bibinfo {author} {\bibfnamefont {L.}~\bibnamefont {Pendrill}}, \
  and\ \bibinfo {author} {\bibfnamefont {B.}~\bibnamefont {Cagnac}},\ }\href
  {\doibase 10.1016/0375-9601(79)90480-8} {\bibfield  {journal} {\bibinfo
  {journal} {Phys. Lett. A}\ }\textbf {\bibinfo {volume} {72}},\ \bibinfo
  {pages} {315} (\bibinfo {year} {1979})}\BibitemShut {NoStop}%
\bibitem [{\citenamefont {Fauth}\ \emph {et~al.}(1987)\citenamefont {Fauth},
  \citenamefont {Walther},\ and\ \citenamefont
  {Werner}}]{Fauth_Walther_Werner_1987}%
  \BibitemOpen
  \bibfield  {author} {\bibinfo {author} {\bibfnamefont {M.}~\bibnamefont
  {Fauth}}, \bibinfo {author} {\bibfnamefont {H.}~\bibnamefont {Walther}}, \
  and\ \bibinfo {author} {\bibfnamefont {E.}~\bibnamefont {Werner}},\ }\href
  {\doibase 10.1007/BF01384997} {\bibfield  {journal} {\bibinfo  {journal} {Z.
  Phys. D}\ }\textbf {\bibinfo {volume} {7}},\ \bibinfo {pages} {293} (\bibinfo
  {year} {1987})}\BibitemShut {NoStop}%
\bibitem [{\citenamefont {Baye}\ \emph {et~al.}(1992)\citenamefont {Baye},
  \citenamefont {Clerbaux},\ and\ \citenamefont
  {Vincke}}]{Baye_Clerbaux_Vincke_1992}%
  \BibitemOpen
  \bibfield  {author} {\bibinfo {author} {\bibfnamefont {D.}~\bibnamefont
  {Baye}}, \bibinfo {author} {\bibfnamefont {N.}~\bibnamefont {Clerbaux}}, \
  and\ \bibinfo {author} {\bibfnamefont {M.}~\bibnamefont {Vincke}},\ }\href
  {\doibase 10.1016/0375-9601(92)90548-Z} {\bibfield  {journal} {\bibinfo
  {journal} {Phys. Lett. A}\ }\textbf {\bibinfo {volume} {166}},\ \bibinfo
  {pages} {135} (\bibinfo {year} {1992})}\BibitemShut {NoStop}%
\bibitem [{\citenamefont {Dzyaloshinskii}(1992)}]{Dzyaloshinskii_1992}%
  \BibitemOpen
  \bibfield  {author} {\bibinfo {author} {\bibfnamefont {I.}~\bibnamefont
  {Dzyaloshinskii}},\ }\href {\doibase 10.1016/0375-9601(92)91056-W} {\bibfield
   {journal} {\bibinfo  {journal} {Phys. Lett. A}\ }\textbf {\bibinfo {volume}
  {165}},\ \bibinfo {pages} {69} (\bibinfo {year} {1992})}\BibitemShut
  {NoStop}%
\bibitem [{\citenamefont {Raithel}\ \emph {et~al.}(1993)\citenamefont
  {Raithel}, \citenamefont {Fauth},\ and\ \citenamefont
  {Walther}}]{Raithel_Fauth_Walther_1993}%
  \BibitemOpen
  \bibfield  {author} {\bibinfo {author} {\bibfnamefont {G.}~\bibnamefont
  {Raithel}}, \bibinfo {author} {\bibfnamefont {M.}~\bibnamefont {Fauth}}, \
  and\ \bibinfo {author} {\bibfnamefont {H.}~\bibnamefont {Walther}},\ }\href
  {\doibase 10.1103/PhysRevA.47.419} {\bibfield  {journal} {\bibinfo  {journal}
  {Phys. Rev. A}\ }\textbf {\bibinfo {volume} {47}},\ \bibinfo {pages} {419}
  (\bibinfo {year} {1993})}\BibitemShut {NoStop}%
\bibitem [{\citenamefont {Schmelcher}\ and\ \citenamefont
  {Cederbaum}(1993)}]{Schmelcher_Cederbaum_1993}%
  \BibitemOpen
  \bibfield  {author} {\bibinfo {author} {\bibfnamefont {P.}~\bibnamefont
  {Schmelcher}}\ and\ \bibinfo {author} {\bibfnamefont {L.~S.}\ \bibnamefont
  {Cederbaum}},\ }\href {\doibase 10.1016/0009-2614(93)87188-9} {\bibfield
  {journal} {\bibinfo  {journal} {Chem. Phys. Lett.}\ }\textbf {\bibinfo
  {volume} {208}},\ \bibinfo {pages} {548} (\bibinfo {year}
  {1993})}\BibitemShut {NoStop}%
\bibitem [{\citenamefont {Dippel}\ \emph {et~al.}(1994)\citenamefont {Dippel},
  \citenamefont {Schmelcher},\ and\ \citenamefont
  {Cederbaum}}]{Dippel_Schmelcher_Cederbaum_1994}%
  \BibitemOpen
  \bibfield  {author} {\bibinfo {author} {\bibfnamefont {O.}~\bibnamefont
  {Dippel}}, \bibinfo {author} {\bibfnamefont {P.}~\bibnamefont {Schmelcher}},
  \ and\ \bibinfo {author} {\bibfnamefont {L.~S.}\ \bibnamefont {Cederbaum}},\
  }\href {\doibase 10.1103/PhysRevA.49.4415} {\bibfield  {journal} {\bibinfo
  {journal} {Phys. Rev. A}\ }\textbf {\bibinfo {volume} {49}},\ \bibinfo
  {pages} {4415} (\bibinfo {year} {1994})}\BibitemShut {NoStop}%
\bibitem [{\citenamefont {Z\"ollner}\ \emph
  {et~al.}(2005{\natexlab{a}})\citenamefont {Z\"ollner}, \citenamefont
  {Meyer},\ and\ \citenamefont {Schmelcher}}]{zoellner2005a}%
  \BibitemOpen
  \bibfield  {author} {\bibinfo {author} {\bibfnamefont {S.}~\bibnamefont
  {Z\"ollner}}, \bibinfo {author} {\bibfnamefont {H.~D.}\ \bibnamefont
  {Meyer}}, \ and\ \bibinfo {author} {\bibfnamefont {P.}~\bibnamefont
  {Schmelcher}},\ }\href@noop {} {\bibfield  {journal} {\bibinfo  {journal}
  {Phys. Rev. A}\ }\textbf {\bibinfo {volume} {72}},\ \bibinfo {pages} {033416}
  (\bibinfo {year} {2005}{\natexlab{a}})}\BibitemShut {NoStop}%
\bibitem [{\citenamefont {Z\"ollner}\ \emph
  {et~al.}(2005{\natexlab{b}})\citenamefont {Z\"ollner}, \citenamefont
  {Meyer},\ and\ \citenamefont {Schmelcher}}]{zoellner2005b}%
  \BibitemOpen
  \bibfield  {author} {\bibinfo {author} {\bibfnamefont {S.}~\bibnamefont
  {Z\"ollner}}, \bibinfo {author} {\bibfnamefont {H.~D.}\ \bibnamefont
  {Meyer}}, \ and\ \bibinfo {author} {\bibfnamefont {P.}~\bibnamefont
  {Schmelcher}},\ }\href@noop {} {\bibfield  {journal} {\bibinfo  {journal}
  {Europhys. Lett.}\ }\textbf {\bibinfo {volume} {71}},\ \bibinfo {pages} {373}
  (\bibinfo {year} {2005}{\natexlab{b}})}\BibitemShut {NoStop}%
\bibitem [{\citenamefont {Schmelcher}(2001)}]{schmelcher2001}%
  \BibitemOpen
  \bibfield  {author} {\bibinfo {author} {\bibfnamefont {P.}~\bibnamefont
  {Schmelcher}},\ }\href@noop {} {\bibfield  {journal} {\bibinfo  {journal}
  {Phys. Rev. A}\ }\textbf {\bibinfo {volume} {64}},\ \bibinfo {pages} {063412}
  (\bibinfo {year} {2001})}\BibitemShut {NoStop}%
\bibitem [{\citenamefont {Averbukh}\ \emph {et~al.}(1999)\citenamefont
  {Averbukh}, \citenamefont {Moiseyev}, \citenamefont {Schmelcher},\ and\
  \citenamefont {Cederbaum}}]{averbukh1999}%
  \BibitemOpen
  \bibfield  {author} {\bibinfo {author} {\bibfnamefont {V.}~\bibnamefont
  {Averbukh}}, \bibinfo {author} {\bibfnamefont {N.}~\bibnamefont {Moiseyev}},
  \bibinfo {author} {\bibfnamefont {P.}~\bibnamefont {Schmelcher}}, \ and\
  \bibinfo {author} {\bibfnamefont {L.~S.}\ \bibnamefont {Cederbaum}},\
  }\href@noop {} {\bibfield  {journal} {\bibinfo  {journal} {Phys. Rev. A}\
  }\textbf {\bibinfo {volume} {59}},\ \bibinfo {pages} {3695} (\bibinfo {year}
  {1999})}\BibitemShut {NoStop}%
\bibitem [{\citenamefont {Niederprüm}\ \emph
  {et~al.}(2016{\natexlab{b}})\citenamefont {Niederprüm}, \citenamefont
  {Thomas}, \citenamefont {Eichert},\ and\ \citenamefont
  {Ott}}]{Niederpruem_remote_spinflips_2016}%
  \BibitemOpen
  \bibfield  {author} {\bibinfo {author} {\bibfnamefont {T.}~\bibnamefont
  {Niederprüm}}, \bibinfo {author} {\bibfnamefont {O.}~\bibnamefont {Thomas}},
  \bibinfo {author} {\bibfnamefont {T.}~\bibnamefont {Eichert}}, \ and\
  \bibinfo {author} {\bibfnamefont {H.}~\bibnamefont {Ott}},\ }\href {\doibase
  10.1103/PhysRevLett.117.123002} {\bibfield  {journal} {\bibinfo  {journal}
  {Phys. Rev. Lett.}\ }\textbf {\bibinfo {volume} {117}},\ \bibinfo {pages}
  {123002} (\bibinfo {year} {2016}{\natexlab{b}})}\BibitemShut {NoStop}%
\bibitem [{\citenamefont {Arimondo}\ \emph {et~al.}(1977)\citenamefont
  {Arimondo}, \citenamefont {Inguscio},\ and\ \citenamefont
  {Violino}}]{Arimondo_hyperfine_1977}%
  \BibitemOpen
  \bibfield  {author} {\bibinfo {author} {\bibfnamefont {E.}~\bibnamefont
  {Arimondo}}, \bibinfo {author} {\bibfnamefont {M.}~\bibnamefont {Inguscio}},
  \ and\ \bibinfo {author} {\bibfnamefont {P.}~\bibnamefont {Violino}},\ }\href
  {\doibase 10.1103/RevModPhys.49.31} {\bibfield  {journal} {\bibinfo
  {journal} {Rev. Mod. Phys.}\ }\textbf {\bibinfo {volume} {49}},\ \bibinfo
  {pages} {31} (\bibinfo {year} {1977})}\BibitemShut {NoStop}%
\bibitem [{\citenamefont {Johnston}\ and\ \citenamefont
  {Burrow}(1982)}]{Johnston_Burrow_1982}%
  \BibitemOpen
  \bibfield  {author} {\bibinfo {author} {\bibfnamefont {A.~R.}\ \bibnamefont
  {Johnston}}\ and\ \bibinfo {author} {\bibfnamefont {P.~D.}\ \bibnamefont
  {Burrow}},\ }\href {\doibase 10.1088/0022-3700/15/20/005} {\bibfield
  {journal} {\bibinfo  {journal} {J. Phys. B: At. Mol. Phys.}\ }\textbf
  {\bibinfo {volume} {15}},\ \bibinfo {pages} {L745} (\bibinfo {year}
  {1982})}\BibitemShut {NoStop}%
\bibitem [{\citenamefont {Scheer}\ \emph {et~al.}(1998)\citenamefont {Scheer},
  \citenamefont {Thøgersen}, \citenamefont {Bilodeau}, \citenamefont {Brodie},
  \citenamefont {Haugen}, \citenamefont {Andersen}, \citenamefont
  {Kristensen},\ and\ \citenamefont
  {Andersen}}]{Scheer_1998_cs_shape_resonance}%
  \BibitemOpen
  \bibfield  {author} {\bibinfo {author} {\bibfnamefont {M.}~\bibnamefont
  {Scheer}}, \bibinfo {author} {\bibfnamefont {J.}~\bibnamefont {Thøgersen}},
  \bibinfo {author} {\bibfnamefont {R.~C.}\ \bibnamefont {Bilodeau}}, \bibinfo
  {author} {\bibfnamefont {C.~A.}\ \bibnamefont {Brodie}}, \bibinfo {author}
  {\bibfnamefont {H.~K.}\ \bibnamefont {Haugen}}, \bibinfo {author}
  {\bibfnamefont {H.~H.}\ \bibnamefont {Andersen}}, \bibinfo {author}
  {\bibfnamefont {P.}~\bibnamefont {Kristensen}}, \ and\ \bibinfo {author}
  {\bibfnamefont {T.}~\bibnamefont {Andersen}},\ }\href {\doibase
  10.1103/PhysRevLett.80.684} {\bibfield  {journal} {\bibinfo  {journal} {Phys.
  Rev. Lett.}\ }\textbf {\bibinfo {volume} {80}},\ \bibinfo {pages} {684}
  (\bibinfo {year} {1998})}\BibitemShut {NoStop}%
\bibitem [{\citenamefont {Markson}\ \emph {et~al.}(2016)\citenamefont
  {Markson}, \citenamefont {Rittenhouse}, \citenamefont {Schmidt},
  \citenamefont {Shaffer},\ and\ \citenamefont
  {Sadeghpour}}]{markson_theory_2016}%
  \BibitemOpen
  \bibfield  {author} {\bibinfo {author} {\bibfnamefont {S.}~\bibnamefont
  {Markson}}, \bibinfo {author} {\bibfnamefont {S.~T.}\ \bibnamefont
  {Rittenhouse}}, \bibinfo {author} {\bibfnamefont {R.}~\bibnamefont
  {Schmidt}}, \bibinfo {author} {\bibfnamefont {J.~P.}\ \bibnamefont
  {Shaffer}}, \ and\ \bibinfo {author} {\bibfnamefont {H.~R.}\ \bibnamefont
  {Sadeghpour}},\ }\href {\doibase 10.1002/cphc.201600932} {\bibfield
  {journal} {\bibinfo  {journal} {ChemPhysChem}\ }\textbf {\bibinfo {volume}
  {17}},\ \bibinfo {pages} {3683} (\bibinfo {year} {2016})}\BibitemShut
  {NoStop}%
\bibitem [{\citenamefont {Camargo}\ \emph
  {et~al.}(2018{\natexlab{b}})\citenamefont {Camargo}, \citenamefont {Schmidt},
  \citenamefont {Whalen}, \citenamefont {Ding}, \citenamefont {Woehl},
  \citenamefont {Yoshida}, \citenamefont {Burgdörfer}, \citenamefont
  {Dunning}, \citenamefont {Sadeghpour}, \citenamefont {Demler},\ and\
  \citenamefont {Killian}}]{Camargo_2018}%
  \BibitemOpen
  \bibfield  {author} {\bibinfo {author} {\bibfnamefont {F.}~\bibnamefont
  {Camargo}}, \bibinfo {author} {\bibfnamefont {R.}~\bibnamefont {Schmidt}},
  \bibinfo {author} {\bibfnamefont {J.}~\bibnamefont {Whalen}}, \bibinfo
  {author} {\bibfnamefont {R.}~\bibnamefont {Ding}}, \bibinfo {author}
  {\bibfnamefont {G.}~\bibnamefont {Woehl}}, \bibinfo {author} {\bibfnamefont
  {S.}~\bibnamefont {Yoshida}}, \bibinfo {author} {\bibfnamefont
  {J.}~\bibnamefont {Burgdörfer}}, \bibinfo {author} {\bibfnamefont
  {F.}~\bibnamefont {Dunning}}, \bibinfo {author} {\bibfnamefont
  {H.}~\bibnamefont {Sadeghpour}}, \bibinfo {author} {\bibfnamefont
  {E.}~\bibnamefont {Demler}}, \ and\ \bibinfo {author} {\bibfnamefont {T.~C.}\
  \bibnamefont {Killian}},\ }\href@noop {} {\bibfield  {journal} {\bibinfo
  {journal} {Phys. Rev. Lett.}\ }\textbf {\bibinfo {volume} {120}},\ \bibinfo
  {pages} {083401} (\bibinfo {year} {2018}{\natexlab{b}})}\BibitemShut
  {NoStop}%
\bibitem [{\citenamefont {Koch}\ \emph {et~al.}(2018)\citenamefont {Koch},
  \citenamefont {Lemeshko},\ and\ \citenamefont
  {Sugny}}]{Koch_Lemeshko_Sugny_2018}%
  \BibitemOpen
  \bibfield  {author} {\bibinfo {author} {\bibfnamefont {C.~P.}\ \bibnamefont
  {Koch}}, \bibinfo {author} {\bibfnamefont {M.}~\bibnamefont {Lemeshko}}, \
  and\ \bibinfo {author} {\bibfnamefont {D.}~\bibnamefont {Sugny}},\ }\href
  {http://arxiv.org/abs/1810.11338} {\bibfield  {journal} {\bibinfo  {journal}
  {arXiv preprint: 1810.11338}\ } (\bibinfo {year} {2018})}\BibitemShut
  {NoStop}%
\bibitem [{\citenamefont {Stapelfeldt}\ and\ \citenamefont
  {Seideman}(2003)}]{Stapelfeldt_Seideman_2003}%
  \BibitemOpen
  \bibfield  {author} {\bibinfo {author} {\bibfnamefont {H.}~\bibnamefont
  {Stapelfeldt}}\ and\ \bibinfo {author} {\bibfnamefont {T.}~\bibnamefont
  {Seideman}},\ }\href {\doibase 10.1103/RevModPhys.75.543} {\bibfield
  {journal} {\bibinfo  {journal} {Rev. Mod. Phys.}\ }\textbf {\bibinfo {volume}
  {75}},\ \bibinfo {pages} {543} (\bibinfo {year} {2003})}\BibitemShut
  {NoStop}%
\bibitem [{\citenamefont {Engel}\ \emph {et~al.}(2018)\citenamefont {Engel},
  \citenamefont {Dieterle}, \citenamefont {Schmid}, \citenamefont {Tomschitz},
  \citenamefont {Veit}, \citenamefont {Zuber}, \citenamefont {Löw},
  \citenamefont {Pfau},\ and\ \citenamefont {Meinert}}]{Engel_2018}%
  \BibitemOpen
  \bibfield  {author} {\bibinfo {author} {\bibfnamefont {F.}~\bibnamefont
  {Engel}}, \bibinfo {author} {\bibfnamefont {T.}~\bibnamefont {Dieterle}},
  \bibinfo {author} {\bibfnamefont {T.}~\bibnamefont {Schmid}}, \bibinfo
  {author} {\bibfnamefont {C.}~\bibnamefont {Tomschitz}}, \bibinfo {author}
  {\bibfnamefont {C.}~\bibnamefont {Veit}}, \bibinfo {author} {\bibfnamefont
  {N.}~\bibnamefont {Zuber}}, \bibinfo {author} {\bibfnamefont
  {R.}~\bibnamefont {Löw}}, \bibinfo {author} {\bibfnamefont {T.}~\bibnamefont
  {Pfau}}, \ and\ \bibinfo {author} {\bibfnamefont {F.}~\bibnamefont
  {Meinert}},\ }\href {\doibase 10.1103/PhysRevLett.121.193401} {\bibfield
  {journal} {\bibinfo  {journal} {Phys. Rev. Lett.}\ }\textbf {\bibinfo
  {volume} {121}},\ \bibinfo {pages} {193401} (\bibinfo {year}
  {2018})}\BibitemShut {NoStop}%
\end{thebibliography}
